\documentclass{SciPost}

\hypersetup{
    colorlinks,
    linkcolor={red!50!black},
    citecolor={blue!50!black},
    urlcolor={blue!80!black}
}

\usepackage[bitstream-charter]{mathdesign}
\DeclareSymbolFont{usualmathcal}{OMS}{cmsy}{m}{n}
\DeclareSymbolFontAlphabet{\mathcal}{usualmathcal}

\fancypagestyle{SPstyle}{
\fancyhf{}
\lhead{\colorbox{scipostblue}
{\bf \color{white} ~SciPost Physics Codebases }}
\rhead{{\bf \color{scipostdeepblue} ~Submission }}

\fancyfoot[C]{\textbf{\thepage}}
}

\fancypagestyle{nostyle}{
\fancyhf{}
\lhead{}
\rhead{}

\fancyfoot[C]{\textbf{\thepage}}
}

\RequirePackage{comment}
\newcommand{\cost}{{\mathcal C}}
\newcommand{\fneg}{{\mathcal F}}
\newcommand{\Sbb}{{\mathbb S}}
\newcommand{\Hbb}{{\mathbb H}}
\newcommand{\Sevts}{$\Sbb$ events}
\newcommand{\Hevts}{$\Hbb$ events}
\newcommand{\Sevt}{$\Sbb$-event}

\newcommand{\twoton}{2\!\rightarrow\!n}
\newcommand{\twototwo}{2\!\rightarrow\!2}

\newcommand{\twotofour}{2\!\rightarrow\!4}
\newcommand{\twotofive}{2\!\rightarrow\!5}
\newcommand{\eebb}{$e^+e^-\!\rightarrow\!b\bar{b}$}
\newcommand{\eebbg}{$e^+e^-\!\rightarrow\!b\bar{b}g$}
\newcommand{\eebbbb}{$e^+e^-\!\rightarrow\!bb\bar{b}\bar{b}$}

\newcommand{\eebbmm}{$e^+e^-\!\rightarrow\!b\bar{b}\mu^-\mu^+$}
\newcommand{\eebbmmg}{$e^+e^-\!\rightarrow\!b\bar{b}\mu^-\mu^+g$}
\newcommand{\eebbmmgg}{$e^+e^-\!\rightarrow\!b\bar{b}\mu^-\mu^+gg$}
\newcommand{\eebbmmuu}{$e^+e^-\!\rightarrow\!b\bar{b}\mu^-\mu^+u\bar{u}$}
\newcommand{\eebbmmdd}{$e^+e^-\!\rightarrow\!b\bar{b}\mu^-\mu^+d\bar{d}$}
\newcommand{\ggttg}{$gg\!\rightarrow\!t\bar{t}g$}
\newcommand{\ggttgg}{$gg\!\rightarrow\!t\bar{t}gg$}
\newcommand{\ggttuu}{$gg\!\rightarrow\!t\bar{t}u\bar{u}$}
\newcommand{\guttgu}{$gu\!\rightarrow\!t\bar{t}gu$}
\newcommand{\pptt}{$pp\!\rightarrow\!t\bar{t}$}
\newcommand{\ppttj}{$pp\!\rightarrow\!t\bar{t}j$}
\newcommand{\ppttjj}{$pp\!\rightarrow\!t\bar{t}jj$}
\newcommand{\epem}{$e^+e^-$}
\newcommand{\alphaS}{\alpha_{S}}
\newcommand{\alphaSmZ}{\alpha_{S}(m_{Z})}

\newcommand{\eebbbbTOP}{\texorpdfstring{\eebbbb}{eebbbb}}
\newcommand{\eebbmmgTOP}{\texorpdfstring{\eebbmmg}{eebbmmg}}
\newcommand{\ppttjjTOP}{\texorpdfstring{\ppttjj}{ppttjj}}
\newcommand{\deltaTOP}{\texorpdfstring{\amcDELTA}{MC@NLO-Delta}}

\newcommand{\mytt}[1]{\textsc{#1}}
\newcommand{\mvf}{\mytt{MadVfold}}
\newcommand{\mgamc}{MG5aMC}
\newcommand{\amcDELTA}{MC@NLO-$\Delta$}
\newcommand{\vfoldftst}{\texttt{vfold5363}}
\newcommand{\vfoldfteo}{\texttt{vfold5381}}
\newcommand{\vfoldfcg}{\texttt{vfold5381cg}}
\newcommand{\vfoldscg}{\texttt{vfold7cg}}
\newcommand{\madmint}{\texttt{madevent\_mintMC}}
\newcommand{\vecsize}{\texttt{vecsize}}

\newcommand{\xsec}{\sigma}
\newcommand{\phis}{\Phi}

\newcommand{\Bterm}{B}
\newcommand{\Vterm}{V}
\newcommand{\Kterm}{K}
\newcommand{\phib}{\phis_B}
\newcommand{\phir}{\phis_R}
\newcommand{\dphib}{d\phib}
\newcommand{\dphir}{d\phir}
\newcommand{\xsecS}{\xsec_\Sbb}
\newcommand{\dxsecSdphib}{\frac{d\xsecS}{\dphib}}
\newcommand{\IKterm}{I_{\Kterm}}

\newcommand{\efo}{$8\!\times\!4\!\times\!1$}
\newcommand{\ooo}{$1\!\times\!1\!\times\!1$}
\newcommand{\tto}{$2\!\times\!2\!\times\!1$}

\newcommand{\dlopen}{\texttt{dlopen}}

\newlength{\tmtwidth}

\usepackage{multirow}

\begin{document}

\pagestyle{nostyle} 

\begin{center}
{\Large \textbf{\color{scipostdeepblue}
{\mvf: 
accelerating NLO 
event generation 
and \\ \vspace*{2mm}
reducing negative weights 
with SIMD vectorization and GPUs}
}}
\end{center}

\begin{center}\textbf{
Andrea Valassi\textsuperscript{1$\star$}
}\end{center}
\begin{center}
{\bf 1} 
CERN, Experimental Physics Department,
Geneva,
Switzerland
\\[\baselineskip]
$\star$ \href{mailto:email1}
{\small andrea.valassi@cern.ch}
\end{center}

\section*{\color{scipostdeepblue}{Abstract}}
\textbf{\boldmath{
NLO simulations are essential
for LHC physics analyses
but are expensive,
as they are not only slow
but also lead to
negative weights,
which imply the need to 
simulate much 
larger samples of events.
Folding is a powerful technique 
to reduce negative 
weights~but~is~itself computationally 
expensive.
In this paper I propose 
``vectorized folding''
as a new idea to speed up
these calculations
using SIMD and GPUs,
and I present 
its CUDACPP-based 
implementation 
for MadGraph
in \mvf,
including its extension
for unfolded~NLO event generation.
\mvf\ supports automatic 
code generation 
of many
physics processes 
in 
\mgamc\ and MadGraph7.
Preliminary results
show overall NLO speedups
over MadGraph7 of
6.9x and 10.3x for AVX512 and GPUs
with \efo\ folding
and of 1.9x and 1.5x 
without folding in \ppttjj.
This work is based on a test-centric,
LLM-assisted software development process,
which is described 
in detail in this paper.
}}

\vspace{\baselineskip}

\vspace{10pt}
\noindent\rule{\textwidth}{1pt}
\tableofcontents
\noindent\rule{\textwidth}{1pt}
\vspace{10pt}

\section{Introduction}
\label{sec:intro}
The simulation of physics processes
at the Next-to-Leading-Order (NLO) accuracy level
is essential in the analysis 
of the data collected by modern
High Energy Physics (HEP) experiments,
notably
those at CERN's 
Large Hadron Collider (LHC).
In this context, 
a widely used tool
is the Madgraph5\_aMC@NLO 
(\mgamc)~\cite{bib:mg5,bib:\mgamc}
physics event generator,
which implements 
NLO simulations 
of QCD processes
using the 
MC@NLO~\cite{bib:mcnlo,bib:mcnlo2}
matching approach.
NLO simulations are however expensive,
for at least two reasons:
to start with, generating 
one unweighted event with NLO accuracy
is intrinsically a complex and slow calculation
that requires a large CPU time;
in addition,
NLO simulations lead to the
generation of events 
with negative weights,
which result in additional costs 
in terms of computing resources
because much larger event
samples must be generated, 
stored, processed and analysed.
For a sample with a fraction $\fneg$
of negative-weight events,
in fact,
a factor 
$\cost(\fneg)\!=\!1/(1\!-\!2\fneg)^2$
more events must be generated~\cite{bib:delta},
with respect to the case when
there are no negative weights,
to reach the same statistical errors
in the simulated 
Monte Carlo (MC) samples.
The factor $\cost(\fneg)$ 
represents the relative cost
associated with a fraction $\fneg$
of events with negative weights.
Reducing this fraction 
is therefore
essential to 
reduce computing costs,
especially in view of the 
High Luminosity LHC (HL-LHC)
programme~\cite{bib:csbs}.
This has motivated a wide spectrum
of research activities\cite{bib:delta,
bib:danziger2020,bib:posresampler,
bib:nnresampler,bib:bornsp,
bib:arcane1,bib:esme,bib:palmer,
bib:bornsp2,bib:avbkk,bib:fvz},
investigating 
different approaches
to try and 
reduce negative weights
in NLO simulations.

The work I present here, 
in particular, started out 
as a spin-off of a 
research with other colleagues
on reducing negative weights
via Machine Learning (ML)
techniques~\cite{bib:avbkk,bib:fvz}.
As in that project,
I focus in the following
on reducing negative weights 
in \Sevts\ in 
MC@NLO-type calculations,
more
specifically in \mgamc\ and 
its upcoming 
successor MadGraph7~\cite{bib:mg7}.
In MC@NLO, 
computing NLO observables
involves the integration 
over two 
separate categories
of kinematic configurations:
\Hevts\ (for Hard MC evolution
including one prior NLO real emission)
and \Sevts\ (for Standard MC evolution
with no prior NLO real emission).
Both $\Hbb$ and $\Sbb$ events 
can have negative weights,
but their fractions,
their origin, and 
the possible ways to reduce them 
are generally 
different~\cite{bib:delta,bib:bornsp}.
For a $\twoton$ process at NLO,
\Sevts\ are $n$-body kinematic 
configurations described 
by a set of
$3n\!-\!4$ 
variables $\phib$,
while \Hevts\ are $(n\!+\!1)$-body 
configurations described 
by $\{\phib,\phir\}$,
where $\phir$ 
is the set of
three additional variables
describing the 
emitted parton.
A typical choice
for the latter
is $\phir=\{\xi,y,\phi\}$,
where the FKS~\cite{bib:fks} 
variables $\xi,y,\phi$ are
related to the energy, 
polar and azimuthal angle
of the emitted parton.
The contribution of \Sevts\ to
cross sections and other
NLO observables 
involves 
a differential cross section that 
is schematically
of the form
\begin{align}
  \left(\!\dxsecSdphib\!\right) 
  = \! 
  \left[
  \Bterm(\phib)
  \!+\! \Vterm(\phib)
  \!+\!\! \int\!\! 
  \Kterm(\phib,\!\phir)
  \dphir\right], 
  \label{eq:sint}
\end{align}
where the Born and Virtual
contributions $\Bterm$ and $\Vterm$
only depend on $\phib$,
but the 
counterterm~$\Kterm$
also depends on $\phir$
and must be integrated over it.
The main source of negative weights
in \Sevts, on which this work and that 
of Refs.~\cite{bib:avbkk,bib:fvz}
have focused,
is that the integral
$\IKterm(\phib)=\int\Kterm(\phib,\!\phir)\dphir$
of the counterterm $\Kterm$ 
over $\dphir$ at a given point $\phib$
cannot be computed analytically
and is instead 
estimated using MC methods,
from the 
value of $\Kterm(\phib,\!\phir)$
for a single, randomly drawn,
FKS configuration $\phir$.
This 
may lead to
negative weights because
$\Kterm(\phib,\!\phir)$ 
can be locally large and negative 
at $\phir$ even when its integral 
$\IKterm(\phib)$ is positive.
In this context,
a well-established strategy 
to reduce negative weights
in \Sevts\ is ``folding'',
which essentially consists
in estimating the integral
$\IKterm(\phib)$ from the average
of $\Kterm(\phib,\!\phir)$
over several FKS ``folds'' $\phir$
of a randomly drawn configuration,
rather than a single one:
\efo\ folding,
for instance, 
involves 8 values of $\xi$,
4 of $y$ and 1 of $\phi$.
Folding was originally introduced
in the MINT integrator~\cite{bib:mint}
designed for the POWHEG-BOX
generator~\cite{bib:powhegbox},
but has also been ported
into the 
MINT version used by
\mgamc~\cite{bib:delta}
for NLO calculations.

Folding is a powerful technique,
but is computationally expensive
as 
at each $\phib$
it involves the 
calculation of $\Kterm(\phib,\!\phir)$
for many values of $\phir$,
and this
internally
implies the calculation of
Born and Real 
matrix elements (MEs)
for various $n$-body
and $(n\!+\!1)$-body
configurations.
The achievable reductions 
in negative weights and
the corresponding increases
in CPU runtime 
for many 
physics processes
are described 
in Refs.~\cite{bib:delta,bib:bornsp}
and I will not include a similar
analysis here. 
In this paper, I 
focus instead on 
software optimizations
that significantly reduce
the computational cost of folding,
making it a 
more attractive
option to reduce negative weights.
The basic idea that was 
the starting point of this work,
and which represents the 
first main contribution
of this paper,
is that folding is a calculation
that lends itself naturally
to data parallelism 
with lockstep processing,
and can thus efficiently exploit
GPUs and SIMD on vector CPUs,
precisely because it involves
the calculation of the same function
$\Kterm(\phib,\!\phir)$
for different data points $\phir$
at each $\phib$: I will 
refer to this new
approach as ``vectorized folding''.
Compared to the ML techniques
in Refs.~\cite{bib:avbkk,bib:fvz},
which require 
process-specific 
investigations
across phase space
for different channels
and FKS sectors,
vectorized folding 
is 
a brute-force
computational approach
that makes no attempt
at understanding the physics
behind the appearance 
of negative weights;
its advantage, on the other hand,
is that its implementation
is largely process agnostic,
or can be delegated
to automatic code generation
where process-specific.

The second, 
related but distinct, 
contribution of this work 
is the implementation
of vectorized folding
in an extension 
of the 
\mgamc\ software,
which I dubbed
the \mvf\ package.
At a technical level,
this work mainly consisted
in refactoring
the MINT Fortran codebase 
used for NLO simulation
in \mgamc, and
in interfacing it
with the 
CUDACPP~\cite{bib:vchep2021,bib:ichep2022,bib:acat2022,bib:chep2023,bib:chep2024,bib:zw2025,bib:kspl}
plugin that had previously 
been developed 
for LO calculations
in \mgamc.
In the software workflow,
this involves
batching together the momenta
of several folds $\phir$
for a given \Sevt\ at $\phib$, 
and computing their 
Born and Real
contributions 
in one go
via 
CUDACPP. In this case,
in practice, a CUDACPP call
processes several folds of the
same event in parallel,
rather than several events. 

While \mvf\ initially targeted
vectorized folding alone,
in its software design
I kept in mind from
the beginning that
eventually this infrastructure
could be reused
for vectorizing NLO 
event generation also 
in the absence of folding.
This was indeed done
after the completion
of the vectorized folding,
and turned out to be
easier than anticipated.
In the software workflow,
in this case,
a CUDACPP call
processes one or more folds
of one or more events in parallel,
combining fold-level
and event-level
data parallelism.
This work, which is 
the third contribution 
of this paper, is conceptually 
related to other 
past and ongoing 
efforts~\cite{bib:zwnlo1,bib:zwnlo2,bib:zwnlo3}
for vectorizing 
NLO event generation
in \mgamc,
but represents 
an independent
piece of work that 
is based
on extending the
software developed for
vectorized folding in \mvf.

The fourth and final
contribution of this paper
is the description of 
the methodology
that was used 
in this work.
After the 
conceptualization of the
vectorized folding idea, 
the completion of
a fully functional
prototype for both folded
and unfolded NLO event generation,
during the initial phase
of the project described
in the first version of this
paper~\cite{bib:mvf1},
only took around six weeks
spread 
over May-September 2026,
while
six more weeks of work
spread over October-November 
were then
needed 
to achieve 
many additional
optimizations 
and software upgrades,
MC@NLO-Delta~\cite{bib:delta}
support and 
the backport
to code generation,
including support for processes
like \pptt\ which
have many physics subprocesses.
Such rapid progress
was only possible
thanks to a 
test-centric development process
assisted by Large Language Models (LLMs)
through the Claude Code~\cite{bib:claude}
user interface.
I acknowledge that,
without an LLM, this work
would certainly have taken me
many months if not years,
or would simply
never have started:
at the same time,
I believe that a well defined
test-centric development process,
combined with 
a clear high-level design
of the overall 
software architecture,
was the key enabling factor
to constrain and 
efficiently exploit
the power 
of LLM technologies
for these software developments.
I will give some details 
further below.

The use of an LLM,
in particular,
has made it possible
to turn what started
as a proof-of-concept prototype
into a software package
which includes code generation 
for a large number 
of physics processes
for MadGraph7
and could now start to 
be tested as one possible element
of a production workflow.
The already promising
performance numbers 
reported in Ref.~\cite{bib:mvf1}
have also further improved:
in the specific example
of a single \eebbbb\ channel,
which I have been using
as a standard candle
for optimization benchmarking,
the overall speedup
of \mvf\ with respect 
to the latest \mgamc\ release 
is now 
24x with GPUs 
with folding
and 3.6x 
without folding.
As discussed 
further below,
overall speedups are 
somewhat lower
with respect to
the upcoming
MadGraph7 software,
as this includes 
several new optimizations
over \mgamc, some of 
which~\cite{bib:pr260,bib:pr305}
directly benefitted from
those developed in \mvf.
The most recent development
has consisted 
in adding support for
and testing \ppttjj,
a physics process that
is directly relevant
to the LHC experiments.
In this case,
for the full event
generation workflow
including several 
partonic subprocesses
and channels therein,
the overall speedups
with respect 
to a recent MadGraph7 version 
are now 6.9x on AVX512 SIMD CPUs 
and 10.3x on GPUs 
with \efo\ folding,
and 1.9x and 1.5x without it.

The outline
of this paper is the following.
In Sec.~\ref{sec:meth},
I briefly describe the basic
design of the software
architecture and development process
for this project:
the aim 
is to 
give enough
information to understand
the results presented later on,
without giving 
too many technical details.
Sections~\ref{sec:fld}
and~\ref{sec:nof}
present the \eebbbb\ timing 
and speedup results
for NLO event generation
with vectorized folding (\efo),
and without folding (\ooo), 
respectively.
In Sec.~\ref{sec:delta}
I give further \eebbbb\ results
for calculations using the 
\amcDELTA\ prescription~\cite{bib:delta},
with and without folding.
Results for \eebbmmg\ are
given in Sec.~\ref{sec:eebbmmg}.
In Sec.~\ref{sec:ppttjj}
I present results for \ppttjj,
a much more complex physics
process directly relevant
to LHC physics analyses.
In Sec.~\ref{sec:conc},
I finally present my conclusions 
about this work.

\section{A test-centric,
LLM-assisted, software development process}
\label{sec:meth}

Designing an architecture
for functional and 
(where relevant) performance testing,
in my opinion,
is an even more important
part of software development
than working on the
actual implementation code.
Testing is the backbone
of development, not 
an optional component
that can be added a posteriori.
This was true before,
but I am even more convinced
of it now, in the new era
of LLM-assisted development.
In a nutshell, the software 
development process
that I have used 
for the project 
described in this paper
is the following:
\begin{enumerate}
\item Build an infrastructure
for quick, repeated,
functional and performance tests.
\item Design a high-level plan
to achieve NLO fold vectorization
(and, later on, NLO event
vectorization) progressively,
in several steps and sub-steps.
The first steps 
focused on providing 
missing functionalities,
the later ones
on squeezing out 
better performance.
\item Ask the LLM to analyse
the codebase, providing 
feedback on the current plan.
Are there technical obstacles 
to providing some of the
required functionalities?
What are the computational 
bottlenecks in the current
timing profiles, and what
can be done to address them?
If necessary, go back to point 2
and review the plan.
\item Ask the LLM to implement,
test and commit the next step
or sub-step of development.
\item Validate the code
developed by the LLM
by carefully rerunning
all functional 
and/or performance tests.
Inspect the code visually, too:
a superficial review
is enough, because 
at this stage
{\em the code is good 
if it passes the tests}.
(Note that some subtle 
issues did not immediately
show up in the tests
that I had set up,
such as some problems 
in parameter cards;
these were however
detected by the LLM,
which often ran more tests 
than those I had asked for.)
Go back to point 4 
in case of problems.
\item Review the overall
progress of developments,
with the help of the LLM.
Proceed to the next step
in point 4 as initially foreseen;
or, go back to point 3 
or point 2
if the plan must be adjusted.
In some cases,
go back to point 1
and adjust the testing
infrastructure.
\item At the end of the project
(or of a large chunk of it),
analyse 
and document the 
progress achieved and
carefully review the new code
before considering it 
for production use.
\end{enumerate}
The advantage of this 
development process
is that it allows
extremely rapid progress.
At the same time,
its disadvantage may be 
a perceived lack of control,
as the implementation details 
in each individual step
may be more difficult to grasp
by a human than by an LLM.
The key to ensuring that 
the process makes sense
lies in the first and second point,
namely the development
of a testing infrastructure
and the design of a high-level plan
for fold (and, later, event) vectorization.
These are briefly described below.

The current status of
development for the project
is now essentially point 7:
significant 
performance improvements have
been achieved
both for fold vectorization
and event vectorization
for selected physics processes,
as discussed 
in the following sections;
in addition, 
these
developments have been
fully backported to
automatic code generation,
using recent versions
of the \mgamc\ and MadGraph7
software code bases,
also (for MadGraph7 only) 
for processes
such as \ppttjj\ which 
involve several subprocesses
with different 
initial and/or final 
state partons
in each of them.
(Only one 
limitation
should be noted: 
\mvf\ does
not yet support processes
where the Born has multiple
coupling combinations,
such as those combining 
electroweak and QCD NLO corrections;
for these processes, 
the CUDACPP modes stop 
with an error, 
and stock MadGraph 
should be used instead.)
This paper is 
meant to document 
the progress achieved 
and the methodology used,
so that it can 
represent
a basis for a 
review of 
the software
before considering it
for production use or
upstream integration.
In this context,
a basic user guide,
available in 
Appendix~\ref{app:guide},
has been prepared
to allow anyone interested
to test \mvf\ and
provide feedback
on its functionalities
and limitations.

\subsection
{MadGraph and CUDACPP 
software version
evolution in \mvf} 
Before entering the details
of the \mvf\ testing 
infrastructure
and step-by-step timeline 
of developments,
it is useful to give
a high-level overview
of the MadGraph and
CUDACPP software versions
that were used 
in different phases
of this project.
In the following sections,
in particular,
performance results
are discussed for the
\mvf\ versions
referred to as
\vfoldftst, \vfoldfteo,
\vfoldfcg\ and \vfoldscg,
which are described below.

\paragraph{Initial developments
in \mgamc\ v3.6.3.}
This work
initially 
started from an 
``FVZ'' branch
of \mgamc,
which was 
built on top of
release v3.6.3
and included
a very large number 
of previous developments
targeting
the research of 
Refs.~\cite{bib:avbkk,bib:fvz}.
Some features of the
FVZ branch were 
very useful for the 
functional testing of MINT0 and MINT1,
because
that code is instrumented
to write out data dumps
with very detailed information
about the momenta and ME contributions
computed for each event and fold,
and those dumps can be used
to identify 
any differences
between backends
or regressions across
subsequent software 
versions over time.
The results presented 
in the first version
of this paper~\cite{bib:mvf1}
were all obtained using
that branch of \mgamc, 
which notably uses 
the release v1.00.02 
of the CUDACPP plugin.

\paragraph{\underline{\vfoldftst} ---
code cleanup in
\mgamc\ v3.6.3.}
After completing
the initial development phase,
the code base was 
extensively cleaned up.
All developments
were rebased on top
of a cleaner \mgamc\ code base,
which is essentially
stock v3.6.3,
with only minimal modifications to
provide optional FVZ data dumps,
while many other changes 
relevant to the FVZ project 
itself were removed.
The \mvf\ development 
history timeline was
also simplified.
In this version of \mvf,
CUDACPP was still an extension 
of the v1.00.02 release.

\paragraph{\underline{\vfoldfteo} ---
upgrade to \mgamc\ v3.8.1.}
As a next step,
\mvf\ was ported to the 
latest \mgamc\ version v3.8.1.
This also 
implied rebasing CUDACPP
extensions over
the latest version
of the plugin
(namely, 
the version tagged v1.01.02,
plus 12 upstream master 
commits from July 2026).
As discussed further below,
this had a positive
impact on performance
because 
the v1.01.02 release
of CUDACPP includes
the helicity 
streams~\cite{bib:kspl}
optimization.

\paragraph{\underline{\vfoldfcg} ---
code generation in \mgamc\ v3.8.1.}
Automatic code
generation was 
first implemented
in \mvf\ using
\mgamc\ v3.8.1.
The process-specific code
used to prepare the results presented
for this version \vfoldfcg\ 
of \mvf\ is
essentially the same as
in \vfoldfteo, but was
produced through code generation.

\paragraph{\underline{\vfoldscg} ---
code generation
upgrade to MadGraph7.}
Finally, the \mvf\ code
generation infrastructure
was ported to 
the new MadGraph7 framework.
(All developments
were built over the latest
main as of 8 Nov 2026 ---
commit fff90b78cd, 
which is more than 2000 commits
after the v0.2.1 tag:
the MadGraph7 repository 
has already moved a lot 
since that time,
but I have
not resynchronized
\mvf\ to those changes.)
In this context,
there was no need to
move to the new
matrix element engine
MadMatrix,
as the modified 
CUDACPP plugin
developed in \mvf\ for 
\mgamc\ could easily be integrated
into MadGraph7
with minimal modifications.
Only automatically generated
code was used for testing
physics processes in this 
version of \mvf.

\subsection{Functional and performance testing}
\label{sec:functest}

\paragraph{Physics processes.}
The \mgamc\ infrastructure
for NLO simulations 
relies on automatic code generation
for the desired physics process
and is based on the MINT integrator.
Once a physics process is specified,
the relevant Fortran code is generated,
notably including the
largely process-agnostic MINT components
and the process-specific
matrix element calculation kernels.
While in the \mgamc\ project,
to a large extent,
only the code-generating code is
hosted in a git repository,
the starting point
of all technical work 
for this project was,
quite simply,
to generate 
all relevant code 
for a few selected physics processes
and include it
in a 
git repository.
As in the case of CUDACPP 
developments~\cite{bib:kspl},
this was a key ingredient 
of the work I present,
which would have been 
impossible otherwise:
I stress, in particular,
that during all of my
initial developments,
notably the 
performance optimizations
for \epem\ processes,
I always asked the LLM
to analyse and modify
the existing Fortran 
(and/or CUDACPP) code
for one given physics process 
at a time, not the code 
generating code.
Each step of development
for the process considered
was then committed to the repository.
It was only much later
that I 
asked the LLM to 
backport 
all developments
to automatic code generation,
first in \mgamc\ and
later also in MadGraph7.
From then onwards,
a few additional enhancements,
notably those targeting $pp$ processes,
were instead developed
directly in code generating code.

More in detail,
in the course 
of this work I considered
several physics processes,
to test specific functional 
or performance issues,
but especially focused 
on the following:
\begin{itemize}
\item \eebb\ at NLO ---
the simple $\twototwo$ process
that I only used in the 
very first developments
focusing on building 
the basic functionality,
as it is fast to build and test;
\item \eebbbb\ at NLO ---
a more complex 
$\twotofour$ process,
which for a long time was
my standard candle
for benchmarking and 
optimizing timing profiles,
as it
involves
slower computations
of Born and Real MEs
with respect to \eebb;
\item \eebbmm\ at NLO ---
another complex 
$\twotofour$ process;
\item \eebbmmg\ at NLO ---
an even more complex 
$\twotofive$ process,
which unlike the previous
two involves
three separate Real 
processes instead 
of a single one;
\item \pptt\ plus
0, 1 or 2 jets at NLO 
(i.e. \pptt, \ppttj, \ppttjj) ---
realistic examples
of LHC processes
with a richer QCD structure,
which include more than one
\texttt{P0\_<subproc>}
subprocess 
each involving 
many Real processes
(example: six for \ggttgg,
nine for \ggttuu) and
the interface to 
parton distribution functions.
\end{itemize}
The first sets of results
I show in the following
refer to \eebbbb\ and
\eebbmmg\ at NLO,
based on 
automatic code generation
in \mgamc\ and/or MadGraph7.
To be precise,
in my tests 
of these processes
I focused on a single
\madmint\ application 
for only one ``channel''
(one ``GF'' subdirectory):
this involved
some changes 
in the relevant 
MadGraph 
Python scripts,
which were another small 
but important ingredient
for providing a faster
iteration cycle with shorter,
but still representative, tests.
The later results
I show for $pp$ processes,
conversely,
refer to the full MadGraph workflow
orchestrating several 
\madmint\ applications
for different channels
in different subprocesses.
More details about the
\mvf\ testing infrastructure,
and instructions 
on how to use it,
are provided in 
Appendix~\ref{app:guide}.

Apart from the 
exceptions described below and 
other settings
explicitly discussed in the text 
(folding, number of events, 
accuracy targets, \amcDELTA), 
all results in this paper
use the default MadGraph
param\_card and run\_card,
notably default SM
masses, widths and couplings, and
center-of-mass energies of 1 TeV
for \epem\ and 
13 TeV
for $pp$ collisions.
For $pp$ collisions,
MadGraph resets 
the strong coupling $\alphaS$ 
in the param\_card 
to the value of the PDF set 
in the run\_card. 
This 
is the default 
NNPDF2.3 NLO set (nn23nlo),
giving
$\alphaSmZ\!=\!0.119$ 
for the $pp$ results of 
Tables~\ref{tab:111CudaPPTTJJ}--\ref{tab:ppttjjsummary}. 
\mgamc\ 3.6.3 applied this reset 
also to \epem\ collisions, 
where no PDF is used; 
as a consequence, the results in
Tables~\ref{tab:fldCuda5363}--\ref{tab:nofCuda7cg} 
also use $\alphaSmZ\!=\!0.119$,
since the MINT1 grids were 
produced with this version 
and the same value 
of $\alphaS$ was kept
in MINT2 for consistency.
Tables~\ref{tab:fldCudaDelta}--\ref{tab:nofCudaEEMMBBG},
conversely,
use the default param\_card 
value $\alphaSmZ\!=\!0.118$
(except in the timing columns 
of Table~\ref{tab:delta} 
taken from Tables~\ref{tab:fldCuda7cg} 
and~\ref{tab:nofCuda7cg}). 
Finally, all results
use \texttt{parton\_shower=PYTHIA8}, 
which selects the Pythia8~\cite{bib:pythia} 
MC counterterms 
for MC@NLO calculations,
and \texttt{reweight\_scale}=False,
which disables scale reweighting.

\paragraph{MINT0, MINT1, MINT2.} 
NLO event generation 
in \mgamc\ is based on MINT
and includes three phases:
MINT0 (setting up grids),
MINT1 (computing upper envelope)
and MINT2 (generating events).
For most of my developments,
I used quick
tests that 
go through all three phases 
(with significantly relaxed
constraints on MINT0 and MINT1
physics accuracy targets,
to speed up the test iteration cycle).
This was 
necessary to
test and debug
the functionalities 
of all MINT phases,
which often go through
different branches of the code.
For 
performance benchmarking
and optimization,
however, 
I modified 
the test infrastructure
to only test MINT2 in
a gridpack-like mode,
using precomputed MINT0 and MINT1
integration grids 
(possibly with default
physics accuracy targets)
as well as pre-built 
libraries and 
binaries if possible.
The results presented
in the next sections,
in particular,
only refer to MINT2
event generation.

\paragraph{Functional testing: 
same random numbers, identical results.}
The most important ingredient
in the functional testing 
for this work was,
once again as previously 
in CUDACPP~\cite{bib:kspl},
the choice to fix 
the random numbers used
in the calculation,
so as to enforce 
the strict reproducibility of results.
In the current testing infrastructure
refined through several iterations,
in particular, two sets of tests
are executed in each development 
step or sub-step.
\begin{enumerate}
\item In each step,
physics results are produced
for a variety of Fortran-based
or CUDACPP-based modes and backends. 
These are described more 
in detail in Sec.~\ref{sec:arch}.
With the exception of the F0 mode,
which represents the 
upstream \mgamc\ release and uses
a different 
sequence of random numbers,
all other modes F1, F2, F and C
use the same random numbers 
and are expected to produce
the same results.
The comparison between 
Fortran modes F1, F2 and F
generally requires 
strictly identical results
(at least in \mgamc,
whereas in MadGraph7
these generally agree
to rounding precision,
with relative differences 
below about 1E-8 
for cross sections 
and 1E-6 for events);
conversely, their comparison
to the CUDACPP C modes
requires compatibility 
at the 1E-3 or 1E-5 level.
This ensures that all 
backends provide the same results:
I note, in particular,
that I do not show any 
physics distributions
in this paper, as I made 
no attempt to validate
the physics output 
of \mvf\ other than
comparing it to the 
Fortran results.
Specifically,
F0 results are assumed
to provide good enough physics
as they represent a \mgamc\ release;
F1, F2 and F results 
differ from those of F0 
in their random number sequences,
and are therefore also assumed 
to provide good enough physics.
C mode results, which are 
the real outcome of this work,
are validated by comparing 
them to those of the F modes.
The total cross sections 
of F0 and F1/F2/F modes 
have also been occasionally 
compared within statistical 
fluctuations in dedicated
longer tests.
\item In each step,
except those involving
profound changes 
in the use of random numbers,
the results for each mode
are also similarly compared
to the corresponding results
from a previous step.
This ensures that there
are no unforeseen code 
regressions.
\end{enumerate}
In the 
testing infrastructure
for detailed time benchmarking,
where only MINT2 is executed,
the main tool for validating 
physics outputs has been 
the comparison of LHE event files.
In the quick
validation tests
where
also MINT0 and MINT1
are 
executed, the FVZ dump 
files~\cite{bib:avbkk,bib:fvz}
produced by MINT1 
have also been
routinely compared.
(Concerning 
the achieved 
compatibility at the
1E-3 or 1E-5 
level between modes F and C,
especially in the 
LHE files,
I should mention that 
this is also, to some extent,
a question of luck. 
Small numerical
differences between
modes or backends,
in fact, might have
been enough to tilt
a hit-or-miss decision
one way or another,
but this was never
identified in my tests.)

\paragraph{Timing profiles and performance testing.}
Performance testing for this work
went through a few iterations,
but eventually converged on 
the use of 
\texttt{log\_MINT2.txt} logfiles,
which by default already
contain a time profile
for the most important
components of MINT2 event generation.
All the tables presented 
in the following sections
have been produced by ad-hoc
scripts parsing these files.
The information contained
in these files was extensively reviewed,
and in a few cases modified
to harmonize the meaning
of time splitup across
different modes. 

\subsection{Software architecture:
batching and reproducibility}
\label{sec:arch}

As hinted above,
the two basic design ideas
on which the \mvf\ software
was gradually built
are essentially batching
and reproducibility:
the aim was that of
batching several folds
(later, events) to
allow the parallel
calculation of MEs
via CUDACPP and
exploit SIMD and GPUs,
while at the same time
ensuring that results
are reproducible,
when starting from the
same random number seeds,
independently of
if and how this batching is done.
As described below,
this was eventually achieved
with the help of the LLM
using a two-walk workflow,
which also allowed
the comparison of different
(e.g. scalar and batched)
computation modes.

\subsubsection{A two-walk 
gather-and-replay workflow}

At the very beginning of this project,
after setting up a well-encapsulated
LLM sandbox using Apptainer containers,
I asked the LLM to do an initial
evaluation of the possibility
to batch ME calculations
for vectorized folding
by offloading them to an external 
vectorized engine like CUDACPP.
The main challenge I saw,
and the main question 
for the LLM, was whether
it would be possible 
to disentangle a reentrant
calculation with well defined
inputs and outputs
from the stateful
calculations currently
involving several
Fortran COMMON blocks~\cite{bib:ecfa2023}.
The software architecture
I had in mind, 
which I proposed to the LLM,
consisted in a 
two-walk design
where all relevant inputs
for the batched calculation
(notably event momenta,
but possibly more) are recorded
during a first pass, 
possibly stopping short
of ME calculation themselves,
while these values are then 
used as inputs 
for the actual batched calculation
delegated to CUDACPP.

As discussed below,
two different variants
(``A'' and ``B'') 
of this two-walk architecture 
were implemented over time,
both of which however
share the same basic workflow:
\begin{itemize}

\item {\bf Fixed batch capacity.}
The scalar processing
of one fold at a time
is transformed into 
a loop over batches of folds.
The maximum number of folds
in each batch is the 
product of the 
folds per event
times parameter ``kbatch'',
which represents the 
maximum number 
of ``kpoints'' in a batch.
A kpoint is simply 
a weighted event,
or a candidate event
for unweighting,
for which all relevant
MEs must be computed.
For a calculation with 
\efo\ folding and kbatch=64,
for example,
batches must be properly
sized to contain 
up to 2048 folds.
(To be precise,
the kbatch parameter
was only introduced
in strategy B,
when vectorized folding
was extended to
vectorized event generation;
the general idea, however,
also applies to strategy A,
where kbatch was implicitly
hardcoded to 1).

\item {\bf Gather walk.}
In a first ``gather'' walk,
all the inputs
that are needed to compute the
Born and Real MEs
for the folds in a batch,
notably momenta
and running couplings,
are determined
from the relevant random numbers
and are then stashed 
in FIFO containers.
Any other quantities
which are also computed 
as a by-product of the
determination of
momenta and couplings
are not saved.

\item{\bf Batched ME evaluation.}
The MEs for all folds
in a batch are computed.
As noted below,
this is done differently 
in mode F,
where internally
each ME is computed 
sequentially in Fortran,
and in mode C,
where MEs are computed 
in parallel with one
or more CUDACPP calls.
More precisely,
in mode C
the folds in the batch
are partitioned in
groups of \vecsize\ elements, where 
the parameter \vecsize\ defines 
how many folds
at a time are passed to CUDACPP;
if there are not enough
folds to fill the last chunk
of \vecsize\ elements,
this is padded 
with dummy data.
Each CUDACPP call,
in turn, may actually 
involve a loop over
smaller groups of folds:
in an AVX2 double-precision
calculation with \vecsize=32,
for instance, 
a batch of 2048 folds
is processed by 64
CUDACPP calls per type of ME,
where the 32 folds 
in each CUDACPP call
are processed in 8 iterations 
over AVX2 vectors of 4 doubles.
The simple description above
applies to Born batching,
as well as to Real batching
when there is a single 
library, but things are more complex
when there are several 
Real libraries
(e.g. in \ppttjj,
where each subprocess
has 3 to 9 
libraries);
in that case,
the Real batch 
is first partitioned 
by 
library,
each chunk padded.

\item {\bf Replay walk.}
In the second ``replay'' walk,
the same processing
which was used to generate
momenta from random numbers
in the gather walk is repeated,
to fill in not only momenta
but also all other relevant
variables that are needed
to compute event weights.
The difference with respect
to the stock scalar workflow
is that MEs are read off
the results of the batched
evaluation, instead of
being computed one by one.

\end{itemize}
The two variants of this workflow
that were implemented over time
mainly differ in the way
random number generation (RNG)
is handled to achieve reproducibility:
\begin{itemize}
    
\item {\bf Old strategy ``A'': 
RANMAR save/restore/redraw.}
The initial implementation
of the two-walk 
workflow was 
the following:
save the RANMAR state
before the gather walk;
draw all relevant random numbers
in the gather walk;
compute MEs in batch;
restore the RANMAR state
before the replay walk;
redraw the same random 
numbers in the same order
during the replay walk.
This strategy 
was simple and suboptimal,
but effective.
In fact,
this first phase of the project
was very useful, quickly resulting
in a successful
proof of concept 
of vectorized folding
with CUDACPP offload
for \eebb.
In particular,
the basic technicalities
of linking CUDACPP 
and Fortran MINT code
were 
easily solved
using a \dlopen-based approach.
Other basic issues
such as the matching
of the Fortran and CUDACPP
conventions 
for particle ordering
and channel numbering
were also fixed.
Since all Fortran
calculations are performed
in double precision,
CUDACPP builds were
configured to use
double precision
\texttt{FPTYPE=d},
rather than the default
mixed precision
\texttt{FPTYPE=m},
where Feynman diagrams
are computed in double
precision but colour sums
in single precision.
(With the default mixed precision, 
in fact, a few lines
in the FVZ data dumps
differed between Fortran 
and CUDACPP by more than 1E-3,
which broke the 
strict reproducibility 
on which functional tests rely:
double precision restored it.)
An uninitialized-variable 
bug 
that was hidden in the Fortran code 
since a long time was fixed.
Eventually, 
not only \eebb\ but
also \eebbbb\ and
\eebbmm\ were fully debugged.
Debugging the two latter
processes
was especially important
to have a first look
at timing profiles and
possible performance optimizations.

\item {\bf New strategy ``B'': 
pre-allocated random number slots.}
While the RANMAR save-and-restore 
strategy was extremely useful
to develop the first 
proof-of-concept prototype of \mvf,
it soon became clear
that its intrinsic overhead 
would prevent large speedups, 
because the gather walk must go 
through every step 
where random numbers are drawn.
I therefore decided to 
move to a new RNG approach
based on pre-allocated
random number slots,
which is not affected by this issue
but is still reproducible
and gives identical results
in scalar and batched processing.
The 
advantage of this approach
is that the gather and replay walks
become decoupled from the RNG sequence,
making it possible 
to use a leaner gather walk
which skips any work that
is not directly relevant
to the stashing of
momenta and couplings.
The main problem with preserving the 
reproducibility of results
when moving from a scalar
and sequential calculation
to one involving batching, 
in fact, is the following.
In a scalar 
calculation, the number of
random numbers drawn for 
a specific fold 
is not a constant.
(I use ``fold'' rather than ``event''
in this paragraph
as this design initially targeted
fold vectorization, but the
idea also applies to
event vectorization.)
This is the case, 
in particular,
if a workflow involves
two sequential operations 
A and B, where A always draws
a random number but B only 
draws one depending
on the results of A.
To address this issue, however,
it is enough to do the following:
for each fold, 
two random numbers 
are preallocated,
one for A and one for B.
Depending on the result
of A for a specific fold, 
the random number preallocated
for B may be used or thrown away,
but the sequence 
of random number drawing
is fully predictable 
independently of whether
A and B are scalar 
or batched for several
folds 
at a time.
Technically, the solution
has consisted
in replacing all relevant
\texttt{ran2()}
calls in the code
by a different function
\texttt{vfold\_newran2(islot)}, 
where a random number slot 
is specified for each call. 
In the gather walk,
the slots of a fold 
are filled once per fold
using \texttt{ran2()}
and saved in a per-fold history; 
in the replay walk, 
the slots of each fold 
are reloaded from this history, 
so that no random numbers are drawn 
and every drawing site 
reads the same value 
as in the gather walk.
Event (kpoint) slots, 
filled once per kpoint, 
are handled in the same way; 
their per-kpoint history was added 
when vectorized folding
was extended to
vectorized event generation.
Where a drawing site is called 
several times within a fold, 
once per FKS sector, 
the first sector 
reads the fold slot,
which contains the value
computed by \texttt{ran2()},
while each further sector 
reads an independent value, 
computed using 
a counter-based RNG 
construction~\cite{bib:philox}, 
by applying two rounds of 
the SplitMix64 mixing 
function~\cite{bib:splitmix}
to the bits of a dedicated 
fold slot combined 
with the slot identifier 
and the sector index:
this requires no additional 
RANMAR draws and is reproducible 
when a fold is replayed.
(This handling of RNG
for different FKS sectors
was introduced 
very late in \mvf\ to
replace a much simpler
implementation where
each FKS sector in a fold
was reusing the same 
random number, which
could lead to unwanted
correlations:
if needed, this counter-based
implementation could be
replaced by one purely
based on RANMAR slots instead).
It should be noted that
there are still a few
bare \texttt{ran2()}
calls in the code,
at locations where
this does not prevent
the reproducibility
of results across
scalar and batched modes.
These bare \texttt{ran2()} calls
are made outside 
the gather/replay walks, 
once per event, 
from the original RANMAR stream; 
in MINT2, the event candidates 
of each type are drawn 
from separate RANMAR streams, 
so that batching them 
does not change the sequence.

\end{itemize}

\subsubsection{Calculation modes}
\label{sec:modes}

The new slot-based RNG design 
also led to a rationalization of 
calculation mode definitions.
In particular,
it is at this stage that 
I introduced the difference
between modes F0, F1 and F,
to 
distinguish different types 
of Fortran-based calculations 
from the CUDACPP-based mode C.
An additional mode F2
was 
added later on,
as new opportunities
for Fortran code optimizations
were identified and implemented.
Initially,
notably
in the first phase
of the project described in 
Ref.~\cite{bib:mvf1}
and in the ``step10'' 
results presented 
in the following sections 
for \vfoldftst\ and \vfoldfteo,
the following modes
were defined.
(The choice is done at runtime
rather than at compile time,
based on the 
environment variable
\texttt{ENV\_VFOLD\_ME\_MODE} 
--- environment variables
also control
other optimizations,
as described in
Appendix~\ref{app:guide}.)
\begin{itemize}

\item {\bf Mode F0: 
default RNG, 
no optimizations,
scalar API, 
sequential Fortran MEs.}
This 
represents
the calculation path
of stock 
\mgamc\ (except for minor
bug fixes which do
not affect timing).
In particular:
random number drawing
uses unmodified
\texttt{ran2} calls;
\mvf\ adds 
no Fortran
optimizations
like Born memo or fused BSF;
MEs for different folds
(or events) are
computed via a scalar API
with sequential Fortran calls.

\item {\bf Mode F1: 
slot RNG, 
no optimizations,
scalar API, 
sequential Fortran MEs.}
Mode F1 differs 
from mode F0 
only in its use of random numbers 
(slot-based RNG and, in MINT2, 
separate RANMAR streams for the 
event candidates of each type).
This mode adds no
optimizations,
and MEs 
are still
computed via a scalar API
with sequential Fortran calls.

\item {\bf Mode F2: 
slot RNG, 
Born memo +
fused BSF,
scalar API, 
sequential Fortran MEs.}
Mode F2 
uses the same RNG as F1, 
but 
it includes
the ``Born memo'' 
(as of step7 in
development, see below)
and ``fused BSF''
(as of step9)
optimizations
in the Fortran code.
MEs are still
computed via a scalar API
with sequential Fortran calls.

\item {\bf Mode F: 
slot RNG, 
Born memo + fused BSF,
batched API, 
sequential Fortran MEs.}
Mode F uses the 
same slot-based RNG and 
Fortran optimizations as F2:
it differs from it
in that it proceeds 
through a batched API,
which, however,
still implements the
ME calculation as a loop
over sequential Fortran calls.
The main interest
of F is that
its comparison to F2
provides an estimate
of the timing overhead
of the batching
of folds (or events),
which is a prerequisite 
for CUDACPP offloading 
in mode C.

\item {\bf Mode C: 
slot RNG, 
Born memo + fused BSF,
batched API, 
parallel CUDACPP MEs
(and other optimizations).}
Mode C represents 
the final outcome 
of this work.
Like mode F,
it uses slot-based RNG,
a Born memo (in Fortran),
fused BSF (in CUDACPP,
as of step9)
and a batched API.
It differs from F
in that the batched API
is implemented
using a truly parallel
calculation of MEs
in CUDACPP.
It also contains
two additional
``sborn'' (step9)
and ``gscale'' (step10)
optimizations
in CUDACPP and Fortran,
respectively.
\end{itemize}
A few additional remarks
about the modes above
are also 
appropriate:
\begin{itemize}

\item {\bf Mode F0*.}
In the tables 
in the following sections,
timing results for mode F0
are given in a column F0*.
In each table,
all 
numbers 
refer to MINT2-only 
event generation,
starting from the
same optimized 
MINT0/MINT1 grids.
This
ensures that the 
same unweighting efficiency
(the same 
hit-or-miss ratio)
is {\em expected} for 
all modes.
However, the use of
different random numbers
in F0 (default RNG) and
in F1/F2/F/C (slot-based RNG)
implies that a different number
of event candidates or ``kpoints''
must be evaluated 
in the two cases
to reach the same
target number
of unweighted events 
(\texttt{NEVENTS} 
in the tables).
The F0 timing values
are therefore rescaled
to the kpoints
of F1/F2/F/C
to allow a direct comparison,
yielding the ``F0*''
timing values.

\item {\bf Vectorization speedup.}
The comparison 
of mode F2 and mode C
was originally meant to provide 
a direct estimate
of (fold and/or event) 
vectorization speedup.
In particular, 
F2 provides a better reference 
for this comparison
than mode F
(or mode C/none),
because it does not
include the overhead 
from batching.
Initially, in fact, 
mode F2 only differed from mode C
in that it used a scalar API
and sequential Fortran MEs
instead of a batched API
and parallel CUDACPP MEs.
Over time, however,
several additional optimizations
have made their way both 
into the F2 and C modes
in different steps of development,
making their comparison
no longer 
as straightforward 
to interpret.

\item 
{\bf Overall \mvf\ speedup.}
The comparison 
of mode F1 and mode C,
conversely,
provides a
non-controversial
estimate
of the overall speedup
achieved in this work
by the \mvf\ software
over stock \mgamc\ or
stock MadGraph7.
This includes 
not only the speedup
coming from 
vectorization alone,
but also that coming
from all other additional
optimizations 
(Born memo, fused BSF, 
sborn, gscale, 
faster timers, 
further memos 
--- enabled or disabled
in different steps 
of development)
in Fortran and CUDACPP code.
\end{itemize}
The description 
above also applies 
to the more recent 
developments
from ``step11'' to ``step17''
(including the results
presented in the
following sections
for \mvf\ versions
\vfoldfcg\ and \vfoldscg),
but a few additional
comments are appropriate
in that case:
\begin{itemize}

\item {\bf CPU timers.}
As of step11, 
modes F2 and above include
one additional optimization:
to reduce the overhead 
of the 
time profiling instrumentation 
via the Fortran 
\texttt{cpu\_time} subroutine,
whose implementation in
\texttt{libgfortran} involves
an expensive 
\texttt{getrusage} syscall,
\mvf\ overloads 
\texttt{\_gfortran\_cpu\_time\_4}
with its own implementation,
which can be configured
to use cheaper alternatives.
In the tests presented
in the following,
the default \texttt{getrusage} 
is used in modes F0 and F1,
while all other modes use
\texttt{clock\_gettime(CLOCK\_MONOTONIC)} 
through the Linux vDSO,
which involves no syscall.
An even faster implementation
using \texttt{rdtsc},
similar to what I had
used for
profiling 
CUDACPP~\cite{bib:kspl},
is also available
but was not used 
in the results I report.
Note that \texttt{clock\_gettime} 
measures wall-clock time, 
while \texttt{getrusage} 
measures CPU time: 
these are essentially the same 
for single-threaded CPU runs, 
but may differ in other cases.

\item {\bf Fortran optimizations}. 
As of “step11”, modes F2, F and C 
include new Fortran optimizations, 
absent in F0 and F1: 
memos which avoid repeating 
the evaluations of 
Born-soft-virtual contributions, of 
cuts and of 
phase-space mappings
for the same kinematics, 
and skipping 
the computation of reweighting 
information 
if not necessary.
Conversely, 
the fused BSF Fortran optimization 
was removed from 
modes F2 and F, 
which now compute 
colour-linked Borns 
in stock Fortran code
(which in MadGraph7
includes a similar optimization:
see below).
In mode C, colour-linked Borns 
are still computed in CUDACPP 
together with the Born 
(fused BSF in CUDACPP), 
and 
are used whenever 
SBORN is served from CUDACPP 
(the “sborn” optimization); 
only the Fortran code 
retrieving 
them from CUDACPP 
was made process-independent. 
In the results for 
\vfoldfcg\ and \vfoldscg, 
modes F2 and F 
thus 
include the Born memo 
and the step11 memos,~but 
no longer fused BSF, 
while mode C also includes 
fused BSF in CUDACPP, 
sborn and gscale.

\item {\bf MadGraph7 optimizations.}
As noted above, the 
difference between
modes F1 and F2 is the
absence in the former
of a few Fortran optimizations
introduced by \mvf.
In \mgamc, where 
these 
were initially developed,
this results 
in a large speedup
of mode F2 over F1,
even without 
any CUDACPP offloading.
The situation 
is different in MadGraph7,
where the new default Fortran
code 
includes various 
optimizations~\cite{bib:pr260,bib:pr305}
that are similar to --- 
and were partly derived from 
--- those reported 
for \mvf\ in Ref.~\cite{bib:mvf1}
(Born memo, fused BSF and gscale).
The speedup of F2 over~F1
in MadGraph7 is therefore
lower than in \mgamc,
as the same
opportunities for
software optimizations
have already been
exploited in 
the new stock~F1.
In the \mvf\ port to MadGraph7, 
these Born optimizations 
are kept in modes F0, F1, F2 and F, 
which therefore use the same 
Fortran Born code as stock MadGraph7.
(To get a rough estimate of
the changes 
with respect to \mgamc,
a new mode F1[u] is also
used in some tests,
where 
some of the stock
MadGraph7 optimizations
are disabled.)
In mode C, conversely, the Born 
and its colour-linked components 
are computed in CUDACPP (with 
the \mvf\ sborn and gscale optimizations), 
so the stock MadGraph7 Born code 
is only used as a fallback. 
As a consequence, mode C 
in MadGraph7 is essentially 
the same as in MG5aMC, 
while the three Fortran modes 
agree with each other 
only to rounding precision 
(see Sec.~\ref{sec:functest}).

\item {\bf Ninja and COLLIER.}
In MadGraph,
Virtual contributions
are computed~\cite{bib:mgnlo} 
via MadLoop~\cite{bib:madloop},
which internally 
delegates part 
of the calculation
to other packages 
including Ninja~\cite{bib:ninja}
and COLLIER~\cite{bib:collier}.
By default,
Ninja is used first,
with COLLIER as 
fallback
in case of issues.
In the tests described 
in the following sections,
in the case of a specific
\eebbmmg\ channel
(which is not the one 
I finally used to report
results) it
was observed that the 
Virtual calculation
was particularly slow
because both Ninja
and COLLIER were
used for many events.
With Ninja first, numerical 
differences between modes F and C, 
or between CUDACPP backends, 
also changed which of 
the two libraries was used 
for some events, making 
results less reproducible.
As a consequence,
a non-default
configuration was
used for this process,
where COLLIER is used first.
For comparison,
in F1 mode both 
options were tested,
the default one
with Ninja first 
and the custom one
with COLLIER first:
these are indicated
as F1[n] and F1[c]
in the tables of results.

\end{itemize}

\subsection{High-level development timeline}
\label{sec:steps}

Much more in detail,
the development and optimization
of the new pre-allocated RNG strategy
ended up including
the following steps.
Initially, 
I had asked the LLM
to implement a simpler 
high-level plan
which only included 
four points
(corresponding to
Step1, Step2, Step3+Step4
and Step5+Step6 
in the list below),
for fold vectorization
in \eebb\ alone:
over time, other steps were added,
and the focus shifted 
to other physics processes,
as the project progressed.
The initial developments
based on \mgamc\ v3.6.3,
for which results were
presented in 
Ref.~\cite{bib:mvf1},
only included steps 
up to step10 in the
following list;
the following steps
up to step17 included,
and the corresponding
results, are presented
here for the first time.
\begin{itemize}

\item {\bf Step0: Fortran-only.}
Only the default as-is Fortran code 
generated by \mgamc\ exists. 
Until further notice,
focus only on fold vectorization in
the \eebb\ physics process.

\item {\bf Step1: mode F0, F, C scaffolding.}
Add environment variables
to define separate modes F0, F and C.
At this stage, however,
they all follow 
the same processing path (mode F0).
Until further notice, test mode C only
against the none, sse4 and avx2 CPU backends.

\item {\bf Step2: slot-based RNG in modes F and C.}
Identify all relevant 
random number drawing sites
in the code.
Replace \texttt{ran2()}
by \texttt{vfold\_newran2(islot)}.
Internally, the latter
goes back to \texttt{ran2()} in mode F0,
while it uses slot-based RNG 
in modes F and C.

\item {\bf Step3: 
batch Born calls 
in modes F and C.}
Pre-draw
random numbers
for a batch 
of the folds in one event,
and pre-compute particle momenta
based on them.
Move the calculation
of Born matrix elements 
to a batched API:
internally,
in each batched-API call
mode F loops over all folds 
and computes them sequentially.
Real matrix elements
are still computed
with a scalar API.
Mode C still 
coincides with mode F
and uses Fortran MEs.
(During this step,
the Virtual calculation
was debugged 
to understand a regression
between the scalar 
step2 and the 
batched 
step3: this resulted in the 
addition of a warm-up fix
for the first call
of the \texttt{BinothLHA} function,
and the definition
of a new RNG slot 
for the helicity choice 
in the Virtual calculation.)

\item {\bf Step4: 
batch Real calls 
in modes F and C.}
Move the calculation
of Real matrix elements 
to a batched API:
internally,
in each batched-API call
mode F loops over all folds 
and computes them sequentially.
Mode C still 
coincides with mode F
and uses Fortran MEs.

\item {\bf Step5: 
offload Born MEs to CUDACPP
in mode C.}
Internally,
in each batched-API call
for Born MEs,
move C mode to 
vectorized ME calls
in CUDACPP,
using \dlopen\ to locate 
the C++-to-Fortran bridge; 
mode F still loops over 
folds and computes them sequentially.

\item {\bf Step6: 
offload Real MEs to CUDACPP
in mode C.}
Internally,
in each batched-API call
for Real MEs,
move C mode to 
vectorized ME calls
in CUDACPP,
using \dlopen\ (one process: \eebbg);
mode F still loops over all folds 
and computes them sequentially.

\item {\bf Step7: 
functional and performance tests
for \eebbbb.}
Backport the \eebb\ code to \eebbbb:
until further notice,
focus on this process alone.
Fix functional bugs 
affecting the new process.
Test CPU performance
and implement some optimizations,
mainly in counterterm calculations.
(During this step,
a lot of progress took place.
Introduced scalar mode F1
to distinguish it from
batched mode F.
Identified performance penalty
from colour-linked Born functions
--- BSF hereafter ---
like \texttt{SB\_SF\_001}.
Fixed 
performance penalty
from unnecessarily repeated
Born evaluations:
introduced mode F2 
with a ``Born memo''
performance fix,
to distinguish it 
from mode F1
without the fix.)

\item {\bf Step8:
functional and performance tests
for \eebbmmg\ and GPUs.}
Backport the \eebbbb\ code to \eebbmmg.
Identify the need for three
separate Real processes from CUDACPP
(\eebbmmgg, \eebbmmuu, \eebbmmdd),
with three separate ``common'' libraries
(the number of couplings needed
is different in the three cases).
Test on AVX512.
Test 
on GPUs:
add minor fixes 
to allow \dlopen\ of 
eight CUDACPP
libraries (1 Born, 3 Real, 4 common)
and prevent crashes during bridge deletion.
A posteriori, 
extend RNG slot definitions
to allow event vectorization
in addition to fold vectorization
(this was needed 
to prepare the step8 
results previously presented 
in Ref.~\cite{bib:mvf1},
but no step8 results
will be reported here).

\item {\bf Step9(1):
extend fold vectorization 
to event vectorization.}
Modify slot-based RNG
to accommodate event batching.
(This change was also backported 
to step8n~\cite{bib:mvf1}.)
The main difference
with fold vectorization,
in this context,
is that event vectorization
must also properly
account for the 
accept-reject decision
on each individual event
(or ``kpoint'' in the
internal code nomenclature:
a phase space point
for which a weight is computed, 
i.e. 
a candidate
to become an unweighted
event in the LHE file
if it passes the 
hit-or-miss decision);
in particular,
the random number used 
in the hit-or-miss decision 
must become an event slot, 
and the event candidates 
of each type ('virt', 'novi' or 'born') 
must be drawn from 
separate RANMAR streams.

\item {\bf Step9(2):
other optimizations for \eebbbb,
including 
CUDACPP CPU extensions.}
Implement the ``fused BSF''
optimization in Fortran:
this consists in evaluating 
all ten colour-linked Borns 
in a single pass over helicities
(an algebraic factorization 
at one phase-space point, 
not a SIMD vectorization).
Later on, reimplement 
fused BSF 
and add the ``sborn''
optimization in CUDACPP C++ code:
using a modified \dlopen\ bridge,
CUDACPP may now return not only 
the MEs but also the BSF,
as well as the 
per-diagram (AMP2)
and per-colour-flow (JAMP2)
helicity sums
of squared amplitudes.
These optimizations
in Fortran and CUDACPP
have been implemented
only in \eebbbb.
Introduce new 
compile-time
knobs to properly size
memory arrays
for GPU offloading.

\item {\bf Step10:
other optimizations for \eebbbb,
including 
CUDACPP GPU extensions.}
Implement the
fused BSF and sborn
optimizations 
in CUDA.
Add the optional ``gscale''
Fortran optimization in C mode, i.e.
skip Born calculations at the
Real-emission coupling,
rescaling those at the
Born coupling instead:
this is 
process-specific
as it is only possible 
when the Born 
has a single 
coupling combination,
which the code checks at runtime.
{\em (This was the final
development step 
at the time of 
Ref.~\cite{bib:mvf1}:
the results previously
presented there 
for \eebbbb\ will
not be reported again
in the following,
where instead the very
similar results 
obtained after cleanup
in version vfold5363
are presented.)}

\item {\bf Step11:
further optimizations, 
\amcDELTA\ tests
and v3.8.1 port
for \eebbbb.}
Clean up the code for 
\mgamc\ v3.6.3
(\mvf\ \vfoldftst).
Implement new vDSO and
\texttt{rdtsc} timers
with reduced overhead.
Implement additional memos (bornsoftvirtual, 
passcuts, phase space) 
and further optimizations 
related to reweighting 
and shower scales. 
Implement event 
vectorization also in MINT0. 
Separately time the 
GPU reset overhead. 
Fix Fortran alignment 
to reduce unwanted 
jitter of timing results. 
Test \amcDELTA\ and 
add fixes, optimizations 
and timers relevant to it.
Port the code to 
\mgamc\ v3.8.1
(\mvf\ \vfoldfteo).
Remove 
fused BSF in Fortran,
avoiding the need to
backport it 
to code generation,
since this is now better
handled by the sborn
optimization in CUDACPP
(this implies a $\sim5\%$
slowdown in mode F2
in \mgamc, but it 
should be noted that a
similar optimization
is now included
in stock MadGraph7).

\item {\bf Step12:
code generation
(except for CUDACPP),
test other processes.}
Backport all 
optimizations
except for those
in CUDACPP,
to automatic
code generation
in \mgamc\ (\mvf\ \vfoldfcg).
Generate and test
\eebbmmg\ and \ggttg.

\item {\bf Step13:
code generation 
in CUDACPP for \eebbbb.}
Backport all CUDACPP
optimizations 
in \eebbbb\ to 
automatic code generation.

\item {\bf Step14:
code generation 
in CUDACPP for \ggttg\ and 
\eebbmmg\ and 
additional bug fixes
and optimizations
for these processes.}
Implement new 
RNG slots for 
previously missed
\texttt{ran2()} calls 
relevant to \eebbmmg.
Fix other bugs that
were not triggered
in \eebbbb.
Handle in CUDACPP
the second component ANS(2) 
of the MadFKS Fortran SBORN;
this is only needed for
physics processes
involving gluon splitting
such as \eebbmmg\ and \ggttg.
Serve Born AMP2 per amplitude, 
as in the Fortran.
This completes 
\mvf\ code generation
in \mgamc\ v3.8.1
(complete \vfoldfcg).
Regenerate and test
\eebbbb, \eebbmmg\ and \ggttg.

\item {\bf Step15:
MadGraph7 port.}
Port the code generation
infrastructure from
\mgamc\ v3.8.1
to MadGraph7 
(complete \vfoldscg).
This includes
only minor changes
in CUDACPP; 
notably, as the 
CUDACPP plugin
does not support 
the flavour grouping 
introduced in MadGraph7, 
this had to be disabled 
when generating CUDACPP code.
Regenerate and test
\eebbbb, \eebbmmg\ and \ggttg.

\item {\bf Step16:
support 
\texttt{P0\_<subproc>}
subprocesses;
further fixes
and optimizations.}
Code generation for 
\pptt\ and \ppttj.
Prepare and test gridpacks
for these processes,
using new ad-hoc tools.
Improve RNG using 
independent random numbers 
for each FKS sector
in a fold (this removes
possible correlations
across FKS sectors).
Switch to diagram-level
helicity filtering
in Fortran and CUDACPP
to make them consistent.
New optimizations:
enable helicity-flip pairing 
in CUDACPP 
(as in MadFKS Fortran);
serve the rotated Born 
of \texttt{get\_mbar} 
from CUDACPP;
optimize some lookups 
which were quadratic 
in the number of folds;
new memo to cache 
$\alphaS$-dependent 
parameters and couplings.

\item {\bf Step17:
further fixes and tweaks for
\ppttjj.}
Move the Fortran Born 
amplitude caches 
from static to 
allocatable arrays, 
sized at runtime 
from the actual batch, 
to bypass the
``\texttt{relocation
truncated to fit}''
link error in
\ppttjj\ builds.
Reuse MINT0 grids
and cache the Fortran
build of the Virtual
for different foldings,
to speed up the
creation of 
\ppttjj\ gridpacks.
Remove unnecessary 
GPU resets at the 
end of each job.

\end{itemize}

\section{Results for 
\eebbbbTOP:
vectorized NLO folding}
\label{sec:fld}

In this section,
timing results 
for \eebbbb\ NLO 
event generation
with vectorized folding
are presented for
different versions
of the \mvf\ software
described in
the previous sections.
In the tables, 
all time numbers 
are in seconds
and refer to the MINT2
generation of 200 unweighted events
with \efo\ folding
(in a single channel GF11.0,
which was selected as 
it is one of those 
with higher cross sections).
Timing profiles are given 
for different modes,
and in mode C 
for different CUDACPP backends.
The value of \vecsize\ in 
that case indicates 
how many MEs are
computed in a single
CUDACPP call:
for each SIMD mode,
for convenience 
this is fixed to
the SIMD vector width
(although a multiple 
of this could also be chosen),
while many GPU configurations
with different \vecsize\ values
are tested, as this corresponds
to the number of GPU threads
for kernel launching in that case.
Two different machines
were used to execute
the results presented
throughout this paper:
\begin{itemize}
\item {\bf ``CUDA'' node}:
a node equipped with 
an Intel Xeon Silver 4216 CPU
and an NVIDIA V100 GPU,
targeting CUDA tests,
but where 
tests of SIMD up to AVX2
were also executed;
\item {\bf ``GOLD'' node}:
a machine equipped with 
an Intel Xeon Gold 6326 CPU,
targeting tests of SIMD
up to AVX512 
--- the interest of this
CPU being that it has
very good performance
in the CUDACPP ``512z'' 
mode,
possibly because it has
two AVX512 FMA 
units~\cite{bib:ichep2022,bib:512z}.
\end{itemize}
For simplicity,
in this section
I only present results
from the ``CUDA'' node;
AVX512 results
on the ``GOLD'' node
will only be shown 
in later sections
for more complex
physics processes.

\subsection{\eebbbbTOP:
vectorized folding 
in \vfoldftst\ 
and \vfoldfteo}
\label{sec:fld363vs381}

First, I present results
for ``step10b'',
the latest development
step reached
when Ref.~\cite{bib:mvf1}
was prepared.
Previous results 
for the initial version
of \mvf\ had been reported there
and will not be repeated here:
instead, two new 
sets of results are presented,
for \mvf\ versions 
\vfoldftst\ and 
\vfoldfteo\ (those
of \vfoldftst\ are 
almost the same
as those of 
Ref.~\cite{bib:mvf1},
since this is just 
a cleaned-up version
of the same code).
These are given in
Tables~\ref{tab:fldCuda5363}
and~\ref{tab:fldCuda5381},
which refer to tests
on the ``CUDA'' node
(Xeon Silver CPU and V100 GPU).

\paragraph{Fortran and C/SIMD modes.}
The two sets of results 
are quite similar to each other 
for Fortran and C/SIMD modes
(although \vfoldfteo\ seems 
consistently 
slower
than \vfoldftst, for
reasons that I did not
investigate). 
In particular, 
the following
general comments 
(many of which also apply
to other tables
presented in later sections)
can be made.
\begin{itemize}

\item 
The total and partial time profiles
of the F0* and F1 modes are 
very similar,
because these 
are essentially identical
(except for minor bug fixes
and the choice of random numbers)
and F0* numbers are rescaled
to refer to the same number
of kpoints as F1.

\item 
F2 mode is much faster than F1,
mainly thanks to the 
``Born memo'' and ``fused BSF''
optimizations in Fortran,
which speed up the Born
and Counterterm 
by around x3 and~x2.

\item 
F mode is slower than F2,
largely due to the overhead
of the two-walk 
batching architecture
(whose cost has been
seen in other tests 
to increase with
the kbatch parameter,
which controls batch sizes).
In particular, 
the slowdown comes
from the Counterterm
(CT\_and\_MCsub), Misc, 
PhaseSpace (PS\_Generation)
and Nbody\_prefactor
contributions,
whose times all 
increase significantly
between F2 and F.

\item 
Mode C/none is significantly
faster than F,
even in the absence 
of any SIMD speedup,
as the Born and
Counterterm contributions
benefit from further speedups
by approximately x20 and x10.
This is essentially
the combined effect
of two optimizations:
the ``gscale'' optimization,
whereby most Born contributions 
are computed by coupling rescaling;
and the ``sborn'' optimization,
whereby, for those Born MEs
that are actually computed, 
this now happens in CUDACPP
which returns
not only the MEs 
but also the colour-linked BSF, 
per-diagram AMP2 
and per-colour-flow JAMP2.

\item 
In mode C/avx2,
a speedup around 4x
over C/none
is expected for ME calculations 
in double precision
(\texttt{FPTYPE=d})
in CUDACPP.
This is indeed observed
for the Real contributions, 
which decrease 
from 116.2s to 25.6s (x4.5)
and from 119.2s to 26.7s (x4.5)
in the two tables.
Conversely, Born times
only exhibit a moderate
decrease between
C/none and C/avx2,
because 
all folds in a given kpoint 
share the same Born
thanks to the Born memo, 
hence there are 
no folds to vectorize:
a decrease
is nevertheless observed 
because
this calculation
is combining fold 
vectorization and 
event vectorization
(the parameter
kbatch=64 indicates
that 64 kpoints
and all their folds
are being processed
in one batch).

\item
The absolute time spent
in computing Virtuals
is essentially flat across
all modes from F1 to C,
as no Fortran optimization
or CUDACPP speedup
can be expected in their case.
In relative terms,
the time spent in computing Virtuals
is very low compared
to Real, Born and other contributions: 
in folded event generation,
Virtuals are only computed
once per kpoint 
(if they are computed at all),
rather than once per fold,
which is one reason
why vectorized folding
is very attractive.

\item 
Overall, 
the speedups observed
for fold vectorization
from F1 to C/avx2
are approximately x6.0
for both \vfoldftst\ (from 
280.9s to 47.1s)
and \vfoldfteo\ (from 
289.2s to 48.6s).
The calculation of Reals
has been significantly 
reduced but is still 
the bottleneck 
in C/avx2 mode,
followed by Misc: 
the Born and Counterterm, 
which were amongst the
largest contributions 
in F1 mode, have instead
almost disappeared.

\end{itemize}

\begin{table}[p]
\begin{center}{
\hspace*{-1cm}
\small
\setlength\tabcolsep{5pt} 
\providecommand{\tmtcell}[1]{\makebox[\tmtwidth][r]{#1}}
\begin{tabular}{lrrrrrrrrrrrrrr}
\multicolumn{15}{@{}l@{}}{\makebox[0pt][l]{{\bf \underline{vfold5363} $|$ epem\_bbbxbx $|$ step10b $|$ NEVENTS=200 $|$ folding=8,4,1 $|$ kbatch=64}}} \\
\multicolumn{15}{@{}l@{}}{\makebox[0pt][l]{\footnotesize {\em bsffused -1: fused single-pass BSF kernel OFF for F0/F1, ON for F2/F/C}}} \\
\multicolumn{15}{@{}l@{}}{\makebox[0pt][l]{\footnotesize {\em sborn 1: SBORN served from cudacpp, in mode C only}}} \\
\multicolumn{15}{@{}l@{}}{\makebox[0pt][l]{\footnotesize {\em gscale 1: Borns at the real-emission coupling $g'$ obtained by rescaling those at $g$, in mode C only}}} \\
\multicolumn{15}{@{}l@{}}{\makebox[0pt][l]{\footnotesize {\em timers: getrusage in all modes}}} \\
\multicolumn{15}{@{}l@{}}{\makebox[0pt][l]{\footnotesize {\em mode F0: 10031 kpoints = 76 virt + 9955 novi $\to$ 200 events (10 virt + 190 novi)}}} \\
\multicolumn{15}{@{}l@{}}{\makebox[0pt][l]{\footnotesize {\em mode F, F1, F2, Cs: 10172 kpoints = 144 virt + 10028 novi $\to$ 200 events (13 virt + 187 novi)}}} \\
\multicolumn{15}{@{}l@{}}{\makebox[0pt][l]{\footnotesize {\em mode F, Cs: Born / Real 183 / 169 batched calls, mean batch 1911.4 / 1958.7}}} \\
\multicolumn{15}{@{}l@{}}{\makebox[0pt][l]{\footnotesize {\em F0*: F0 rescaled to the kpoints of F (separately for virt and novi)}}} \\
\multicolumn{15}{@{}l@{}}{\makebox[0pt][l]{\footnotesize {\em ml\_reduction: ninja for all}}} \\
\hline
mode & \tmtcell{F0*} & \tmtcell{F1} & \tmtcell{F2} & \tmtcell{F} & \tmtcell{C} & \tmtcell{C} & \tmtcell{C} & \tmtcell{C} & \tmtcell{C} & \tmtcell{C} & \tmtcell{C} & \tmtcell{C} & \tmtcell{C} & \tmtcell{C} \\
cudacpp\_backend & \tmtcell{-} & \tmtcell{-} & \tmtcell{-} & \tmtcell{-} & \tmtcell{none} & \tmtcell{sse4} & \tmtcell{avx2} & \tmtcell{cuda} & \tmtcell{cuda} & \tmtcell{cuda} & \tmtcell{cuda} & \tmtcell{cuda} & \tmtcell{cuda} & \tmtcell{cuda} \\
cudacpp\_vecsize & \tmtcell{-} & \tmtcell{-} & \tmtcell{-} & \tmtcell{-} & \tmtcell{1} & \tmtcell{2} & \tmtcell{4} & \tmtcell{32} & \tmtcell{64} & \tmtcell{128} & \tmtcell{256} & \tmtcell{512} & \tmtcell{1024} & \tmtcell{2048} \\
\hline
TOTAL & \tmtcell{281.0} & \tmtcell{280.9} & \tmtcell{195.1} & \tmtcell{221.8} & \tmtcell{138.0} & \tmtcell{80.6} & \tmtcell{47.1} & \tmtcell{90.6} & \tmtcell{57.5} & \tmtcell{41.7} & \tmtcell{34.2} & \tmtcell{30.3} & \tmtcell{26.6} & \tmtcell{25.8} \\
Born & \tmtcell{90.1} & \tmtcell{90.0} & \tmtcell{27.7} & \tmtcell{26.8} & \tmtcell{1.5} & \tmtcell{1.2} & \tmtcell{1.0} & \tmtcell{1.2} & \tmtcell{1.0} & \tmtcell{1.1} & \tmtcell{1.1} & \tmtcell{1.2} & \tmtcell{1.2} & \tmtcell{1.2} \\
Real & \tmtcell{149.2} & \tmtcell{149.1} & \tmtcell{140.6} & \tmtcell{141.8} & \tmtcell{116.2} & \tmtcell{59.3} & \tmtcell{25.6} & \tmtcell{68.4} & \tmtcell{35.6} & \tmtcell{19.8} & \tmtcell{12.1} & \tmtcell{8.1} & \tmtcell{4.5} & \tmtcell{3.7} \\
Virtual & \tmtcell{0.7} & \tmtcell{0.6} & \tmtcell{0.6} & \tmtcell{0.6} & \tmtcell{0.6} & \tmtcell{0.6} & \tmtcell{0.6} & \tmtcell{0.6} & \tmtcell{0.6} & \tmtcell{0.6} & \tmtcell{0.6} & \tmtcell{0.7} & \tmtcell{0.6} & \tmtcell{0.6} \\
PS\_Generation & \tmtcell{1.6} & \tmtcell{1.6} & \tmtcell{1.5} & \tmtcell{2.9} & \tmtcell{2.8} & \tmtcell{2.8} & \tmtcell{3.0} & \tmtcell{2.8} & \tmtcell{2.8} & \tmtcell{2.8} & \tmtcell{2.8} & \tmtcell{2.8} & \tmtcell{2.8} & \tmtcell{2.8} \\
CT\_and\_MCsub & \tmtcell{29.8} & \tmtcell{29.8} & \tmtcell{15.4} & \tmtcell{32.6} & \tmtcell{3.3} & \tmtcell{3.2} & \tmtcell{3.3} & \tmtcell{3.3} & \tmtcell{3.2} & \tmtcell{3.3} & \tmtcell{3.3} & \tmtcell{3.3} & \tmtcell{3.3} & \tmtcell{3.2} \\
Nbody\_prefactor & \tmtcell{1.1} & \tmtcell{1.1} & \tmtcell{1.1} & \tmtcell{4.3} & \tmtcell{1.2} & \tmtcell{1.2} & \tmtcell{1.2} & \tmtcell{1.3} & \tmtcell{1.2} & \tmtcell{1.2} & \tmtcell{1.3} & \tmtcell{1.3} & \tmtcell{1.2} & \tmtcell{1.2} \\
Misc & \tmtcell{8.6} & \tmtcell{8.7} & \tmtcell{8.2} & \tmtcell{12.8} & \tmtcell{12.3} & \tmtcell{12.2} & \tmtcell{12.3} & \tmtcell{13.0} & \tmtcell{12.9} & \tmtcell{12.9} & \tmtcell{13.0} & \tmtcell{13.0} & \tmtcell{12.9} & \tmtcell{12.9} \\
\hline
\end{tabular}

\caption{Timing results (\vfoldftst) 
for generating 200 
\eebbbb\ events (channel GF11)
with \efo\ folding
using step10b in MINT2,
on the ``CUDA'' node.}
\label{tab:fldCuda5363}
\vspace*{1.2cm}
\hspace*{-1cm}
\small
\setlength\tabcolsep{5pt} 
\providecommand{\tmtcell}[1]{\makebox[\tmtwidth][r]{#1}}
\begin{tabular}{lrrrrrrrrrrrrrr}
\multicolumn{15}{@{}l@{}}{\makebox[0pt][l]{{\bf \underline{vfold5381} $|$ epem\_bbbxbx $|$ step10b $|$ NEVENTS=200 $|$ folding=8,4,1 $|$ kbatch=64}}} \\
\multicolumn{15}{@{}l@{}}{\makebox[0pt][l]{\footnotesize {\em bsffused -1: fused single-pass BSF kernel OFF for F0/F1, ON for F2/F/C}}} \\
\multicolumn{15}{@{}l@{}}{\makebox[0pt][l]{\footnotesize {\em sborn 1: SBORN served from cudacpp, in mode C only}}} \\
\multicolumn{15}{@{}l@{}}{\makebox[0pt][l]{\footnotesize {\em gscale 1: Borns at the real-emission coupling $g'$ obtained by rescaling those at $g$, in mode C only}}} \\
\multicolumn{15}{@{}l@{}}{\makebox[0pt][l]{\footnotesize {\em timers: getrusage in all modes}}} \\
\multicolumn{15}{@{}l@{}}{\makebox[0pt][l]{\footnotesize {\em mode F0: 10031 kpoints = 76 virt + 9955 novi $\to$ 200 events (10 virt + 190 novi)}}} \\
\multicolumn{15}{@{}l@{}}{\makebox[0pt][l]{\footnotesize {\em mode F, F1, F2, Cs: 10172 kpoints = 144 virt + 10028 novi $\to$ 200 events (13 virt + 187 novi)}}} \\
\multicolumn{15}{@{}l@{}}{\makebox[0pt][l]{\footnotesize {\em mode F, Cs: Born / Real 183 / 169 batched calls, mean batch 1911.4 / 1958.7}}} \\
\multicolumn{15}{@{}l@{}}{\makebox[0pt][l]{\footnotesize {\em F0*: F0 rescaled to the kpoints of F (separately for virt and novi)}}} \\
\multicolumn{15}{@{}l@{}}{\makebox[0pt][l]{\footnotesize {\em ml\_reduction: ninja for all}}} \\
\hline
mode & \tmtcell{F0*} & \tmtcell{F1} & \tmtcell{F2} & \tmtcell{F} & \tmtcell{C} & \tmtcell{C} & \tmtcell{C} & \tmtcell{C} & \tmtcell{C} & \tmtcell{C} & \tmtcell{C} & \tmtcell{C} & \tmtcell{C} & \tmtcell{C} \\
cudacpp\_backend & \tmtcell{-} & \tmtcell{-} & \tmtcell{-} & \tmtcell{-} & \tmtcell{none} & \tmtcell{sse4} & \tmtcell{avx2} & \tmtcell{cuda} & \tmtcell{cuda} & \tmtcell{cuda} & \tmtcell{cuda} & \tmtcell{cuda} & \tmtcell{cuda} & \tmtcell{cuda} \\
cudacpp\_vecsize & \tmtcell{-} & \tmtcell{-} & \tmtcell{-} & \tmtcell{-} & \tmtcell{1} & \tmtcell{2} & \tmtcell{4} & \tmtcell{32} & \tmtcell{64} & \tmtcell{128} & \tmtcell{256} & \tmtcell{512} & \tmtcell{1024} & \tmtcell{2048} \\
\hline
TOTAL & \tmtcell{287.6} & \tmtcell{289.2} & \tmtcell{203.8} & \tmtcell{228.4} & \tmtcell{141.6} & \tmtcell{84.8} & \tmtcell{48.6} & \tmtcell{28.4} & \tmtcell{25.9} & \tmtcell{24.8} & \tmtcell{24.7} & \tmtcell{24.3} & \tmtcell{24.2} & \tmtcell{24.2} \\
Born & \tmtcell{92.7} & \tmtcell{93.3} & \tmtcell{28.7} & \tmtcell{27.7} & \tmtcell{1.5} & \tmtcell{1.3} & \tmtcell{1.1} & \tmtcell{1.1} & \tmtcell{1.0} & \tmtcell{1.0} & \tmtcell{1.1} & \tmtcell{1.1} & \tmtcell{1.1} & \tmtcell{1.1} \\
Real & \tmtcell{152.5} & \tmtcell{153.2} & \tmtcell{147.3} & \tmtcell{145.8} & \tmtcell{119.2} & \tmtcell{61.7} & \tmtcell{26.7} & \tmtcell{5.9} & \tmtcell{3.5} & \tmtcell{2.3} & \tmtcell{1.8} & \tmtcell{1.6} & \tmtcell{1.6} & \tmtcell{1.5} \\
Virtual & \tmtcell{0.7} & \tmtcell{0.7} & \tmtcell{0.7} & \tmtcell{0.7} & \tmtcell{0.7} & \tmtcell{0.7} & \tmtcell{0.7} & \tmtcell{0.7} & \tmtcell{0.7} & \tmtcell{0.7} & \tmtcell{0.7} & \tmtcell{0.7} & \tmtcell{0.7} & \tmtcell{0.7} \\
PS\_Generation & \tmtcell{1.7} & \tmtcell{1.7} & \tmtcell{1.6} & \tmtcell{3.0} & \tmtcell{2.9} & \tmtcell{3.0} & \tmtcell{2.9} & \tmtcell{2.9} & \tmtcell{2.9} & \tmtcell{2.9} & \tmtcell{3.1} & \tmtcell{2.9} & \tmtcell{2.9} & \tmtcell{3.0} \\
CT\_and\_MCsub & \tmtcell{29.8} & \tmtcell{29.9} & \tmtcell{15.8} & \tmtcell{33.4} & \tmtcell{3.4} & \tmtcell{3.6} & \tmtcell{3.3} & \tmtcell{3.3} & \tmtcell{3.3} & \tmtcell{3.3} & \tmtcell{3.4} & \tmtcell{3.4} & \tmtcell{3.3} & \tmtcell{3.4} \\
Nbody\_prefactor & \tmtcell{1.2} & \tmtcell{1.2} & \tmtcell{1.2} & \tmtcell{4.4} & \tmtcell{1.3} & \tmtcell{1.3} & \tmtcell{1.2} & \tmtcell{1.3} & \tmtcell{1.3} & \tmtcell{1.3} & \tmtcell{1.3} & \tmtcell{1.3} & \tmtcell{1.3} & \tmtcell{1.3} \\
Misc & \tmtcell{9.0} & \tmtcell{9.1} & \tmtcell{8.6} & \tmtcell{13.4} & \tmtcell{12.7} & \tmtcell{13.3} & \tmtcell{12.7} & \tmtcell{13.3} & \tmtcell{13.1} & \tmtcell{13.3} & \tmtcell{13.4} & \tmtcell{13.4} & \tmtcell{13.3} & \tmtcell{13.3} \\
\hline
\end{tabular}

\caption{Timing results (\vfoldfteo)
for generating 200 
\eebbbb\ events (channel GF11)
with \efo\ folding
using step10b in MINT2,
on the ``CUDA'' node.}
\label{tab:fldCuda5381}
}\end{center}
\end{table}

\begin{table}[p]
\begin{center}{
\hspace*{-1cm}
\small
\setlength\tabcolsep{5pt} 
\providecommand{\tmtcell}[1]{\makebox[\tmtwidth][r]{#1}}
\begin{tabular}{lrrrrrrrrrrrrrr}
\multicolumn{15}{@{}l@{}}{\makebox[0pt][l]{{\bf \underline{vfold5381cg} $|$ epem\_bbbxbx $|$ step14t $|$ NEVENTS=200 $|$ folding=8,4,1 $|$ kbatch=64}}} \\
\multicolumn{15}{@{}l@{}}{\makebox[0pt][l]{\footnotesize {\em sborn 1: SBORN served from cudacpp, in mode C only}}} \\
\multicolumn{15}{@{}l@{}}{\makebox[0pt][l]{\footnotesize {\em gscale 1: Borns at the real-emission coupling $g'$ obtained by rescaling those at $g$, in mode C only}}} \\
\multicolumn{15}{@{}l@{}}{\makebox[0pt][l]{\footnotesize {\em timers: getrusage in F0*, F1 (a syscall per timer call), clock\_gettime in all other modes}}} \\
\multicolumn{15}{@{}l@{}}{\makebox[0pt][l]{\footnotesize {\em mode F0: 10031 kpoints = 76 virt + 9955 novi $\to$ 200 events (10 virt + 190 novi)}}} \\
\multicolumn{15}{@{}l@{}}{\makebox[0pt][l]{\footnotesize {\em mode F, F1, F2, Cs: 10172 kpoints = 144 virt + 10028 novi $\to$ 200 events (13 virt + 187 novi)}}} \\
\multicolumn{15}{@{}l@{}}{\makebox[0pt][l]{\footnotesize {\em mode F, Cs: Born / Real 160 / 157 batched calls, mean batch 2047.7 / 2047.1}}} \\
\multicolumn{15}{@{}l@{}}{\makebox[0pt][l]{\footnotesize {\em F0*: F0 rescaled to the kpoints of F (separately for virt and novi)}}} \\
\multicolumn{15}{@{}l@{}}{\makebox[0pt][l]{\footnotesize {\em ml\_reduction: ninja for all}}} \\
\hline
mode & \tmtcell{F0*} & \tmtcell{F1} & \tmtcell{F2} & \tmtcell{F} & \tmtcell{C} & \tmtcell{C} & \tmtcell{C} & \tmtcell{C} & \tmtcell{C} & \tmtcell{C} & \tmtcell{C} & \tmtcell{C} & \tmtcell{C} & \tmtcell{C} \\
cudacpp\_backend & \tmtcell{-} & \tmtcell{-} & \tmtcell{-} & \tmtcell{-} & \tmtcell{none} & \tmtcell{sse4} & \tmtcell{avx2} & \tmtcell{cuda} & \tmtcell{cuda} & \tmtcell{cuda} & \tmtcell{cuda} & \tmtcell{cuda} & \tmtcell{cuda} & \tmtcell{cuda} \\
cudacpp\_vecsize & \tmtcell{-} & \tmtcell{-} & \tmtcell{-} & \tmtcell{-} & \tmtcell{1} & \tmtcell{2} & \tmtcell{4} & \tmtcell{32} & \tmtcell{64} & \tmtcell{128} & \tmtcell{256} & \tmtcell{512} & \tmtcell{1024} & \tmtcell{2048} \\
\hline
TOTAL & \tmtcell{284.2} & \tmtcell{284.5} & \tmtcell{185.2} & \tmtcell{187.7} & \tmtcell{126.8} & \tmtcell{69.1} & \tmtcell{36.2} & \tmtcell{17.3} & \tmtcell{13.7} & \tmtcell{12.6} & \tmtcell{11.9} & \tmtcell{11.8} & \tmtcell{11.9} & \tmtcell{11.8} \\
Born & \tmtcell{91.3} & \tmtcell{91.5} & \tmtcell{26.4} & \tmtcell{25.8} & \tmtcell{1.2} & \tmtcell{0.9} & \tmtcell{0.8} & \tmtcell{0.8} & \tmtcell{0.7} & \tmtcell{0.7} & \tmtcell{0.7} & \tmtcell{0.7} & \tmtcell{0.8} & \tmtcell{0.8} \\
Real & \tmtcell{150.7} & \tmtcell{151.0} & \tmtcell{141.8} & \tmtcell{140.3} & \tmtcell{115.7} & \tmtcell{58.7} & \tmtcell{25.7} & \tmtcell{5.4} & \tmtcell{3.1} & \tmtcell{2.0} & \tmtcell{1.4} & \tmtcell{1.3} & \tmtcell{1.2} & \tmtcell{1.2} \\
Virtual & \tmtcell{0.7} & \tmtcell{0.6} & \tmtcell{0.7} & \tmtcell{0.7} & \tmtcell{0.7} & \tmtcell{0.7} & \tmtcell{0.7} & \tmtcell{0.7} & \tmtcell{0.7} & \tmtcell{0.7} & \tmtcell{0.7} & \tmtcell{0.7} & \tmtcell{0.7} & \tmtcell{0.7} \\
PS\_Generation & \tmtcell{1.8} & \tmtcell{1.7} & \tmtcell{0.7} & \tmtcell{1.3} & \tmtcell{1.3} & \tmtcell{1.3} & \tmtcell{1.3} & \tmtcell{1.3} & \tmtcell{1.3} & \tmtcell{1.3} & \tmtcell{1.2} & \tmtcell{1.3} & \tmtcell{1.3} & \tmtcell{1.3} \\
CT\_and\_MCsub & \tmtcell{29.4} & \tmtcell{29.5} & \tmtcell{11.0} & \tmtcell{11.2} & \tmtcell{1.8} & \tmtcell{1.7} & \tmtcell{1.7} & \tmtcell{1.7} & \tmtcell{1.7} & \tmtcell{1.7} & \tmtcell{1.6} & \tmtcell{1.7} & \tmtcell{1.7} & \tmtcell{1.7} \\
Nbody\_prefactor & \tmtcell{1.2} & \tmtcell{1.2} & \tmtcell{0.2} & \tmtcell{2.7} & \tmtcell{0.2} & \tmtcell{0.2} & \tmtcell{0.2} & \tmtcell{0.2} & \tmtcell{0.2} & \tmtcell{0.2} & \tmtcell{0.2} & \tmtcell{0.2} & \tmtcell{0.2} & \tmtcell{0.2} \\
Bridge\_crt\_del & \tmtcell{0.0} & \tmtcell{0.0} & \tmtcell{0.0} & \tmtcell{0.0} & \tmtcell{0.2} & \tmtcell{0.3} & \tmtcell{0.4} & \tmtcell{1.9} & \tmtcell{0.6} & \tmtcell{0.6} & \tmtcell{0.6} & \tmtcell{0.6} & \tmtcell{0.6} & \tmtcell{0.6} \\
Misc & \tmtcell{9.0} & \tmtcell{9.0} & \tmtcell{4.4} & \tmtcell{5.8} & \tmtcell{5.8} & \tmtcell{5.5} & \tmtcell{5.5} & \tmtcell{5.5} & \tmtcell{5.5} & \tmtcell{5.5} & \tmtcell{5.4} & \tmtcell{5.5} & \tmtcell{5.5} & \tmtcell{5.4} \\
\hline
\end{tabular}

\caption{Timing results (\vfoldfcg) 
for generating 200 
\eebbbb\ events (channel GF11)
with \efo\ folding
using step14t in MINT2,
on the ``CUDA'' node.}
\label{tab:fldCuda5381cg}
\vspace*{1.2cm}
\hspace*{-1.4cm}
\small
\setlength\tabcolsep{5pt} 
\providecommand{\tmtcell}[1]{\makebox[\tmtwidth][r]{#1}}
\begin{tabular}{lrrrrrrrrrrrrrrr}
\multicolumn{16}{@{}l@{}}{\makebox[0pt][l]{{\bf \underline{vfold7cg} $|$ epem\_bbbxbx $|$ step15b $|$ NEVENTS=200 $|$ folding=8,4,1 $|$ kbatch=64}}} \\
\multicolumn{16}{@{}l@{}}{\makebox[0pt][l]{\footnotesize {\em sborn 1: SBORN served from cudacpp, in mode C only}}} \\
\multicolumn{16}{@{}l@{}}{\makebox[0pt][l]{\footnotesize {\em gscale 1: Borns at the real-emission coupling $g'$ obtained by rescaling those at $g$, in mode C only}}} \\
\multicolumn{16}{@{}l@{}}{\makebox[0pt][l]{\footnotesize {\em timers: getrusage in F0*, F1, F1[u] (a syscall per timer call), clock\_gettime in all other modes}}} \\
\multicolumn{16}{@{}l@{}}{\makebox[0pt][l]{\footnotesize {\em mode F0: 10031 kpoints = 76 virt + 9955 novi $\to$ 200 events (10 virt + 190 novi)}}} \\
\multicolumn{16}{@{}l@{}}{\makebox[0pt][l]{\footnotesize {\em mode F, F1, F2, Cs: 10172 kpoints = 144 virt + 10028 novi $\to$ 200 events (13 virt + 187 novi)}}} \\
\multicolumn{16}{@{}l@{}}{\makebox[0pt][l]{\footnotesize {\em mode F, Cs: Born / Real 160 / 157 batched calls, mean batch 2047.7 / 2047.1}}} \\
\multicolumn{16}{@{}l@{}}{\makebox[0pt][l]{\footnotesize {\em F0*: F0 rescaled to the kpoints of F (separately for virt and novi)}}} \\
\multicolumn{16}{@{}l@{}}{\makebox[0pt][l]{\footnotesize {\em F1[u]: F1 without two stock MadGraph7 optimizations (cache and $\alpha\sb{S}$ rescaling of Born amplitudes)}}} \\
\multicolumn{16}{@{}l@{}}{\makebox[0pt][l]{\footnotesize {\em ml\_reduction: ninja for all}}} \\
\hline
mode & \tmtcell{F0*} & \tmtcell{F1} & \tmtcell{F1[u]} & \tmtcell{F2} & \tmtcell{F} & \tmtcell{C} & \tmtcell{C} & \tmtcell{C} & \tmtcell{C} & \tmtcell{C} & \tmtcell{C} & \tmtcell{C} & \tmtcell{C} & \tmtcell{C} & \tmtcell{C} \\
cudacpp\_backend & \tmtcell{-} & \tmtcell{-} & \tmtcell{-} & \tmtcell{-} & \tmtcell{-} & \tmtcell{none} & \tmtcell{sse4} & \tmtcell{avx2} & \tmtcell{cuda} & \tmtcell{cuda} & \tmtcell{cuda} & \tmtcell{cuda} & \tmtcell{cuda} & \tmtcell{cuda} & \tmtcell{cuda} \\
cudacpp\_vecsize & \tmtcell{-} & \tmtcell{-} & \tmtcell{-} & \tmtcell{-} & \tmtcell{-} & \tmtcell{1} & \tmtcell{2} & \tmtcell{4} & \tmtcell{32} & \tmtcell{64} & \tmtcell{128} & \tmtcell{256} & \tmtcell{512} & \tmtcell{1024} & \tmtcell{2048} \\
\hline
TOTAL & \tmtcell{156.3} & \tmtcell{156.0} & \tmtcell{237.8} & \tmtcell{151.9} & \tmtcell{158.6} & \tmtcell{126.3} & \tmtcell{69.1} & \tmtcell{36.0} & \tmtcell{16.1} & \tmtcell{13.8} & \tmtcell{12.5} & \tmtcell{12.1} & \tmtcell{12.3} & \tmtcell{11.8} & \tmtcell{12.3} \\
Born & \tmtcell{3.6} & \tmtcell{3.6} & \tmtcell{83.2} & \tmtcell{6.8} & \tmtcell{7.4} & \tmtcell{1.2} & \tmtcell{0.9} & \tmtcell{0.8} & \tmtcell{0.8} & \tmtcell{0.7} & \tmtcell{0.7} & \tmtcell{0.7} & \tmtcell{0.7} & \tmtcell{0.8} & \tmtcell{0.8} \\
Real & \tmtcell{136.4} & \tmtcell{136.0} & \tmtcell{137.8} & \tmtcell{136.1} & \tmtcell{136.4} & \tmtcell{115.7} & \tmtcell{58.7} & \tmtcell{25.6} & \tmtcell{5.3} & \tmtcell{3.1} & \tmtcell{2.0} & \tmtcell{1.4} & \tmtcell{1.3} & \tmtcell{1.2} & \tmtcell{1.2} \\
Virtual & \tmtcell{0.7} & \tmtcell{0.6} & \tmtcell{0.6} & \tmtcell{0.7} & \tmtcell{0.7} & \tmtcell{0.7} & \tmtcell{0.7} & \tmtcell{0.7} & \tmtcell{0.7} & \tmtcell{0.7} & \tmtcell{0.7} & \tmtcell{0.7} & \tmtcell{0.7} & \tmtcell{0.7} & \tmtcell{0.7} \\
PS\_Generation & \tmtcell{1.5} & \tmtcell{1.6} & \tmtcell{1.6} & \tmtcell{0.7} & \tmtcell{1.3} & \tmtcell{1.2} & \tmtcell{1.2} & \tmtcell{1.3} & \tmtcell{1.3} & \tmtcell{1.3} & \tmtcell{1.2} & \tmtcell{1.3} & \tmtcell{1.3} & \tmtcell{1.2} & \tmtcell{1.3} \\
CT\_and\_MCsub & \tmtcell{5.0} & \tmtcell{4.9} & \tmtcell{5.0} & \tmtcell{3.2} & \tmtcell{4.5} & \tmtcell{1.9} & \tmtcell{1.9} & \tmtcell{2.0} & \tmtcell{1.9} & \tmtcell{1.9} & \tmtcell{1.9} & \tmtcell{1.9} & \tmtcell{2.0} & \tmtcell{1.9} & \tmtcell{2.0} \\
Nbody\_prefactor & \tmtcell{1.1} & \tmtcell{1.1} & \tmtcell{1.1} & \tmtcell{0.2} & \tmtcell{2.8} & \tmtcell{0.2} & \tmtcell{0.2} & \tmtcell{0.2} & \tmtcell{0.2} & \tmtcell{0.2} & \tmtcell{0.2} & \tmtcell{0.2} & \tmtcell{0.2} & \tmtcell{0.2} & \tmtcell{0.2} \\
Bridge\_crt\_del & \tmtcell{0.0} & \tmtcell{0.0} & \tmtcell{0.0} & \tmtcell{0.0} & \tmtcell{0.0} & \tmtcell{0.1} & \tmtcell{0.1} & \tmtcell{0.1} & \tmtcell{0.6} & \tmtcell{0.5} & \tmtcell{0.5} & \tmtcell{0.5} & \tmtcell{0.5} & \tmtcell{0.5} & \tmtcell{0.5} \\
Misc & \tmtcell{8.0} & \tmtcell{8.1} & \tmtcell{8.3} & \tmtcell{4.3} & \tmtcell{5.6} & \tmtcell{5.4} & \tmtcell{5.4} & \tmtcell{5.4} & \tmtcell{5.4} & \tmtcell{5.4} & \tmtcell{5.3} & \tmtcell{5.4} & \tmtcell{5.6} & \tmtcell{5.3} & \tmtcell{5.6} \\
\hline
\end{tabular}

\caption{Timing results 
(\vfoldscg)
for generating 200 
\eebbbb\ events (channel GF11)
with \efo\ folding
using step15b in MINT2,
on the ``CUDA'' node.}
\label{tab:fldCuda7cg}
}\end{center}
\end{table}

\paragraph{C/cuda modes.}
The main difference 
between 
\vfoldftst\ and \vfoldfteo, however,
and the interest 
of the latter,
can be seen by comparing
their time profiles on GPU.
In particular:

\begin{itemize}

\item 
The best GPU results 
are almost a factor 2 faster
than the AVX2 ones
for both 
\vfoldftst\ and \vfoldfteo.
Overall, 
the speedups observed
for fold vectorization
from F1 to the fastest 
C/cuda are approximately 
x10.9 (from 280.9s to 25.8s)
and x12.0 (from 289.2s to 24.2s)
for these two software versions.

\item 
The big difference
between the two versions
lies in the time spent 
in the calculation of Reals.
In both cases,
Reals take around 150s
and are by far the 
largest contribution 
in mode F1.
Large speedups are 
then achieved by offloading
this calculation to the GPU;
not surprisingly, 
these speedups 
are more pronounced
as more work is given to the GPU,
namely as the CUDACPP 
\vecsize\ is increased 
from 32 to 2048
(the total number 
of GPU threads:
in \mvf, 
where each thread processes
one fold, 
the maximum 
number of folds that
can be processed 
in parallel on the GPU
is the product 
of the kpoint 
batch capacity kbatch
times the number 
of folds per kpoint
--- here $64\times32=2048$).
However,
\vfoldftst\ only
reduces this time to 3.7s,
achieving this
if 2048 threads are used
(while, for instance,
Reals take up 35.6s
if 64 threads are used);
\vfoldfteo, conversely,
reduces the Real contribution
to 1.5s with 2048 threads
(i.e. a factor 100 
faster than in F1),
and achieves a very good
performance also with a
lower number of threads
(for instance, 
3.5s with 64 threads).
This is 
achieved thanks to
the helicity streams 
optimization, 
which was added to CUDACPP 
between the v3.6.3
and v3.8.1 releases of \mgamc,
and was shown~\cite{bib:kspl}
to increase peak throughput
and achieve it with 
much fewer threads.

\item
The GPU speedup of Reals 
is so large 
that other components,
notably Misc 
at around~13s,
become the 
bottlenecks
of the calculation
and the prime target 
of further optimizations,
notably those described
in the following 
Sec.~\ref{sec:fld5vs7}.

\end{itemize}

\subsection{\eebbbbTOP:
vectorized folding 
in \vfoldfcg\ and \vfoldscg}
\label{sec:fld5vs7}

Next, I present results
obtained with the 
most recent optimizations
developed and 
tested for \eebbbb,
in \mvf\ versions 
\vfoldfcg\ (step14t) 
and \vfoldscg\ (step15b).
These are given in
Tables~\ref{tab:fldCuda5381cg}
and~\ref{tab:fldCuda7cg}.
These two tables
are directly comparable
to each other and to
the previous two
Tables~\ref{tab:fldCuda5363}
and~\ref{tab:fldCuda5381},
because they refer 
to the generation
of exactly the same events
from the same random numbers,
on the same ``CUDA'' node.

\paragraph{Further 
\mvf\ optimizations
in \vfoldfcg.}
The comparison
of Tables~\ref{tab:fldCuda5381}
and~\ref{tab:fldCuda5381cg},
in particular,
shows that significant speedups 
between \mvf\ step10 and step14 
have been achieved by the 
step11
optimizations 
described
in Sec.~\ref{sec:steps}
(faster timers;
bornsoftvirtual,
passcuts and 
phase space memo;
reweighting and 
shower scales ---
note also that the fused BSF
implementation was instead removed,
but this has no negative impact
in mode C as colour-linked Borns 
are still served by CUDACPP).
More in detail:
\begin{itemize}

\item 
The Misc component,
which at around 13s
was the main bottleneck
and primary target
for these further
optimizations,
is now reduced to
approximately 5.5s.
The Counterterm
and PhaseSpace contributions
have also 
decreased by around 
a factor~2
in modes F2 to~C.
This is not surprising,
as many step11 optimizations,
notably the lean gather walk,
were introduced
to reduce the overhead
of the two-walk
infrastructure.

\item 
The time to create 
and delete the CUDACPP bridge,
which was previously
accounted for in Misc
and is now 
measured separately,
is approximately 0.6s
when a GPU is used
(except for a probable
outlier at 1.9s).
This is a one-off cost
that is mainly due to
the GPU hardware 
initialization
during these tests,
which does not increase
as larger samples of events
are generated:
therefore,
it could 
make sense
to remove it when
computing speedups
between different modes.
A non-zero but smaller
time is also spent
in Bridge creation/deletion
in CPU/SIMD modes, indicating
that this cannot be completely
attributed to GPU initialization.

\item 
The total generation
time in \vfoldfcg\ is
reduced to 36.2s for C/avx2 
and to 
11.8s (11.2s if one removes 
the GPU reset)
for C/cuda 
with 512 \vecsize:
with respect to
284.5s for F1,
these are 
overall speedups
of x7.9 and x24.1 (x25.4)
over stock \mgamc.

\end{itemize}
    
\paragraph{MadGraph7 
optimizations
in \vfoldscg.}

The comparison
of Tables~\ref{tab:fldCuda7cg}
and~\ref{tab:fldCuda5381cg},
conversely,
shows that 
significant speedups 
in stock Fortran modes
have been achieved by the
optimizations~\cite{bib:pr260,bib:pr305}
introduced in MadGraph7,
which are similar to
(and were partly derived from)
those reported for 
\mvf\ in Ref.~\cite{bib:mvf1}.
In particular:
\begin{itemize}

\item 
The most visible
change is that 
the Born contribution
in stock F1 mode
drops from 91.5s
in \vfoldfcg\ to
3.6s in \vfoldscg.
For comparison,
in mode F1[u],
where the two main optimizations in
stock MadGraph7 
have been disabled,
this increases back to 83.2s.
Note also
that F2 Born is now slower than F1, 
which indicates that
the MadGraph7 stock
optimizations are more 
effective than the 
\mvf\ Born memo for
Fortran modes.

\item 
The Counterterms
also drop significantly
from 29.5s to 4.9s,
while the Real contribution 
is reduced
from 151.0s to 136.0s.
This reduction is likely
to come from optimizations
other than those disabled
in mode F1[u], as these
two contributions (unlike Born)
do not increase much
in that mode.

\item
In all CUDACPP modes
from C/none to C/cuda,
all contributions
are instead 
essentially unchanged
between \vfoldfcg\ and
\vfoldscg. 

\item 
The overall \mvf\ speedups
in \vfoldscg\ are 
therefore lower
than in \vfoldfcg,
but still significant:
the total generation
times are 
36.0s for C/avx2 
and 11.8s for C/cuda 
with 1024 \vecsize:
with respect to
the much lower
156.0s for F1,
however, these now
represent overall speedups
of x4.3 and x13.2
over stock MadGraph7.

\end{itemize}

\section{Results for 
\eebbbbTOP:
vectorized NLO 
event generation}
\label{sec:nof}

In this section,
I present
timing results 
for \eebbbb\ NLO 
vectorized event generation
without folding
for different versions
of the \mvf\ software.
In the tables, 
all time numbers 
are in seconds
and refer to the MINT2
generation of 200 unweighted events
with \ooo\ folding
in a single channel GF11.0.
All results
come from tests executed
on the same ``CUDA'' machine
where the numbers of
Sec.~\ref{sec:fld} 
were derived
for \efo\ folding.
As discussed later on
in Sec.~\ref{sec:fldVsNof},
the tables presented here
can be directly compared
to those in Sec.~\ref{sec:fld},
if one takes into account
the large difference 
in their folding configurations.

\begin{table}[p]
\begin{center}{
\hspace*{-1cm}
\small
\setlength\tabcolsep{5pt} 
\providecommand{\tmtcell}[1]{\makebox[\tmtwidth][r]{#1}}
\begin{tabular}{lrrrrrrrrrrrrr}
\multicolumn{14}{@{}l@{}}{\makebox[0pt][l]{{\bf \underline{vfold5363} $|$ epem\_bbbxbx $|$ step10b $|$ NEVENTS=200 $|$ folding=1,1,1 $|$ kbatch=512}}} \\
\multicolumn{14}{@{}l@{}}{\makebox[0pt][l]{\footnotesize {\em bsffused -1: fused single-pass BSF kernel OFF for F0/F1, ON for F2/F/C}}} \\
\multicolumn{14}{@{}l@{}}{\makebox[0pt][l]{\footnotesize {\em sborn 1: SBORN served from cudacpp, in mode C only}}} \\
\multicolumn{14}{@{}l@{}}{\makebox[0pt][l]{\footnotesize {\em gscale 1: Borns at the real-emission coupling $g'$ obtained by rescaling those at $g$, in mode C only}}} \\
\multicolumn{14}{@{}l@{}}{\makebox[0pt][l]{\footnotesize {\em timers: getrusage in all modes}}} \\
\multicolumn{14}{@{}l@{}}{\makebox[0pt][l]{\footnotesize {\em mode F0: 7675 kpoints = 306 virt + 7369 novi $\to$ 200 events (10 virt + 190 novi)}}} \\
\multicolumn{14}{@{}l@{}}{\makebox[0pt][l]{\footnotesize {\em mode F, F1, F2, Cs: 8146 kpoints = 287 virt + 7859 novi $\to$ 200 events (9 virt + 191 novi)}}} \\
\multicolumn{14}{@{}l@{}}{\makebox[0pt][l]{\footnotesize {\em mode F, Cs: Born / Real 34 / 25 batched calls, mean batch 409.2 / 424.6}}} \\
\multicolumn{14}{@{}l@{}}{\makebox[0pt][l]{\footnotesize {\em F0*: F0 rescaled to the kpoints of F (separately for virt and novi)}}} \\
\multicolumn{14}{@{}l@{}}{\makebox[0pt][l]{\footnotesize {\em ml\_reduction: ninja for all}}} \\
\hline
mode & \tmtcell{F0*} & \tmtcell{F1} & \tmtcell{F2} & \tmtcell{F} & \tmtcell{C} & \tmtcell{C} & \tmtcell{C} & \tmtcell{C} & \tmtcell{C} & \tmtcell{C} & \tmtcell{C} & \tmtcell{C} & \tmtcell{C} \\
cudacpp\_backend & \tmtcell{-} & \tmtcell{-} & \tmtcell{-} & \tmtcell{-} & \tmtcell{none} & \tmtcell{sse4} & \tmtcell{avx2} & \tmtcell{cuda} & \tmtcell{cuda} & \tmtcell{cuda} & \tmtcell{cuda} & \tmtcell{cuda} & \tmtcell{cuda} \\
cudacpp\_vecsize & \tmtcell{-} & \tmtcell{-} & \tmtcell{-} & \tmtcell{-} & \tmtcell{1} & \tmtcell{2} & \tmtcell{4} & \tmtcell{32} & \tmtcell{64} & \tmtcell{128} & \tmtcell{256} & \tmtcell{512} & \tmtcell{1024} \\
\hline
TOTAL & \tmtcell{8.08} & \tmtcell{8.08} & \tmtcell{6.63} & \tmtcell{8.94} & \tmtcell{6.25} & \tmtcell{4.13} & \tmtcell{2.92} & \tmtcell{4.89} & \tmtcell{3.63} & \tmtcell{3.06} & \tmtcell{2.82} & \tmtcell{2.63} & \tmtcell{2.71} \\
Born & \tmtcell{2.20} & \tmtcell{2.20} & \tmtcell{1.20} & \tmtcell{1.69} & \tmtcell{0.70} & \tmtcell{0.39} & \tmtcell{0.21} & \tmtcell{0.32} & \tmtcell{0.19} & \tmtcell{0.13} & \tmtcell{0.12} & \tmtcell{0.09} & \tmtcell{0.10} \\
Real & \tmtcell{3.62} & \tmtcell{3.62} & \tmtcell{3.49} & \tmtcell{4.57} & \tmtcell{3.70} & \tmtcell{1.91} & \tmtcell{0.85} & \tmtcell{2.31} & \tmtcell{1.23} & \tmtcell{0.71} & \tmtcell{0.50} & \tmtcell{0.32} & \tmtcell{0.32} \\
Virtual & \tmtcell{1.17} & \tmtcell{1.17} & \tmtcell{1.18} & \tmtcell{1.18} & \tmtcell{1.21} & \tmtcell{1.20} & \tmtcell{1.22} & \tmtcell{1.23} & \tmtcell{1.21} & \tmtcell{1.21} & \tmtcell{1.20} & \tmtcell{1.21} & \tmtcell{1.25} \\
PS\_Generation & \tmtcell{0.05} & \tmtcell{0.05} & \tmtcell{0.04} & \tmtcell{0.10} & \tmtcell{0.10} & \tmtcell{0.09} & \tmtcell{0.10} & \tmtcell{0.10} & \tmtcell{0.09} & \tmtcell{0.09} & \tmtcell{0.09} & \tmtcell{0.10} & \tmtcell{0.10} \\
CT\_and\_MCsub & \tmtcell{0.73} & \tmtcell{0.73} & \tmtcell{0.42} & \tmtcell{0.86} & \tmtcell{0.09} & \tmtcell{0.09} & \tmtcell{0.09} & \tmtcell{0.09} & \tmtcell{0.09} & \tmtcell{0.09} & \tmtcell{0.08} & \tmtcell{0.09} & \tmtcell{0.09} \\
Nbody\_prefactor & \tmtcell{0.03} & \tmtcell{0.03} & \tmtcell{0.03} & \tmtcell{0.10} & \tmtcell{0.04} & \tmtcell{0.04} & \tmtcell{0.04} & \tmtcell{0.04} & \tmtcell{0.04} & \tmtcell{0.04} & \tmtcell{0.04} & \tmtcell{0.04} & \tmtcell{0.04} \\
Misc & \tmtcell{0.28} & \tmtcell{0.29} & \tmtcell{0.28} & \tmtcell{0.45} & \tmtcell{0.43} & \tmtcell{0.42} & \tmtcell{0.42} & \tmtcell{0.81} & \tmtcell{0.78} & \tmtcell{0.79} & \tmtcell{0.78} & \tmtcell{0.79} & \tmtcell{0.81} \\
\hline
\end{tabular}

\caption{Timing results (\vfoldftst)
for generating 200 
\eebbbb\ events (channel GF11)
with \ooo\ folding
using step10b in MINT2,
on the ``CUDA'' node.}
\label{tab:nofCuda5363}
\vspace*{1.2cm}
\hspace*{-1cm}
\small
\setlength\tabcolsep{5pt} 
\providecommand{\tmtcell}[1]{\makebox[\tmtwidth][r]{#1}}
\begin{tabular}{lrrrrrrrrrrrrr}
\multicolumn{14}{@{}l@{}}{\makebox[0pt][l]{{\bf \underline{vfold5381} $|$ epem\_bbbxbx $|$ step10c $|$ NEVENTS=200 $|$ folding=1,1,1 $|$ kbatch=512}}} \\
\multicolumn{14}{@{}l@{}}{\makebox[0pt][l]{\footnotesize {\em bsffused -1: fused single-pass BSF kernel OFF for F0/F1, ON for F2/F/C}}} \\
\multicolumn{14}{@{}l@{}}{\makebox[0pt][l]{\footnotesize {\em sborn 1: SBORN served from cudacpp, in mode C only}}} \\
\multicolumn{14}{@{}l@{}}{\makebox[0pt][l]{\footnotesize {\em gscale 1: Borns at the real-emission coupling $g'$ obtained by rescaling those at $g$, in mode C only}}} \\
\multicolumn{14}{@{}l@{}}{\makebox[0pt][l]{\footnotesize {\em timers: getrusage in all modes}}} \\
\multicolumn{14}{@{}l@{}}{\makebox[0pt][l]{\footnotesize {\em mode F0: 7675 kpoints = 306 virt + 7369 novi $\to$ 200 events (10 virt + 190 novi)}}} \\
\multicolumn{14}{@{}l@{}}{\makebox[0pt][l]{\footnotesize {\em mode F, F1, F2, Cs: 8146 kpoints = 287 virt + 7859 novi $\to$ 200 events (9 virt + 191 novi)}}} \\
\multicolumn{14}{@{}l@{}}{\makebox[0pt][l]{\footnotesize {\em mode F, Cs: Born / Real 34 / 25 batched calls, mean batch 409.2 / 424.6}}} \\
\multicolumn{14}{@{}l@{}}{\makebox[0pt][l]{\footnotesize {\em F0*: F0 rescaled to the kpoints of F (separately for virt and novi)}}} \\
\multicolumn{14}{@{}l@{}}{\makebox[0pt][l]{\footnotesize {\em ml\_reduction: ninja for all}}} \\
\hline
mode & \tmtcell{F0*} & \tmtcell{F1} & \tmtcell{F2} & \tmtcell{F} & \tmtcell{C} & \tmtcell{C} & \tmtcell{C} & \tmtcell{C} & \tmtcell{C} & \tmtcell{C} & \tmtcell{C} & \tmtcell{C} & \tmtcell{C} \\
cudacpp\_backend & \tmtcell{-} & \tmtcell{-} & \tmtcell{-} & \tmtcell{-} & \tmtcell{none} & \tmtcell{sse4} & \tmtcell{avx2} & \tmtcell{cuda} & \tmtcell{cuda} & \tmtcell{cuda} & \tmtcell{cuda} & \tmtcell{cuda} & \tmtcell{cuda} \\
cudacpp\_vecsize & \tmtcell{-} & \tmtcell{-} & \tmtcell{-} & \tmtcell{-} & \tmtcell{1} & \tmtcell{2} & \tmtcell{4} & \tmtcell{32} & \tmtcell{64} & \tmtcell{128} & \tmtcell{256} & \tmtcell{512} & \tmtcell{1024} \\
\hline
TOTAL & \tmtcell{8.34} & \tmtcell{8.36} & \tmtcell{7.02} & \tmtcell{9.17} & \tmtcell{6.43} & \tmtcell{4.28} & \tmtcell{2.99} & \tmtcell{2.69} & \tmtcell{2.57} & \tmtcell{2.47} & \tmtcell{2.46} & \tmtcell{2.45} & \tmtcell{2.49} \\
Born & \tmtcell{2.28} & \tmtcell{2.29} & \tmtcell{1.27} & \tmtcell{1.73} & \tmtcell{0.72} & \tmtcell{0.41} & \tmtcell{0.21} & \tmtcell{0.19} & \tmtcell{0.12} & \tmtcell{0.09} & \tmtcell{0.07} & \tmtcell{0.06} & \tmtcell{0.07} \\
Real & \tmtcell{3.73} & \tmtcell{3.73} & \tmtcell{3.70} & \tmtcell{4.67} & \tmtcell{3.79} & \tmtcell{1.97} & \tmtcell{0.85} & \tmtcell{0.22} & \tmtcell{0.14} & \tmtcell{0.10} & \tmtcell{0.08} & \tmtcell{0.08} & \tmtcell{0.10} \\
Virtual & \tmtcell{1.22} & \tmtcell{1.23} & \tmtcell{1.24} & \tmtcell{1.23} & \tmtcell{1.25} & \tmtcell{1.25} & \tmtcell{1.28} & \tmtcell{1.25} & \tmtcell{1.26} & \tmtcell{1.24} & \tmtcell{1.26} & \tmtcell{1.26} & \tmtcell{1.26} \\
PS\_Generation & \tmtcell{0.05} & \tmtcell{0.05} & \tmtcell{0.05} & \tmtcell{0.10} & \tmtcell{0.10} & \tmtcell{0.10} & \tmtcell{0.10} & \tmtcell{0.10} & \tmtcell{0.10} & \tmtcell{0.10} & \tmtcell{0.10} & \tmtcell{0.10} & \tmtcell{0.10} \\
CT\_and\_MCsub & \tmtcell{0.74} & \tmtcell{0.74} & \tmtcell{0.43} & \tmtcell{0.87} & \tmtcell{0.09} & \tmtcell{0.09} & \tmtcell{0.09} & \tmtcell{0.09} & \tmtcell{0.09} & \tmtcell{0.09} & \tmtcell{0.09} & \tmtcell{0.09} & \tmtcell{0.09} \\
Nbody\_prefactor & \tmtcell{0.03} & \tmtcell{0.03} & \tmtcell{0.03} & \tmtcell{0.10} & \tmtcell{0.04} & \tmtcell{0.04} & \tmtcell{0.04} & \tmtcell{0.04} & \tmtcell{0.04} & \tmtcell{0.04} & \tmtcell{0.04} & \tmtcell{0.04} & \tmtcell{0.04} \\
Misc & \tmtcell{0.30} & \tmtcell{0.30} & \tmtcell{0.30} & \tmtcell{0.47} & \tmtcell{0.44} & \tmtcell{0.44} & \tmtcell{0.43} & \tmtcell{0.83} & \tmtcell{0.83} & \tmtcell{0.82} & \tmtcell{0.82} & \tmtcell{0.83} & \tmtcell{0.83} \\
\hline
\end{tabular}

\caption{Timing results (\vfoldfteo)
for generating 200 
\eebbbb\ events (channel GF11)
with \ooo\ folding
using step10c in MINT2,
on the ``CUDA'' node.}
\label{tab:nofCuda5381}
}\end{center}
\end{table}

\begin{table}[p]
\begin{center}{
\hspace*{-1cm}
\small
\setlength\tabcolsep{5pt} 
\providecommand{\tmtcell}[1]{\makebox[\tmtwidth][r]{#1}}
\begin{tabular}{lrrrrrrrrrrrrr}
\multicolumn{14}{@{}l@{}}{\makebox[0pt][l]{{\bf \underline{vfold5381cg} $|$ epem\_bbbxbx $|$ step14t $|$ NEVENTS=200 $|$ folding=1,1,1 $|$ kbatch=512}}} \\
\multicolumn{14}{@{}l@{}}{\makebox[0pt][l]{\footnotesize {\em sborn 1: SBORN served from cudacpp, in mode C only}}} \\
\multicolumn{14}{@{}l@{}}{\makebox[0pt][l]{\footnotesize {\em gscale 1: Borns at the real-emission coupling $g'$ obtained by rescaling those at $g$, in mode C only}}} \\
\multicolumn{14}{@{}l@{}}{\makebox[0pt][l]{\footnotesize {\em timers: getrusage in F0*, F1 (a syscall per timer call), clock\_gettime in all other modes}}} \\
\multicolumn{14}{@{}l@{}}{\makebox[0pt][l]{\footnotesize {\em mode F0: 7675 kpoints = 306 virt + 7369 novi $\to$ 200 events (10 virt + 190 novi)}}} \\
\multicolumn{14}{@{}l@{}}{\makebox[0pt][l]{\footnotesize {\em mode F, F1, F2, Cs: 8146 kpoints = 287 virt + 7859 novi $\to$ 200 events (9 virt + 191 novi)}}} \\
\multicolumn{14}{@{}l@{}}{\makebox[0pt][l]{\footnotesize {\em mode F, Cs: Born / Real 17 / 16 batched calls, mean batch 511.9 / 511.8}}} \\
\multicolumn{14}{@{}l@{}}{\makebox[0pt][l]{\footnotesize {\em F0*: F0 rescaled to the kpoints of F (separately for virt and novi)}}} \\
\multicolumn{14}{@{}l@{}}{\makebox[0pt][l]{\footnotesize {\em ml\_reduction: ninja for all}}} \\
\hline
mode & \tmtcell{F0*} & \tmtcell{F1} & \tmtcell{F2} & \tmtcell{F} & \tmtcell{C} & \tmtcell{C} & \tmtcell{C} & \tmtcell{C} & \tmtcell{C} & \tmtcell{C} & \tmtcell{C} & \tmtcell{C} & \tmtcell{C} \\
cudacpp\_backend & \tmtcell{-} & \tmtcell{-} & \tmtcell{-} & \tmtcell{-} & \tmtcell{none} & \tmtcell{sse4} & \tmtcell{avx2} & \tmtcell{cuda} & \tmtcell{cuda} & \tmtcell{cuda} & \tmtcell{cuda} & \tmtcell{cuda} & \tmtcell{cuda} \\
cudacpp\_vecsize & \tmtcell{-} & \tmtcell{-} & \tmtcell{-} & \tmtcell{-} & \tmtcell{1} & \tmtcell{2} & \tmtcell{4} & \tmtcell{32} & \tmtcell{64} & \tmtcell{128} & \tmtcell{256} & \tmtcell{512} & \tmtcell{1024} \\
\hline
TOTAL & \tmtcell{8.26} & \tmtcell{8.23} & \tmtcell{7.15} & \tmtcell{7.14} & \tmtcell{5.11} & \tmtcell{3.43} & \tmtcell{2.47} & \tmtcell{2.50} & \tmtcell{2.41} & \tmtcell{2.34} & \tmtcell{2.32} & \tmtcell{2.29} & \tmtcell{2.32} \\
Born & \tmtcell{2.26} & \tmtcell{2.25} & \tmtcell{1.28} & \tmtcell{1.24} & \tmtcell{0.46} & \tmtcell{0.25} & \tmtcell{0.13} & \tmtcell{0.11} & \tmtcell{0.07} & \tmtcell{0.05} & \tmtcell{0.04} & \tmtcell{0.04} & \tmtcell{0.04} \\
Real & \tmtcell{3.69} & \tmtcell{3.69} & \tmtcell{3.66} & \tmtcell{3.58} & \tmtcell{2.96} & \tmtcell{1.50} & \tmtcell{0.65} & \tmtcell{0.16} & \tmtcell{0.10} & \tmtcell{0.08} & \tmtcell{0.06} & \tmtcell{0.06} & \tmtcell{0.07} \\
Virtual & \tmtcell{1.21} & \tmtcell{1.20} & \tmtcell{1.24} & \tmtcell{1.24} & \tmtcell{1.27} & \tmtcell{1.26} & \tmtcell{1.25} & \tmtcell{1.27} & \tmtcell{1.29} & \tmtcell{1.27} & \tmtcell{1.27} & \tmtcell{1.26} & \tmtcell{1.26} \\
PS\_Generation & \tmtcell{0.05} & \tmtcell{0.05} & \tmtcell{0.03} & \tmtcell{0.04} & \tmtcell{0.04} & \tmtcell{0.04} & \tmtcell{0.04} & \tmtcell{0.04} & \tmtcell{0.04} & \tmtcell{0.04} & \tmtcell{0.04} & \tmtcell{0.04} & \tmtcell{0.04} \\
CT\_and\_MCsub & \tmtcell{0.73} & \tmtcell{0.72} & \tmtcell{0.73} & \tmtcell{0.70} & \tmtcell{0.08} & \tmtcell{0.08} & \tmtcell{0.08} & \tmtcell{0.08} & \tmtcell{0.08} & \tmtcell{0.08} & \tmtcell{0.09} & \tmtcell{0.08} & \tmtcell{0.08} \\
Nbody\_prefactor & \tmtcell{0.03} & \tmtcell{0.03} & \tmtcell{0.00} & \tmtcell{0.07} & \tmtcell{0.00} & \tmtcell{0.00} & \tmtcell{0.00} & \tmtcell{0.00} & \tmtcell{0.00} & \tmtcell{0.00} & \tmtcell{0.00} & \tmtcell{0.00} & \tmtcell{0.00} \\
Bridge\_crt\_del & \tmtcell{0.00} & \tmtcell{0.00} & \tmtcell{0.00} & \tmtcell{0.00} & \tmtcell{0.04} & \tmtcell{0.05} & \tmtcell{0.07} & \tmtcell{0.57} & \tmtcell{0.57} & \tmtcell{0.57} & \tmtcell{0.57} & \tmtcell{0.57} & \tmtcell{0.58} \\
Misc & \tmtcell{0.30} & \tmtcell{0.30} & \tmtcell{0.22} & \tmtcell{0.27} & \tmtcell{0.24} & \tmtcell{0.24} & \tmtcell{0.24} & \tmtcell{0.25} & \tmtcell{0.24} & \tmtcell{0.24} & \tmtcell{0.24} & \tmtcell{0.24} & \tmtcell{0.24} \\
\hline
\end{tabular}

\caption{Timing results (\vfoldfcg)
for generating 200 
\eebbbb\ events (channel GF11)
with \ooo\ folding
using step14t in MINT2,
on the ``CUDA'' node.}
\label{tab:nofCuda5381cg}
\vspace*{1.2cm}
\hspace*{-1cm}
\small
\setlength\tabcolsep{5pt} 
\providecommand{\tmtcell}[1]{\makebox[\tmtwidth][r]{#1}}
\begin{tabular}{lrrrrrrrrrrrrrr}
\multicolumn{15}{@{}l@{}}{\makebox[0pt][l]{{\bf \underline{vfold7cg} $|$ epem\_bbbxbx $|$ step15b $|$ NEVENTS=200 $|$ folding=1,1,1 $|$ kbatch=512}}} \\
\multicolumn{15}{@{}l@{}}{\makebox[0pt][l]{\footnotesize {\em sborn 1: SBORN served from cudacpp, in mode C only}}} \\
\multicolumn{15}{@{}l@{}}{\makebox[0pt][l]{\footnotesize {\em gscale 1: Borns at the real-emission coupling $g'$ obtained by rescaling those at $g$, in mode C only}}} \\
\multicolumn{15}{@{}l@{}}{\makebox[0pt][l]{\footnotesize {\em timers: getrusage in F0*, F1, F1[u] (a syscall per timer call), clock\_gettime in all other modes}}} \\
\multicolumn{15}{@{}l@{}}{\makebox[0pt][l]{\footnotesize {\em mode F0: 7675 kpoints = 306 virt + 7369 novi $\to$ 200 events (10 virt + 190 novi)}}} \\
\multicolumn{15}{@{}l@{}}{\makebox[0pt][l]{\footnotesize {\em mode F, F1, F2, Cs: 8146 kpoints = 287 virt + 7859 novi $\to$ 200 events (9 virt + 191 novi)}}} \\
\multicolumn{15}{@{}l@{}}{\makebox[0pt][l]{\footnotesize {\em mode F, Cs: Born / Real 17 / 16 batched calls, mean batch 511.9 / 511.8}}} \\
\multicolumn{15}{@{}l@{}}{\makebox[0pt][l]{\footnotesize {\em F0*: F0 rescaled to the kpoints of F (separately for virt and novi)}}} \\
\multicolumn{15}{@{}l@{}}{\makebox[0pt][l]{\footnotesize {\em F1[u]: F1 without two stock MadGraph7 optimizations (cache and $\alpha\sb{S}$ rescaling of Born amplitudes)}}} \\
\multicolumn{15}{@{}l@{}}{\makebox[0pt][l]{\footnotesize {\em ml\_reduction: ninja for all}}} \\
\hline
mode & \tmtcell{F0*} & \tmtcell{F1} & \tmtcell{F1[u]} & \tmtcell{F2} & \tmtcell{F} & \tmtcell{C} & \tmtcell{C} & \tmtcell{C} & \tmtcell{C} & \tmtcell{C} & \tmtcell{C} & \tmtcell{C} & \tmtcell{C} & \tmtcell{C} \\
cudacpp\_backend & \tmtcell{-} & \tmtcell{-} & \tmtcell{-} & \tmtcell{-} & \tmtcell{-} & \tmtcell{none} & \tmtcell{sse4} & \tmtcell{avx2} & \tmtcell{cuda} & \tmtcell{cuda} & \tmtcell{cuda} & \tmtcell{cuda} & \tmtcell{cuda} & \tmtcell{cuda} \\
cudacpp\_vecsize & \tmtcell{-} & \tmtcell{-} & \tmtcell{-} & \tmtcell{-} & \tmtcell{-} & \tmtcell{1} & \tmtcell{2} & \tmtcell{4} & \tmtcell{32} & \tmtcell{64} & \tmtcell{128} & \tmtcell{256} & \tmtcell{512} & \tmtcell{1024} \\
\hline
TOTAL & \tmtcell{5.57} & \tmtcell{5.65} & \tmtcell{7.18} & \tmtcell{5.62} & \tmtcell{6.02} & \tmtcell{5.03} & \tmtcell{3.41} & \tmtcell{2.43} & \tmtcell{2.46} & \tmtcell{2.33} & \tmtcell{2.29} & \tmtcell{2.28} & \tmtcell{2.25} & \tmtcell{2.27} \\
Born & \tmtcell{0.58} & \tmtcell{0.58} & \tmtcell{2.06} & \tmtcell{0.63} & \tmtcell{0.74} & \tmtcell{0.45} & \tmtcell{0.25} & \tmtcell{0.13} & \tmtcell{0.12} & \tmtcell{0.08} & \tmtcell{0.05} & \tmtcell{0.04} & \tmtcell{0.04} & \tmtcell{0.04} \\
Real & \tmtcell{3.33} & \tmtcell{3.37} & \tmtcell{3.39} & \tmtcell{3.36} & \tmtcell{3.48} & \tmtcell{2.92} & \tmtcell{1.50} & \tmtcell{0.65} & \tmtcell{0.16} & \tmtcell{0.10} & \tmtcell{0.08} & \tmtcell{0.06} & \tmtcell{0.06} & \tmtcell{0.07} \\
Virtual & \tmtcell{1.19} & \tmtcell{1.23} & \tmtcell{1.25} & \tmtcell{1.23} & \tmtcell{1.22} & \tmtcell{1.24} & \tmtcell{1.24} & \tmtcell{1.23} & \tmtcell{1.26} & \tmtcell{1.24} & \tmtcell{1.24} & \tmtcell{1.25} & \tmtcell{1.25} & \tmtcell{1.25} \\
PS\_Generation & \tmtcell{0.04} & \tmtcell{0.04} & \tmtcell{0.05} & \tmtcell{0.02} & \tmtcell{0.04} & \tmtcell{0.04} & \tmtcell{0.04} & \tmtcell{0.04} & \tmtcell{0.04} & \tmtcell{0.04} & \tmtcell{0.04} & \tmtcell{0.04} & \tmtcell{0.04} & \tmtcell{0.04} \\
CT\_and\_MCsub & \tmtcell{0.13} & \tmtcell{0.13} & \tmtcell{0.13} & \tmtcell{0.18} & \tmtcell{0.21} & \tmtcell{0.10} & \tmtcell{0.10} & \tmtcell{0.10} & \tmtcell{0.10} & \tmtcell{0.10} & \tmtcell{0.10} & \tmtcell{0.10} & \tmtcell{0.10} & \tmtcell{0.10} \\
Nbody\_prefactor & \tmtcell{0.03} & \tmtcell{0.03} & \tmtcell{0.03} & \tmtcell{0.00} & \tmtcell{0.07} & \tmtcell{0.00} & \tmtcell{0.01} & \tmtcell{0.00} & \tmtcell{0.00} & \tmtcell{0.01} & \tmtcell{0.01} & \tmtcell{0.00} & \tmtcell{0.00} & \tmtcell{0.01} \\
Bridge\_crt\_del & \tmtcell{0.00} & \tmtcell{0.00} & \tmtcell{0.00} & \tmtcell{0.00} & \tmtcell{0.00} & \tmtcell{0.04} & \tmtcell{0.04} & \tmtcell{0.04} & \tmtcell{0.55} & \tmtcell{0.54} & \tmtcell{0.55} & \tmtcell{0.54} & \tmtcell{0.55} & \tmtcell{0.54} \\
Misc & \tmtcell{0.27} & \tmtcell{0.28} & \tmtcell{0.28} & \tmtcell{0.20} & \tmtcell{0.26} & \tmtcell{0.24} & \tmtcell{0.24} & \tmtcell{0.23} & \tmtcell{0.24} & \tmtcell{0.23} & \tmtcell{0.23} & \tmtcell{0.24} & \tmtcell{0.23} & \tmtcell{0.23} \\
\hline
\end{tabular}

\caption{Timing results (\vfoldscg)
for generating 200 
\eebbbb\ events (channel GF11)
with \ooo\ folding
using step15b in MINT2,
on the ``CUDA'' node.}
\label{tab:nofCuda7cg}
}\end{center}
\end{table}

\subsection{\eebbbbTOP:
no-folding event generation 
in \vfoldftst\ and \vfoldfteo}
\label{sec:nof363vs381}

As in the previous 
Sec.~\ref{sec:fld},
I first present results
for step10b and step10c
in \mvf\ versions 
\vfoldftst\ and 
\vfoldfteo.
These are given in
Tables~\ref{tab:nofCuda5363}
and~\ref{tab:nofCuda5381},
respectively.
(Note: step10c uses 
the same software 
as step10b, with
an extra fix
ensuring that 
MINT2 \ooo\ tests
reuse the 
\vfoldftst\ MINT0/1
grids, so that the
two tables refer 
to exactly 
the same events.)
Many of the observations
previously made for 
vectorized folding
still apply,
but there are some 
important differences:
\begin{itemize}
    
\item 
There is now a
more pronounced relative
speedup of Born
both via AVX2
and on the GPU.
The GPU speedup of Reals 
is a bit less striking
(but the benefit
of helicity streams
in \vfoldfteo\ with 
respect to \vfoldftst\ is
still clearly visible).

\item 
GPU throughputs 
improve as the CUDACPP 
\vecsize\ increases 
from 32 to 512,
the maximum event 
capacity fixed 
by kbatch=512,
but they slowly degrade
beyond it at \vecsize=1024,
as the unnecessarily large 
CUDACPP batch is padded
with dummy data.

\item
Unlike in vectorized folding,
Virtuals are
now the bottleneck
of the calculation 
after accelerating 
Real and Born MEs 
with SIMD or a GPU.
The Misc contribution
remains large
(especially on GPU
where in these step10 tables
it also includes the
hardware initialization),
but is no longer the
main issue. 

\item
The overall speedups 
achieved in \vfoldftst\ and
\vfoldfteo\ are quite similar.
For the latter,
the total time
decreases from
8.36s in F1
to 2.99s in C/avx2 (x2.8)
and 2.45s in C/cuda
with \vecsize\ 512 (x3.4).

\end{itemize}

\subsection{\eebbbbTOP:
no-folding event generation 
in \vfoldfcg\ and \vfoldscg}
\label{sec:nof5vs7}

Next, I present results
obtained with the 
most recent optimizations
developed and 
tested for \eebbbb,
in \mvf\ versions 
\vfoldfcg\ (step14t) 
and \vfoldscg\ (step15b).
These are given in
Tables~\ref{tab:nofCuda5381cg}
and~\ref{tab:nofCuda7cg}.
Again, these two tables
are directly comparable
to each other and to
the previous two
Tables~\ref{tab:nofCuda5363}
and~\ref{tab:nofCuda5381},
because they refer 
to the generation
of exactly the same events
from the same random numbers,
on the same ``CUDA'' node.
A few notable similarities
and differences with patterns
previously observed 
include the following:
\begin{itemize}

\item Performance
in CUDACPP C modes is slightly better
in step14 than it was in step10.
However, the effect 
is much less noticeable
than it was in the case
of vectorized folding,
largely because 
step11 optimizations
mainly reduce
the overhead of the
two-walk infrastructure,
which is less 
important when each event
has only 1 fold instead of~32.
A notable reduction
of the order of 20\%
is observed in several C/SIMD
contributions, which might
be an effect of the faster
profiling timers.

\item 
As in the case of 
vectorized folding,
the impact of the new
optimizations in stock MadGraph7 
is particularly large for Born,
but it is also visible
for Counterterms and Real.
(Note also that the removal
of the \mvf\ fused BSF
implementation is visible
in the F2 Counterterm in 
Table~\ref{tab:nofCuda5381cg},
which is now the same 
as the F1 value;
in any case, this optimization
is no longer needed 
in modes F2 and F, where
stock MadGraph7 provides a
better one, nor in C
modes, where colour-linked
Borns are served by CUDACPP.)

\item 
The 0.55s fixed cost
is now a substantial
fraction (20 to 25\%) 
of the total time
spent in GPU modes,
for these short tests
where only 200 
unweighted events
are generated.
It makes even more sense 
to remove it when analysing
speedups between 
different modes.

\item 
Virtuals are
now very clearly 
the bottleneck
of the calculation.
Once the GPU fixed cost
is accounted for separately,
the relative weight 
of the Misc contribution
is instead much lower
in the overall time profile.

\item
The overall speedups 
achieved in 
\vfoldfcg\ and
\vfoldscg\ differ 
because of the large improvements 
in stock MadGraph7.
In the \mgamc\ codebase,
the total time
decreases from 8.23s
to 2.47s (3.3x) in AVX2
and to 2.29s (3.6x) ---
or 1.72s (4.8x) if one
subtracts the 0.57s fixed cost ---
on the GPU.
In the MadGraph7 codebase,
the total time
decreases from 5.65s
to 2.43s (2.3x) in AVX2
and to 2.25s (2.5x) ---
or 1.70s (3.3x) if one
subtracts the 0.55s fixed cost ---
on the GPU.

\end{itemize}

\subsection{\eebbbbTOP:
comparison of
folding and no-folding 
performance}
\label{sec:fldVsNof}

At this point,
it is interesting
to compare the timing numbers
without folding 
in this section 
to those with \efo\ folding
in the previous 
Sec.~\ref{sec:fld}.
The two sets of numbers
can be directly compared,
in the sense that
they both refer 
to the generation
of 200 unweighted events.
It should be kept in mind,
however, that the 
unweighting efficiencies
are different:
with \efo\ folding,
10172 kpoints had to be computed
to generate 200 events,
while with \ooo\ folding
only 8146 kpoints were needed.
In the case of 
the Fortran-based MadGraph7 calculation
with the \mvf\ slot-based RNG,
described by column F1 
in Tables~\ref{tab:nofCuda7cg}
and~\ref{tab:fldCuda7cg},
using \efo\ folding increases
event generation times
from 5.65s to 156.0s (28x). 
Naively,
the relative cost
of \efo\ folding
might have been expected to be
even larger,
taking into account that
up to 32 times more work
is performed for each kpoint,
compounded with the fact
that around 25\% more kpoints
are needed due to a worse
unweighting efficiency.
The relative ratio
was indeed slightly worse in \mgamc,
before the new MadGraph7 optimizations,
which mainly benefit folding:
in Tables~\ref{tab:nofCuda5381cg}
and~\ref{tab:fldCuda5381cg},
in particular, F1 times
used to be 8.23s and 284.5s (35x). 

Things are different
if one looks at
the best results
achieved 
in \mvf\ on GPU
(column Ccuda
with \vecsize\ 512 or 1024
in the same tables). 
In that case,
the move from \ooo\ to
\efo\ folding only leads
to an increase
from 2.25s to 11.8s (5.2x),
or, if one subtracts
the 0.55s GPU fixed cost,
from 1.70s to 11.3s (6.6x).
This shows that the
initial main goal 
of this whole exercise,
namely reducing the
computational cost of folding,
has been achieved.
The point here is that
\mvf\ achieves very good
speedups both 
with \efo\ folding
and without any folding,
but those achievable
with folding are much larger,
because there was much more
that could be optimized away
to start with,
or, looking at this from
another angle,
because the relative weight
of Virtual calculations
is much lower with folding
than without it.

Since the aim 
of folding 
is reducing negative weights,
it is also useful 
to mention 
its effect
in this respect
for the \eebbbb\ process analysed 
in this section:
in the GF11.0 channel,
the cross section and
absolute cross section
are $I\!\sim\!2.105\!\times\!10^{-4}$~pb
and $J\!\sim\!6.079\!\times\!10^{-4}$~pb
without folding,
and $I\!\sim\!2.136\!\times\!10^{-4}$~pb
and $J\!\sim\!4.863\!\times\!10^{-4}$~pb
with \efo\ folding.
(These are the central values
of Table~\ref{tab:delta},
further discussed in 
Sec.~\ref{sec:delta},
where 
errors are also given.
Single MINT1 runs fluctuate 
by more than their quoted errors, 
so the table quotes 
means over independent seeds.)
This implies 
a reduction in the negative weight
fraction $\fneg\!=\!1/2\,(1-I/J)$
from 33\% to 28\%,
and a 
reduction
in the relative cost
$\cost(\fneg)$ 
from 8.3 to 5.2.
When comparing
these against
the increased costs
in event generation times,
one should however remember that
in most cases folding mainly
targets the reduction
of downstream costs
(detector simulation,
event reconstruction,
analysis and storage).
As previously mentioned, 
a more comprehensive analysis 
of the effect of folding 
on negative weights is
beyond the scope of this paper;
nevertheless, in the following
a few more examples will be given,
for illustrative purposes only.

\begin{table}[p]
\begin{center}{
\hspace*{-1cm}
\small
\setlength\tabcolsep{5pt} 
\providecommand{\tmtcell}[1]{\makebox[\tmtwidth][r]{#1}}
\begin{tabular}{lrrrrrrrrrrrrrr}
\multicolumn{15}{@{}l@{}}{\makebox[0pt][l]{{\bf \underline{vfold7cg} $|$ epem\_bbbxbx $|$ step15b $|$ NEVENTS=200 $|$ folding=8,4,1 $|$ MC@NLO-$\Delta$ $|$ kbatch=64}}} \\
\multicolumn{15}{@{}l@{}}{\makebox[0pt][l]{\footnotesize {\em sborn 1: SBORN served from cudacpp, in mode C only}}} \\
\multicolumn{15}{@{}l@{}}{\makebox[0pt][l]{\footnotesize {\em gscale 1: Borns at the real-emission coupling $g'$ obtained by rescaling those at $g$, in mode C only}}} \\
\multicolumn{15}{@{}l@{}}{\makebox[0pt][l]{\footnotesize {\em timers: getrusage in F0*, F1 (a syscall per timer call), clock\_gettime in all other modes}}} \\
\multicolumn{15}{@{}l@{}}{\makebox[0pt][l]{\footnotesize {\em mode F0: 9777 kpoints = 668 virt + 9109 novi $\to$ 200 events (19 virt + 181 novi)}}} \\
\multicolumn{15}{@{}l@{}}{\makebox[0pt][l]{\footnotesize {\em mode F, F1, F2, Cs: 8106 kpoints = 440 virt + 7666 novi $\to$ 200 events (17 virt + 183 novi)}}} \\
\multicolumn{15}{@{}l@{}}{\makebox[0pt][l]{\footnotesize {\em mode F, Cs: Born / Real 127 / 120 batched calls, mean batch 2047.6 / 2047.1}}} \\
\multicolumn{15}{@{}l@{}}{\makebox[0pt][l]{\footnotesize {\em F0*: F0 rescaled to the kpoints of F (separately for virt and novi)}}} \\
\multicolumn{15}{@{}l@{}}{\makebox[0pt][l]{\footnotesize {\em ml\_reduction: ninja for all}}} \\
\hline
mode & \tmtcell{F0*} & \tmtcell{F1} & \tmtcell{F2} & \tmtcell{F} & \tmtcell{C} & \tmtcell{C} & \tmtcell{C} & \tmtcell{C} & \tmtcell{C} & \tmtcell{C} & \tmtcell{C} & \tmtcell{C} & \tmtcell{C} & \tmtcell{C} \\
cudacpp\_backend & \tmtcell{-} & \tmtcell{-} & \tmtcell{-} & \tmtcell{-} & \tmtcell{none} & \tmtcell{sse4} & \tmtcell{avx2} & \tmtcell{cuda} & \tmtcell{cuda} & \tmtcell{cuda} & \tmtcell{cuda} & \tmtcell{cuda} & \tmtcell{cuda} & \tmtcell{cuda} \\
cudacpp\_vecsize & \tmtcell{-} & \tmtcell{-} & \tmtcell{-} & \tmtcell{-} & \tmtcell{1} & \tmtcell{2} & \tmtcell{4} & \tmtcell{32} & \tmtcell{64} & \tmtcell{128} & \tmtcell{256} & \tmtcell{512} & \tmtcell{1024} & \tmtcell{2048} \\
\hline
TOTAL & \tmtcell{172.6} & \tmtcell{174.3} & \tmtcell{170.0} & \tmtcell{171.8} & \tmtcell{145.0} & \tmtcell{101.8} & \tmtcell{77.3} & \tmtcell{64.2} & \tmtcell{61.8} & \tmtcell{61.1} & \tmtcell{60.7} & \tmtcell{60.6} & \tmtcell{60.0} & \tmtcell{60.3} \\
Born & \tmtcell{3.1} & \tmtcell{3.1} & \tmtcell{5.8} & \tmtcell{5.9} & \tmtcell{1.0} & \tmtcell{0.8} & \tmtcell{0.7} & \tmtcell{0.7} & \tmtcell{0.6} & \tmtcell{0.6} & \tmtcell{0.6} & \tmtcell{0.6} & \tmtcell{0.6} & \tmtcell{0.7} \\
Real & \tmtcell{106.0} & \tmtcell{107.5} & \tmtcell{106.8} & \tmtcell{104.5} & \tmtcell{88.4} & \tmtcell{44.9} & \tmtcell{19.7} & \tmtcell{4.2} & \tmtcell{2.5} & \tmtcell{1.6} & \tmtcell{1.2} & \tmtcell{1.1} & \tmtcell{1.0} & \tmtcell{1.0} \\
Virtual & \tmtcell{1.8} & \tmtcell{1.8} & \tmtcell{1.8} & \tmtcell{1.9} & \tmtcell{1.9} & \tmtcell{1.9} & \tmtcell{1.9} & \tmtcell{1.9} & \tmtcell{1.9} & \tmtcell{1.9} & \tmtcell{1.9} & \tmtcell{1.9} & \tmtcell{1.9} & \tmtcell{1.9} \\
PS\_Generation & \tmtcell{1.7} & \tmtcell{1.7} & \tmtcell{1.2} & \tmtcell{1.6} & \tmtcell{1.5} & \tmtcell{1.6} & \tmtcell{1.6} & \tmtcell{1.6} & \tmtcell{1.6} & \tmtcell{1.6} & \tmtcell{1.6} & \tmtcell{1.7} & \tmtcell{1.6} & \tmtcell{1.6} \\
CT\_and\_MCsub & \tmtcell{7.0} & \tmtcell{6.6} & \tmtcell{5.1} & \tmtcell{6.1} & \tmtcell{4.1} & \tmtcell{4.1} & \tmtcell{4.1} & \tmtcell{4.1} & \tmtcell{4.1} & \tmtcell{4.1} & \tmtcell{4.1} & \tmtcell{4.1} & \tmtcell{4.1} & \tmtcell{4.1} \\
Delta & \tmtcell{42.5} & \tmtcell{42.8} & \tmtcell{42.3} & \tmtcell{41.4} & \tmtcell{40.3} & \tmtcell{40.6} & \tmtcell{41.3} & \tmtcell{43.3} & \tmtcell{42.6} & \tmtcell{42.9} & \tmtcell{42.7} & \tmtcell{42.7} & \tmtcell{42.4} & \tmtcell{42.7} \\
Nbody\_prefactor & \tmtcell{0.9} & \tmtcell{0.9} & \tmtcell{0.2} & \tmtcell{2.5} & \tmtcell{0.2} & \tmtcell{0.2} & \tmtcell{0.2} & \tmtcell{0.2} & \tmtcell{0.2} & \tmtcell{0.2} & \tmtcell{0.2} & \tmtcell{0.2} & \tmtcell{0.2} & \tmtcell{0.2} \\
Bridge\_crt\_del & \tmtcell{0.0} & \tmtcell{0.0} & \tmtcell{0.0} & \tmtcell{0.0} & \tmtcell{0.0} & \tmtcell{0.0} & \tmtcell{0.0} & \tmtcell{0.6} & \tmtcell{0.6} & \tmtcell{0.5} & \tmtcell{0.5} & \tmtcell{0.5} & \tmtcell{0.5} & \tmtcell{0.6} \\
Misc & \tmtcell{9.6} & \tmtcell{9.8} & \tmtcell{6.7} & \tmtcell{7.8} & \tmtcell{7.6} & \tmtcell{7.7} & \tmtcell{7.8} & \tmtcell{7.7} & \tmtcell{7.7} & \tmtcell{7.7} & \tmtcell{7.8} & \tmtcell{7.8} & \tmtcell{7.7} & \tmtcell{7.6} \\
\hline
\end{tabular}

\caption{Timing results (\vfoldscg)
for generating 200 
\eebbbb\ events (channel GF11)
with \efo\ folding and 
\amcDELTA\ using step15b in MINT2,
on the ``CUDA'' node.}
\label{tab:fldCudaDelta}
\vspace*{1.2cm}
\hspace*{-1cm}
\small
\setlength\tabcolsep{5pt} 
\providecommand{\tmtcell}[1]{\makebox[\tmtwidth][r]{#1}}
\begin{tabular}{lrrrrrrrrrrrrr}
\multicolumn{14}{@{}l@{}}{\makebox[0pt][l]{{\bf \underline{vfold7cg} $|$ epem\_bbbxbx $|$ step15b $|$ NEVENTS=200 $|$ folding=1,1,1 $|$ MC@NLO-$\Delta$ $|$ kbatch=512}}} \\
\multicolumn{14}{@{}l@{}}{\makebox[0pt][l]{\footnotesize {\em sborn 1: SBORN served from cudacpp, in mode C only}}} \\
\multicolumn{14}{@{}l@{}}{\makebox[0pt][l]{\footnotesize {\em gscale 1: Borns at the real-emission coupling $g'$ obtained by rescaling those at $g$, in mode C only}}} \\
\multicolumn{14}{@{}l@{}}{\makebox[0pt][l]{\footnotesize {\em timers: getrusage in F0*, F1 (a syscall per timer call), clock\_gettime in all other modes}}} \\
\multicolumn{14}{@{}l@{}}{\makebox[0pt][l]{\footnotesize {\em mode F0: 8302 kpoints = 223 virt + 8079 novi $\to$ 200 events (11 virt + 189 novi)}}} \\
\multicolumn{14}{@{}l@{}}{\makebox[0pt][l]{\footnotesize {\em mode F, F1, F2, Cs: 8758 kpoints = 472 virt + 8286 novi $\to$ 200 events (11 virt + 189 novi)}}} \\
\multicolumn{14}{@{}l@{}}{\makebox[0pt][l]{\footnotesize {\em mode F, Cs: Born / Real 18 / 17 batched calls, mean batch 511.9 / 511.7}}} \\
\multicolumn{14}{@{}l@{}}{\makebox[0pt][l]{\footnotesize {\em F0*: F0 rescaled to the kpoints of F (separately for virt and novi)}}} \\
\multicolumn{14}{@{}l@{}}{\makebox[0pt][l]{\footnotesize {\em ml\_reduction: ninja for all}}} \\
\hline
mode & \tmtcell{F0*} & \tmtcell{F1} & \tmtcell{F2} & \tmtcell{F} & \tmtcell{C} & \tmtcell{C} & \tmtcell{C} & \tmtcell{C} & \tmtcell{C} & \tmtcell{C} & \tmtcell{C} & \tmtcell{C} & \tmtcell{C} \\
cudacpp\_backend & \tmtcell{-} & \tmtcell{-} & \tmtcell{-} & \tmtcell{-} & \tmtcell{none} & \tmtcell{sse4} & \tmtcell{avx2} & \tmtcell{cuda} & \tmtcell{cuda} & \tmtcell{cuda} & \tmtcell{cuda} & \tmtcell{cuda} & \tmtcell{cuda} \\
cudacpp\_vecsize & \tmtcell{-} & \tmtcell{-} & \tmtcell{-} & \tmtcell{-} & \tmtcell{1} & \tmtcell{2} & \tmtcell{4} & \tmtcell{32} & \tmtcell{64} & \tmtcell{128} & \tmtcell{256} & \tmtcell{512} & \tmtcell{1024} \\
\hline
TOTAL & \tmtcell{8.60} & \tmtcell{8.52} & \tmtcell{8.58} & \tmtcell{8.99} & \tmtcell{7.92} & \tmtcell{6.14} & \tmtcell{5.14} & \tmtcell{5.19} & \tmtcell{5.07} & \tmtcell{4.97} & \tmtcell{4.99} & \tmtcell{4.99} & \tmtcell{5.00} \\
Born & \tmtcell{0.63} & \tmtcell{0.64} & \tmtcell{0.70} & \tmtcell{0.79} & \tmtcell{0.48} & \tmtcell{0.27} & \tmtcell{0.14} & \tmtcell{0.12} & \tmtcell{0.08} & \tmtcell{0.06} & \tmtcell{0.05} & \tmtcell{0.04} & \tmtcell{0.05} \\
Real & \tmtcell{3.58} & \tmtcell{3.58} & \tmtcell{3.60} & \tmtcell{3.73} & \tmtcell{3.14} & \tmtcell{1.58} & \tmtcell{0.70} & \tmtcell{0.18} & \tmtcell{0.11} & \tmtcell{0.08} & \tmtcell{0.07} & \tmtcell{0.06} & \tmtcell{0.08} \\
Virtual & \tmtcell{2.02} & \tmtcell{1.93} & \tmtcell{1.95} & \tmtcell{1.97} & \tmtcell{2.00} & \tmtcell{1.98} & \tmtcell{2.00} & \tmtcell{2.00} & \tmtcell{2.02} & \tmtcell{1.99} & \tmtcell{2.00} & \tmtcell{2.02} & \tmtcell{2.02} \\
PS\_Generation & \tmtcell{0.06} & \tmtcell{0.06} & \tmtcell{0.05} & \tmtcell{0.06} & \tmtcell{0.06} & \tmtcell{0.06} & \tmtcell{0.06} & \tmtcell{0.06} & \tmtcell{0.06} & \tmtcell{0.06} & \tmtcell{0.06} & \tmtcell{0.06} & \tmtcell{0.06} \\
CT\_and\_MCsub & \tmtcell{0.24} & \tmtcell{0.22} & \tmtcell{0.28} & \tmtcell{0.32} & \tmtcell{0.20} & \tmtcell{0.20} & \tmtcell{0.20} & \tmtcell{0.20} & \tmtcell{0.20} & \tmtcell{0.20} & \tmtcell{0.20} & \tmtcell{0.20} & \tmtcell{0.20} \\
Delta & \tmtcell{1.64} & \tmtcell{1.63} & \tmtcell{1.64} & \tmtcell{1.60} & \tmtcell{1.59} & \tmtcell{1.60} & \tmtcell{1.58} & \tmtcell{1.68} & \tmtcell{1.65} & \tmtcell{1.63} & \tmtcell{1.66} & \tmtcell{1.67} & \tmtcell{1.65} \\
Nbody\_prefactor & \tmtcell{0.03} & \tmtcell{0.03} & \tmtcell{0.01} & \tmtcell{0.09} & \tmtcell{0.01} & \tmtcell{0.01} & \tmtcell{0.01} & \tmtcell{0.01} & \tmtcell{0.01} & \tmtcell{0.01} & \tmtcell{0.01} & \tmtcell{0.01} & \tmtcell{0.01} \\
Bridge\_crt\_del & \tmtcell{0.00} & \tmtcell{0.00} & \tmtcell{0.00} & \tmtcell{0.00} & \tmtcell{0.04} & \tmtcell{0.04} & \tmtcell{0.05} & \tmtcell{0.54} & \tmtcell{0.54} & \tmtcell{0.54} & \tmtcell{0.54} & \tmtcell{0.54} & \tmtcell{0.55} \\
Misc & \tmtcell{0.41} & \tmtcell{0.43} & \tmtcell{0.36} & \tmtcell{0.43} & \tmtcell{0.41} & \tmtcell{0.41} & \tmtcell{0.40} & \tmtcell{0.40} & \tmtcell{0.40} & \tmtcell{0.40} & \tmtcell{0.40} & \tmtcell{0.40} & \tmtcell{0.40} \\
\hline
\end{tabular}

\caption{Timing results (\vfoldscg)
for generating 200 
\eebbbb\ events (channel GF11)
with \ooo\ folding and
\amcDELTA\ using step15b in MINT2,
on the ``CUDA'' node.}
\label{tab:nofCudaDelta}
}\end{center}
\end{table}

\begin{table}[t]
\hspace*{-1cm}
\centering
\begin{tabular}{|cc|cccc|cc|}
\hline
folding & MC@NLO-$\Delta$
 & $I$ [$10^{-4}$\,pb] & $J$ [$10^{-4}$\,pb] & $F$ [\%] & $C(F)$
 & F1 & C/cuda \\
\hline
\ooo & no  & $2.105 \pm 0.034$ & $6.079 \pm 0.036$ & $32.7 \pm 0.3$ & $8.34 \pm 0.29$ &   5.65s &  2.25s \\
\ooo & yes & $2.072 \pm 0.058$ & $5.468 \pm 0.059$ & $31.1 \pm 0.6$ & $6.96 \pm 0.42$ &   8.52s &  4.97s \\
\efo & no  & $2.136 \pm 0.041$ & $4.863 \pm 0.044$ & $28.0 \pm 0.5$ & $5.18 \pm 0.22$ & 156.0s  & 11.8s  \\
\efo & yes & $2.110 \pm 0.040$ & $4.000 \pm 0.037$ & $23.6 \pm 0.6$ & $3.60 \pm 0.15$ & 174.3s  & 60.0s  \\
\hline
\end{tabular}
\caption{Summary for \eebbbb\ in
channel GF11.0, with and without
folding and \amcDELTA. 
Cross section $I$, 
absolute cross section $J$, 
negative weight fraction $\fneg=1/2(1-I/J)$ 
and relative cost $\cost(\fneg)=(J/I)^2$ 
are weighted means over 
independent MINT1 results 
(16 runs for the first row,
6 runs otherwise; 
default req\_acc=-1 and nevents=100, 
vfold5381 step11t). 
Times refer to 
the MINT2 
generation of 200 unweighted events 
in modes F1 and in 
the fastest C/cuda configuration 
(vfold7cg step15b, ``CUDA'' node), 
from Tables~\ref{tab:nofCuda7cg}, 
\ref{tab:nofCudaDelta}, 
\ref{tab:fldCuda7cg} 
and~\ref{tab:fldCudaDelta}
for the four rows, respectively.
Note that the MINT2 tests 
of Tables~\ref{tab:fldCuda5363}--\ref{tab:fldCuda7cg} 
(\efo\ folding)
and Tables~\ref{tab:nofCuda5363}--\ref{tab:nofCuda7cg} 
(\ooo\ folding)
reuse the MINT1 grids 
of two integration runs (one per folding) 
with the same random seed 
and hence the same MINT0 grids; 
in both, the integral $I$ is 
about two standard deviations 
above the mean values 
of Table~\ref{tab:delta}: 
this does not affect 
the timing comparisons, 
since all modes use the same grids.
}
\label{tab:delta}
\end{table}

\section{Results for 
\eebbbbTOP\ with \deltaTOP}
\label{sec:delta}

In this section,
I present timing results 
for \eebbbb\ NLO 
vectorized event generation
in the \vfoldscg\ version of \mvf,
when \amcDELTA\ is enabled.
Time numbers in seconds 
for \efo\ and \ooo\ folding
are shown
in Tables~\ref{tab:fldCudaDelta}
and~\ref{tab:nofCudaDelta},
respectively,
which once again
refer to the MINT2
generation of 200 unweighted events
in a single channel GF11.0
on the ``CUDA'' machine.
In this respect,
these \vfoldscg\ timing results
are directly comparable
to those previously given
in Tables~\ref{tab:fldCuda7cg}
and~\ref{tab:nofCuda7cg},
for \efo\ and \ooo\ folding.
(One should keep in mind
however that the number
of kpoints required
to generate 200 events
seems quite different
with and without \amcDELTA,
either because the 
unweighting efficiencies
are not the same,
or because of statistical
fluctuations in the 
unweighting process.)
The total times spent 
in the F1 mode and 
in the fastest C/cuda mode
(that with most optimal \vecsize)
from those four tables
are also summarized
in Table~\ref{tab:delta},
which also gives
the cross section,
absolute cross section,
negative weight fraction $\fneg$
and corresponding cost $\cost(\fneg)$ for
these four configurations.
The following comments can be made:

\begin{itemize}

\item While both folding 
and \amcDELTA\ achieve
good reductions in negative weights
when taken individually,
combining them achieves
an even better result
than one may have expected.
Moving to \efo\ folding 
with \amcDELTA\ from 
\ooo\ folding without it,
in fact, reduces the 
cost $\cost(\fneg)$
of negative weights
from 8.34 to 3.60 (2.3x),
while the product
of the two individual
reductions would give
$\cost(\fneg)\sim4.3$.

\item The vectorization
speedups achieved 
for Born and Real
with \amcDELTA\ are
very similar to those
observed without it:
in other words,
\mvf\ may be used 
to speed up
NLO event generation,
with or without folding,
also when \amcDELTA\ is enabled.
Using \amcDELTA, however,
implies additional 
computational costs,
especially that
of the contribution
indicated as ``Delta''
in the timing tables,
which is a major
performance bottleneck
in the \mvf\ C/cuda results (and 
is essentially
flat across all modes,
because it is presently
not vectorized):
in fact, this accounts
for 1.63s out of 4.97s 
with \ooo\ folding,
and for 42.4s out of 60.0s
with \efo\ folding.
In the tests described
in the tables,
the ``Delta'' calculation
was dominated ($\sim$90\%)
by calls to the
Pythia 8.317 
library, linked with MadGraph7.
This is because,
as previously noted, 
\mvf\ \vfoldscg\ is based 
on the MadGraph7 codebase
as of 8 Nov 2026.
Only a few days later,
however, this part
of the \amcDELTA\ calculation
was replaced
by a native Fortran 
port~\cite{bib:pr397}
inside MadGraph7.
As a consequence,
these results are already
somewhat obsolete,
and it is possible
that different timing profiles
would be observed
with the new MadGraph7
implementation.
If, on the other hand,
it was found that 
the ``Delta'' component
is still the bottleneck
in vectorized folding
with \amcDELTA,
the fact that this is 
now part of the MadGraph7
repository would certainly
be helpful for its
possible optimization.

\end{itemize}

\begin{table}[p]
\begin{center}{
\hspace*{-1.4cm}
\small
\setlength\tabcolsep{5pt} 
\providecommand{\tmtcell}[1]{\makebox[\tmtwidth][r]{#1}}
\begin{tabular}{lrrrrrrrrrrrrrrr}
\multicolumn{16}{@{}l@{}}{\makebox[0pt][l]{{\bf \underline{vfold7cg} $|$ epem\_bbxmummupg $|$ step15e $|$ NEVENTS=200 $|$ folding=8,4,1 $|$ kbatch=64}}} \\
\multicolumn{16}{@{}l@{}}{\makebox[0pt][l]{\footnotesize {\em sborn 1: SBORN served from cudacpp, in mode C only}}} \\
\multicolumn{16}{@{}l@{}}{\makebox[0pt][l]{\footnotesize {\em gscale 1: Borns at the real-emission coupling $g'$ obtained by rescaling those at $g$, in mode C only}}} \\
\multicolumn{16}{@{}l@{}}{\makebox[0pt][l]{\footnotesize {\em timers: getrusage in F0*, F1[n], F1[c] (a syscall per timer call), clock\_gettime in all other modes}}} \\
\multicolumn{16}{@{}l@{}}{\makebox[0pt][l]{\footnotesize {\em mode F0: 14656 kpoints = 5713 virt + 8943 novi $\to$ 200 events (40 virt + 160 novi)}}} \\
\multicolumn{16}{@{}l@{}}{\makebox[0pt][l]{\footnotesize {\em mode F, F1, F2, Cs: 14667 kpoints = 4568 virt + 10099 novi $\to$ 200 events (30 virt + 170 novi)}}} \\
\multicolumn{16}{@{}l@{}}{\makebox[0pt][l]{\footnotesize {\em mode F, Cs: Born / Real 230 / 158 batches, mean batch 1604.6 / 3512.2 (mode C: each Real batch is split over 3 libraries)}}} \\
\multicolumn{16}{@{}l@{}}{\makebox[0pt][l]{\footnotesize {\em F0*: F0 rescaled to the kpoints of F (separately for virt and novi)}}} \\
\hline
mode & \tmtcell{F0*} & \tmtcell{F1[n]} & \tmtcell{F1[c]} & \tmtcell{F2} & \tmtcell{F} & \tmtcell{C} & \tmtcell{C} & \tmtcell{C} & \tmtcell{C} & \tmtcell{C} & \tmtcell{C} & \tmtcell{C} & \tmtcell{C} & \tmtcell{C} & \tmtcell{C} \\
cudacpp\_backend & \tmtcell{-} & \tmtcell{-} & \tmtcell{-} & \tmtcell{-} & \tmtcell{-} & \tmtcell{none} & \tmtcell{sse4} & \tmtcell{avx2} & \tmtcell{cuda} & \tmtcell{cuda} & \tmtcell{cuda} & \tmtcell{cuda} & \tmtcell{cuda} & \tmtcell{cuda} & \tmtcell{cuda} \\
cudacpp\_vecsize & \tmtcell{-} & \tmtcell{-} & \tmtcell{-} & \tmtcell{-} & \tmtcell{-} & \tmtcell{1} & \tmtcell{2} & \tmtcell{4} & \tmtcell{32} & \tmtcell{64} & \tmtcell{128} & \tmtcell{256} & \tmtcell{512} & \tmtcell{1024} & \tmtcell{2048} \\
ml\_reduction & \tmtcell{collier} & \tmtcell{ninja} & \tmtcell{collier} & \tmtcell{collier} & \tmtcell{collier} & \tmtcell{collier} & \tmtcell{collier} & \tmtcell{collier} & \tmtcell{collier} & \tmtcell{collier} & \tmtcell{collier} & \tmtcell{collier} & \tmtcell{collier} & \tmtcell{collier} & \tmtcell{collier} \\
\hline
TOTAL & \tmtcell{466.6} & \tmtcell{467.4} & \tmtcell{460.2} & \tmtcell{454.3} & \tmtcell{500.8} & \tmtcell{361.2} & \tmtcell{221.9} & \tmtcell{145.0} & \tmtcell{97.3} & \tmtcell{94.3} & \tmtcell{91.0} & \tmtcell{90.2} & \tmtcell{90.4} & \tmtcell{90.4} & \tmtcell{91.2} \\
Born & \tmtcell{35.9} & \tmtcell{35.8} & \tmtcell{35.4} & \tmtcell{36.9} & \tmtcell{44.6} & \tmtcell{3.1} & \tmtcell{2.4} & \tmtcell{2.0} & \tmtcell{1.9} & \tmtcell{1.8} & \tmtcell{1.7} & \tmtcell{1.7} & \tmtcell{1.8} & \tmtcell{2.0} & \tmtcell{2.3} \\
Real & \tmtcell{333.7} & \tmtcell{333.1} & \tmtcell{329.9} & \tmtcell{333.5} & \tmtcell{331.1} & \tmtcell{273.1} & \tmtcell{135.0} & \tmtcell{58.1} & \tmtcell{9.5} & \tmtcell{6.2} & \tmtcell{4.1} & \tmtcell{3.1} & \tmtcell{2.7} & \tmtcell{2.7} & \tmtcell{2.9} \\
Virtual & \tmtcell{43.5} & \tmtcell{46.2} & \tmtcell{43.2} & \tmtcell{43.8} & \tmtcell{44.0} & \tmtcell{44.1} & \tmtcell{43.7} & \tmtcell{43.7} & \tmtcell{44.4} & \tmtcell{44.6} & \tmtcell{44.6} & \tmtcell{44.4} & \tmtcell{44.9} & \tmtcell{44.7} & \tmtcell{44.7} \\
PS\_Generation & \tmtcell{5.0} & \tmtcell{4.9} & \tmtcell{4.9} & \tmtcell{4.5} & \tmtcell{7.3} & \tmtcell{7.2} & \tmtcell{7.1} & \tmtcell{7.4} & \tmtcell{7.3} & \tmtcell{7.4} & \tmtcell{7.1} & \tmtcell{7.2} & \tmtcell{7.2} & \tmtcell{7.2} & \tmtcell{7.4} \\
CT\_and\_MCsub & \tmtcell{9.1} & \tmtcell{8.9} & \tmtcell{8.8} & \tmtcell{8.7} & \tmtcell{16.9} & \tmtcell{3.3} & \tmtcell{3.4} & \tmtcell{3.4} & \tmtcell{3.2} & \tmtcell{3.3} & \tmtcell{3.2} & \tmtcell{3.3} & \tmtcell{3.2} & \tmtcell{3.2} & \tmtcell{3.2} \\
Nbody\_prefactor & \tmtcell{15.2} & \tmtcell{15.5} & \tmtcell{15.3} & \tmtcell{14.1} & \tmtcell{39.6} & \tmtcell{14.7} & \tmtcell{14.6} & \tmtcell{14.5} & \tmtcell{14.8} & \tmtcell{14.6} & \tmtcell{14.2} & \tmtcell{14.3} & \tmtcell{14.3} & \tmtcell{14.4} & \tmtcell{14.4} \\
Bridge\_crt\_del & \tmtcell{0.0} & \tmtcell{0.0} & \tmtcell{0.0} & \tmtcell{0.0} & \tmtcell{0.0} & \tmtcell{0.1} & \tmtcell{0.1} & \tmtcell{0.1} & \tmtcell{0.6} & \tmtcell{0.6} & \tmtcell{0.6} & \tmtcell{0.6} & \tmtcell{0.6} & \tmtcell{0.6} & \tmtcell{0.6} \\
Misc & \tmtcell{24.2} & \tmtcell{23.1} & \tmtcell{22.7} & \tmtcell{12.8} & \tmtcell{17.3} & \tmtcell{15.7} & \tmtcell{15.6} & \tmtcell{15.8} & \tmtcell{15.7} & \tmtcell{16.0} & \tmtcell{15.4} & \tmtcell{15.7} & \tmtcell{15.6} & \tmtcell{15.6} & \tmtcell{15.7} \\
\hline
\end{tabular}

\caption{Timing results (\vfoldscg)
for generating 200 
\eebbmmg\ events (channel GF58)
with \efo\ folding
using step15e in MINT2,
on the ``CUDA'' node
(step15e uses the same
software as step15c,
with only some
changes in the scripts).}
\label{tab:fldCudaEEMMBBG}
\vspace*{1.2cm}
\hspace*{-1.4cm}
\small
\setlength\tabcolsep{5pt} 
\providecommand{\tmtcell}[1]{\makebox[\tmtwidth][r]{#1}}
\begin{tabular}{lrrrrrrrrrrrrrr}
\multicolumn{15}{@{}l@{}}{\makebox[0pt][l]{{\bf \underline{vfold7cg} $|$ epem\_bbxmummupg $|$ step15c $|$ NEVENTS=200 $|$ folding=1,1,1 $|$ kbatch=512}}} \\
\multicolumn{15}{@{}l@{}}{\makebox[0pt][l]{\footnotesize {\em sborn 1: SBORN served from cudacpp, in mode C only}}} \\
\multicolumn{15}{@{}l@{}}{\makebox[0pt][l]{\footnotesize {\em gscale 1: Borns at the real-emission coupling $g'$ obtained by rescaling those at $g$, in mode C only}}} \\
\multicolumn{15}{@{}l@{}}{\makebox[0pt][l]{\footnotesize {\em timers: getrusage in F0*, F1[n], F1[c] (a syscall per timer call), clock\_gettime in all other modes}}} \\
\multicolumn{15}{@{}l@{}}{\makebox[0pt][l]{\footnotesize {\em mode F0: 11705 kpoints = 1639 virt + 10066 novi $\to$ 200 events (27 virt + 173 novi)}}} \\
\multicolumn{15}{@{}l@{}}{\makebox[0pt][l]{\footnotesize {\em mode F, F1, F2, Cs: 13422 kpoints = 2341 virt + 11081 novi $\to$ 200 events (27 virt + 173 novi)}}} \\
\multicolumn{15}{@{}l@{}}{\makebox[0pt][l]{\footnotesize {\em mode F, Cs: Born / Real 27 / 22 batches, mean batch 399.0 / 869.6 (mode C: each Real batch is split over 3 libraries)}}} \\
\multicolumn{15}{@{}l@{}}{\makebox[0pt][l]{\footnotesize {\em F0*: F0 rescaled to the kpoints of F (separately for virt and novi)}}} \\
\hline
mode & \tmtcell{F0*} & \tmtcell{F1[n]} & \tmtcell{F1[c]} & \tmtcell{F2} & \tmtcell{F} & \tmtcell{C} & \tmtcell{C} & \tmtcell{C} & \tmtcell{C} & \tmtcell{C} & \tmtcell{C} & \tmtcell{C} & \tmtcell{C} & \tmtcell{C} \\
cudacpp\_backend & \tmtcell{-} & \tmtcell{-} & \tmtcell{-} & \tmtcell{-} & \tmtcell{-} & \tmtcell{none} & \tmtcell{sse4} & \tmtcell{avx2} & \tmtcell{cuda} & \tmtcell{cuda} & \tmtcell{cuda} & \tmtcell{cuda} & \tmtcell{cuda} & \tmtcell{cuda} \\
cudacpp\_vecsize & \tmtcell{-} & \tmtcell{-} & \tmtcell{-} & \tmtcell{-} & \tmtcell{-} & \tmtcell{1} & \tmtcell{2} & \tmtcell{4} & \tmtcell{32} & \tmtcell{64} & \tmtcell{128} & \tmtcell{256} & \tmtcell{512} & \tmtcell{1024} \\
ml\_reduction & \tmtcell{collier} & \tmtcell{ninja} & \tmtcell{collier} & \tmtcell{collier} & \tmtcell{collier} & \tmtcell{collier} & \tmtcell{collier} & \tmtcell{collier} & \tmtcell{collier} & \tmtcell{collier} & \tmtcell{collier} & \tmtcell{collier} & \tmtcell{collier} & \tmtcell{collier} \\
\hline
TOTAL & \tmtcell{40.2} & \tmtcell{40.8} & \tmtcell{39.3} & \tmtcell{40.1} & \tmtcell{42.9} & \tmtcell{36.8} & \tmtcell{31.0} & \tmtcell{27.9} & \tmtcell{27.2} & \tmtcell{26.8} & \tmtcell{26.8} & \tmtcell{26.9} & \tmtcell{26.8} & \tmtcell{26.8} \\
Born & \tmtcell{2.5} & \tmtcell{2.6} & \tmtcell{2.6} & \tmtcell{3.2} & \tmtcell{4.1} & \tmtcell{1.6} & \tmtcell{0.8} & \tmtcell{0.4} & \tmtcell{0.2} & \tmtcell{0.2} & \tmtcell{0.1} & \tmtcell{0.1} & \tmtcell{0.1} & \tmtcell{0.1} \\
Real & \tmtcell{11.5} & \tmtcell{11.4} & \tmtcell{11.4} & \tmtcell{11.5} & \tmtcell{11.6} & \tmtcell{9.6} & \tmtcell{4.7} & \tmtcell{2.1} & \tmtcell{0.5} & \tmtcell{0.4} & \tmtcell{0.3} & \tmtcell{0.3} & \tmtcell{0.3} & \tmtcell{0.4} \\
Virtual & \tmtcell{24.2} & \tmtcell{24.8} & \tmtcell{23.3} & \tmtcell{23.6} & \tmtcell{23.4} & \tmtcell{23.8} & \tmtcell{23.6} & \tmtcell{23.6} & \tmtcell{24.1} & \tmtcell{23.8} & \tmtcell{24.0} & \tmtcell{24.1} & \tmtcell{23.9} & \tmtcell{24.0} \\
PS\_Generation & \tmtcell{0.2} & \tmtcell{0.2} & \tmtcell{0.2} & \tmtcell{0.2} & \tmtcell{0.3} & \tmtcell{0.3} & \tmtcell{0.3} & \tmtcell{0.3} & \tmtcell{0.3} & \tmtcell{0.3} & \tmtcell{0.3} & \tmtcell{0.3} & \tmtcell{0.3} & \tmtcell{0.3} \\
CT\_and\_MCsub & \tmtcell{0.3} & \tmtcell{0.3} & \tmtcell{0.3} & \tmtcell{0.5} & \tmtcell{0.8} & \tmtcell{0.2} & \tmtcell{0.2} & \tmtcell{0.2} & \tmtcell{0.2} & \tmtcell{0.2} & \tmtcell{0.2} & \tmtcell{0.2} & \tmtcell{0.2} & \tmtcell{0.2} \\
Nbody\_prefactor & \tmtcell{0.5} & \tmtcell{0.5} & \tmtcell{0.5} & \tmtcell{0.5} & \tmtcell{1.4} & \tmtcell{0.5} & \tmtcell{0.5} & \tmtcell{0.5} & \tmtcell{0.5} & \tmtcell{0.5} & \tmtcell{0.5} & \tmtcell{0.5} & \tmtcell{0.5} & \tmtcell{0.5} \\
Bridge\_crt\_del & \tmtcell{0.0} & \tmtcell{0.0} & \tmtcell{0.0} & \tmtcell{0.0} & \tmtcell{0.0} & \tmtcell{0.1} & \tmtcell{0.1} & \tmtcell{0.1} & \tmtcell{0.6} & \tmtcell{0.6} & \tmtcell{0.6} & \tmtcell{0.6} & \tmtcell{0.6} & \tmtcell{0.6} \\
Misc & \tmtcell{1.0} & \tmtcell{1.0} & \tmtcell{1.0} & \tmtcell{0.7} & \tmtcell{1.2} & \tmtcell{0.8} & \tmtcell{0.8} & \tmtcell{0.8} & \tmtcell{0.8} & \tmtcell{0.8} & \tmtcell{0.8} & \tmtcell{0.8} & \tmtcell{0.8} & \tmtcell{0.8} \\
\hline
\end{tabular}

\caption{Timing results (\vfoldscg)
for generating 200 
\eebbmmg\ events (channel GF58)
with \ooo\ folding
using step15c in MINT2,
on the ``CUDA'' node.}
\label{tab:nofCudaEEMMBBG}
}\end{center}
\end{table}

\section{Results for \eebbmmgTOP}
\label{sec:eebbmmg}

In this section,
I present timing results 
for a different 
physics process, 
\eebbmmg\ at NLO,
in the \vfoldscg\ version of \mvf.
Time numbers in seconds 
for \efo\ and \ooo\ folding
are shown
in Tables~\ref{tab:fldCudaEEMMBBG}
and~\ref{tab:nofCudaEEMMBBG},
respectively.
These refer to the MINT2
generation of 200 unweighted events
in a single channel GF58.0
(chosen because it was one
of those with higher cross sections)
on the ``CUDA'' machine.
A few recurring patterns
and notable differences
from the previous tests
are worth pointing out:
\begin{itemize}

\item 
The Virtual contribution
is by far the bottleneck
for this process,
not only without folding
but also with \efo\ folding.
In the fastest 
C/cuda configuration,
Virtuals represent 
44.4s out of 90.2s 
with \efo\ folding
and 24.0s out of 26.8s
with \ooo\ folding.
It is therefore
not true in general
that Virtuals are negligible 
in vectorized folding.

\item
As mentioned in 
Sec.~\ref{sec:modes},
it was found in the tests
of another channel
that the Virtual
calculation was 
particularly slow
because both Ninja
(the first-line solution)
and COLLIER 
(the backup solution)
were used for many events.
As a consequence,
all tests were reconfigured
to use COLLIER 
as the primary solution
in all modes.
For comparison,
however, 
the tests in F1 mode 
were executed
both with COLLIER 
and with Ninja:
the results,
listed in columns
F1[n] and F1[c]
in the tables,
show that in this
particular channel
COLLIER is indeed
faster, but only 
by a few percent.

\item
In addition to Virtuals,
Nbody\_prefactor is
a new bottleneck
in \efo\ vectorized folding.
In the fastest 
C/cuda configuration,
this represents
14.3s out of 90.2s.
This time is 
essentially flat
in all other modes except F
(where it is much larger
due to the two-walk overhead),
indicating that this
contribution does not
benefit from vectorization.

\item 
The \mvf\ speedup 
of Born and Real,
however, is just as good
as previously in \eebbbb.
(This is especially
important since 
this process has a
much more complex
internal structure
than \eebbbb,
notably because 
it has a gluon 
in the Born and 
three Real libraries.) 
Thanks to this,
the total generation time
is reduced from 460.2s
to 90.2s (5.1x) 
with \efo\ folding
and from 39.3s
to 26.8s (1.5x)
with \ooo\ folding.

\end{itemize}

\section{Results for \ppttjjTOP}
\label{sec:ppttjj}

In this section,
I finally present 
timing results 
for the most complex 
physics process
studied in this paper, 
\ppttjj\ at NLO,
in the \vfoldscg\ version 
of \mvf.
Time numbers in seconds 
for the MINT2 generation 
of 1000 unweighted events
with three different foldings,
on the ``CUDA'' node
and on the ``GOLD'' node,
are given in the following
Tables~\ref{tab:111CudaPPTTJJ} 
and~\ref{tab:111GoldPPTTJJ} (\ooo\ folding),
\ref{tab:221CudaPPTTJJ} 
and~\ref{tab:221GoldPPTTJJ} (\tto\ folding),
\ref{tab:841CudaPPTTJJ} 
and~\ref{tab:841GoldPPTTJJ} (\efo\ folding).
With respect to the
results previously presented
for \epem\ processes,
there are some important
differences in the
internal structure
of this process,
but also in the way 
it was tested:
\begin{itemize}

\item 
Unlike the \epem\ processes
described so far,
which involved a single
partonic subprocess,
\ppttjj\ includes eighteen 
\texttt{SubProcesses/P0\_<subproc>}
directories,
one for each of 
the possible subprocesses
with different initial 
and final state partons
that contribute to 
this process 
(such as \ggttgg\ or \guttgu).
As previously mentioned,
these subprocesses 
involve different
numbers of Real processes
(example: six for \ggttgg,
nine for \ggttuu).
The calculation also
requires the interface to 
parton distribution functions.
All the relevant code
was automatically generated.

\item
The tests presented
in this section,
unlike those described
in the previous sections
that referred to a single
channel of one subprocess,
include the contributions
of many \texttt{GFn.0} 
channels in many 
\texttt{SubProcesses/P0\_<subproc>}
subprocesses.
In principle, \ppttjj\ involves
448 such channels overall.
However, the tests I describe,
which refer to the MINT2 generation
of a limited number of events,
involve only around
150 channels out of 448;
this is because the MadGraph 
software automatically
decides how many channels
should be executed as they
may produce at least 
one unweighted event.

\item 
Each test refers 
to the execution 
of a ``gridpack'',
i.e. a tarball that
includes pre-built libraries
and executables
as well as MINT0/MINT1
integration grids,
previously computed
with loose physics 
accuracy targets
(req\_acc=0.1).
The infrastructure used
to prepare these gridpacks
was developed in \mvf:
in fact, MadGraph itself provides
the infrastructure to produce
gridpacks for LO simulations,
but not for NLO simulations.
With respect to the default
LO gridpacks in MadGraph,
one notable difference
in the \mvf\ NLO gridpacks
is that they support
multiple backends,
as they include pre-built
libraries and executables
for both the Fortran
and CUDACPP computation
of matrix elements,
with a runtime choice
controlled by 
environment variables.

\item 
Unlike the results
given in previous sections,
which included time profiles
obtained by the instrumentation
of a single \madmint\ Fortran executable,
the results given in this section
refer to the full MadGraph workflow,
where a separate \madmint\ executable
is launched in each channel,
with the overall orchestration
managed by the MadGraph
Python/bash 
event generation infrastructure.
As a consequence,
time profiles
include the sum 
over all channels
of each individual 
Fortran contribution
(e.g. Born or Real),
as well as an estimate
of the time spent
in Python outside bash jobs
(before the first job,
between jobs,
after the last jobs)
and of the time
spent in each bash job
and not accounted for
in the Fortran timers.
It should be stressed,
in particular,
that {\em each test
was executed 
in single-core mode},
even if the CUDA 
and GOLD nodes
have 4 and 32 
logical cores, 
respectively.
In particular,
on the CUDA node
in practice the CPU and GPU
were waiting for each other
and were not doing work
at the same time.
This is most certainly
not the way these jobs
would be executed in production,
but I felt that this
would provide a more 
reliable way to
compare the time profiles
of different configurations.

\item 
Finally, for lack of space
several header lines 
have been removed 
from the six tables,
as in any case
the same comments
apply to all of them.
In particular:
all Virtuals are computed
with Ninja first;
as in previous tables,
the F1[u] mode is the same as F1,
without two stock MadGraph7 
optimizations (cache and 
$\alphaS$ rescaling 
of Born amplitudes);
for this single-core run,
``WALL TOTAL'' equals ``FORTRAN TOTAL'' 
plus ``Job start/exit'' plus 
``Python outside jobs'';
where timers are based 
on getrusage (which measures CPU time)
``Job start/exit'' also includes 
the difference between 
wall and CPU time;
finally,
F0 times are not 
rescaled to F0* here.

\end{itemize}
A few observations
on the results themselves
can now be given:

\begin{table}[p]
\begin{center}{
\hspace*{-2.3cm}
\small
\setlength\tabcolsep{5pt} 
\providecommand{\tmtcell}[1]{\makebox[\tmtwidth][r]{#1}}
\providecommand{\tmtcellw}[1]{\makebox[\dimexpr\tmtwidth+0.5em\relax][r]{#1}}
\providecommand{\tmtcellww}[1]{\makebox[\dimexpr\tmtwidth+1.0em\relax][r]{#1}}
\begin{tabular}{lrrrrrrrrrrrrrrr}
\multicolumn{16}{@{}l@{}}{\makebox[0pt][l]{{\bf pp\_ttxjj $|$ gridpack $|$ NEVENTS=1000 $|$ folding=1,1,1 $|$ kbatch=512}}} \\
\multicolumn{16}{@{}l@{}}{\makebox[0pt][l]{\footnotesize {\em sborn 1: SBORN served from cudacpp, in mode C only}}} \\
\multicolumn{16}{@{}l@{}}{\makebox[0pt][l]{\footnotesize {\em gscale 1: Borns at the real-emission coupling $g'$ obtained by rescaling those at $g$, in mode C only}}} \\
\multicolumn{16}{@{}l@{}}{\makebox[0pt][l]{\footnotesize {\em timers: getrusage in F0, F1, F1u (a syscall per timer call), clock\_gettime in all other modes}}} \\
\multicolumn{16}{@{}l@{}}{\makebox[0pt][l]{\footnotesize {\em mode F0: 80917 kpoints = 2591 virt + 78326 novi $\to$ 1000 events (42 virt + 958 novi)}}} \\
\multicolumn{16}{@{}l@{}}{\makebox[0pt][l]{\footnotesize {\em mode F, F1, F2, Cs: 81539 kpoints = 3061 virt + 78478 novi $\to$ 1000 events (40 virt + 960 novi)}}} \\
\multicolumn{16}{@{}l@{}}{\makebox[0pt][l]{\footnotesize {\em mode F, Cs: Born / Real 281 / 253 batches, mean batch 288.6 / 1033.3 (mode C: each Real batch is split over 3 to 9 libraries)}}} \\
\multicolumn{16}{@{}l@{}}{\makebox[0pt][l]{\footnotesize {\em MINT2 jobs: 152 channels run (those with at least one event) out of 448}}} \\
\hline
mode & \tmtcellw{F0} & \tmtcellw{F1} & \tmtcellw{F1u} & \tmtcellw{F2} & \tmtcellw{F} & \tmtcellw{C} & \tmtcell{C} & \tmtcell{C} & \tmtcell{C} & \tmtcell{C} & \tmtcell{C} & \tmtcell{C} & \tmtcell{C} & \tmtcell{C} & \tmtcell{C} \\
cudacpp\_backend & \tmtcellw{-} & \tmtcellw{-} & \tmtcellw{-} & \tmtcellw{-} & \tmtcellw{-} & \tmtcellw{none} & \tmtcell{sse4} & \tmtcell{avx2} & \tmtcell{cuda} & \tmtcell{cuda} & \tmtcell{cuda} & \tmtcell{cuda} & \tmtcell{cuda} & \tmtcell{cuda} & \tmtcell{cuda} \\
cudacpp\_vecsize & \tmtcellw{-} & \tmtcellw{-} & \tmtcellw{-} & \tmtcellw{-} & \tmtcellw{-} & \tmtcellw{1} & \tmtcell{2} & \tmtcell{4} & \tmtcell{32} & \tmtcell{64} & \tmtcell{128} & \tmtcell{256} & \tmtcell{512} & \tmtcell{1024} & \tmtcell{2048} \\
\hline
WALL TOTAL & \tmtcellw{1011.0} & \tmtcellw{1008.9} & \tmtcellw{1284.8} & \tmtcellw{1013.6} & \tmtcellw{1224.9} & \tmtcellw{1043.8} & \tmtcell{801.6} & \tmtcell{626.8} & \tmtcell{728.0} & \tmtcell{710.0} & \tmtcell{700.6} & \tmtcell{699.2} & \tmtcell{694.2} & \tmtcell{704.3} & \tmtcell{724.0} \\
FORTRAN TOTAL & \tmtcellw{1002.0} & \tmtcellw{999.7} & \tmtcellw{1274.9} & \tmtcellw{1008.2} & \tmtcellw{1219.8} & \tmtcellw{1038.7} & \tmtcell{796.2} & \tmtcell{621.7} & \tmtcell{674.6} & \tmtcell{656.7} & \tmtcell{647.3} & \tmtcell{645.7} & \tmtcell{640.3} & \tmtcell{649.7} & \tmtcell{668.0} \\
- Born & \tmtcellw{66.8} & \tmtcellw{70.5} & \tmtcellw{249.1} & \tmtcellw{70.4} & \tmtcellw{102.7} & \tmtcellw{78.4} & \tmtcell{57.7} & \tmtcell{23.2} & \tmtcell{42.0} & \tmtcell{29.1} & \tmtcell{21.6} & \tmtcell{18.3} & \tmtcell{12.3} & \tmtcell{13.4} & \tmtcell{15.1} \\
- Real & \tmtcellw{409.2} & \tmtcellw{415.5} & \tmtcellw{413.3} & \tmtcellw{415.7} & \tmtcellw{571.6} & \tmtcellw{470.6} & \tmtcell{249.0} & \tmtcell{108.7} & \tmtcell{69.7} & \tmtcell{65.0} & \tmtcell{63.2} & \tmtcell{64.1} & \tmtcell{65.9} & \tmtcell{72.7} & \tmtcell{89.1} \\
- Virtual & \tmtcellw{443.9} & \tmtcellw{428.8} & \tmtcellw{428.6} & \tmtcellw{430.6} & \tmtcellw{430.6} & \tmtcellw{430.6} & \tmtcell{430.5} & \tmtcell{430.8} & \tmtcell{432.5} & \tmtcell{432.3} & \tmtcell{432.2} & \tmtcell{433.2} & \tmtcell{431.9} & \tmtcell{432.9} & \tmtcell{433.1} \\
- PS\_Generation & \tmtcellw{1.6} & \tmtcellw{1.7} & \tmtcellw{1.7} & \tmtcellw{1.6} & \tmtcellw{3.1} & \tmtcellw{3.0} & \tmtcell{3.0} & \tmtcell{3.0} & \tmtcell{3.0} & \tmtcell{3.0} & \tmtcell{3.0} & \tmtcell{3.0} & \tmtcell{3.0} & \tmtcell{3.0} & \tmtcell{3.0} \\
- CT\_and\_MCsub & \tmtcellw{27.1} & \tmtcellw{29.2} & \tmtcellw{128.1} & \tmtcellw{38.4} & \tmtcellw{54.3} & \tmtcellw{2.1} & \tmtcell{2.1} & \tmtcell{2.1} & \tmtcell{2.1} & \tmtcell{2.1} & \tmtcell{2.1} & \tmtcell{2.1} & \tmtcell{2.1} & \tmtcell{2.1} & \tmtcell{2.1} \\
- Nbody\_prefactor & \tmtcellw{0.4} & \tmtcellw{0.4} & \tmtcellw{0.4} & \tmtcellw{0.1} & \tmtcellw{3.3} & \tmtcellw{0.1} & \tmtcell{0.1} & \tmtcell{0.1} & \tmtcell{0.1} & \tmtcell{0.1} & \tmtcell{0.1} & \tmtcell{0.1} & \tmtcell{0.1} & \tmtcell{0.1} & \tmtcell{0.1} \\
- Bridge\_crt\_del & \tmtcellw{0.0} & \tmtcellw{0.0} & \tmtcellw{0.0} & \tmtcellw{0.0} & \tmtcellw{0.0} & \tmtcellw{0.7} & \tmtcell{0.7} & \tmtcell{0.7} & \tmtcell{71.8} & \tmtcell{71.7} & \tmtcell{71.7} & \tmtcell{71.6} & \tmtcell{71.6} & \tmtcell{71.8} & \tmtcell{72.0} \\
- Misc & \tmtcellw{52.9} & \tmtcellw{53.6} & \tmtcellw{53.7} & \tmtcellw{51.3} & \tmtcellw{54.0} & \tmtcellw{53.2} & \tmtcell{53.1} & \tmtcell{53.1} & \tmtcell{53.4} & \tmtcell{53.5} & \tmtcell{53.5} & \tmtcell{53.4} & \tmtcell{53.4} & \tmtcell{53.7} & \tmtcell{53.5} \\
Job start/exit & \tmtcellw{4.9} & \tmtcellw{5.1} & \tmtcellw{5.8} & \tmtcellw{0.9} & \tmtcellw{1.0} & \tmtcellw{0.9} & \tmtcell{0.9} & \tmtcell{0.9} & \tmtcell{49.2} & \tmtcell{49.1} & \tmtcell{49.1} & \tmtcell{49.3} & \tmtcell{49.8} & \tmtcell{50.4} & \tmtcell{51.7} \\
Python outside jobs & \tmtcellw{4.1} & \tmtcellw{4.1} & \tmtcellw{4.1} & \tmtcellw{4.5} & \tmtcellw{4.1} & \tmtcellw{4.2} & \tmtcell{4.5} & \tmtcell{4.2} & \tmtcell{4.2} & \tmtcell{4.2} & \tmtcell{4.2} & \tmtcell{4.2} & \tmtcell{4.2} & \tmtcell{4.2} & \tmtcell{4.2} \\
\hline
\end{tabular}

\caption{Timing results (\vfoldscg)
for generating 1000 
\ppttjj\ events
with \ooo\ folding
using step17g in MINT2,
on the ``CUDA'' node.}
\label{tab:111CudaPPTTJJ}
\vspace*{1.2cm}
\hspace*{-0.8cm}
\small
\setlength\tabcolsep{5pt} 
\providecommand{\tmtcell}[1]{\makebox[\tmtwidth][r]{#1}}
\begin{tabular}{lrrrrrrrrrr}
\multicolumn{11}{@{}l@{}}{\makebox[0pt][l]{{\bf pp\_ttxjj $|$ gridpack $|$ NEVENTS=1000 $|$ folding=1,1,1 $|$ kbatch=512}}} \\
\multicolumn{11}{@{}l@{}}{\makebox[0pt][l]{\footnotesize {\em sborn 1: SBORN served from cudacpp, in mode C only}}} \\
\multicolumn{11}{@{}l@{}}{\makebox[0pt][l]{\footnotesize {\em gscale 1: Borns at the real-emission coupling $g'$ obtained by rescaling those at $g$, in mode C only}}} \\
\multicolumn{11}{@{}l@{}}{\makebox[0pt][l]{\footnotesize {\em timers: getrusage in F0, F1, F1u (a syscall per timer call), clock\_gettime in all other modes}}} \\
\multicolumn{11}{@{}l@{}}{\makebox[0pt][l]{\footnotesize {\em mode F0: 80917 kpoints = 2591 virt + 78326 novi $\to$ 1000 events (42 virt + 958 novi)}}} \\
\multicolumn{11}{@{}l@{}}{\makebox[0pt][l]{\footnotesize {\em mode F, F1, F2, Cs: 81539 kpoints = 3061 virt + 78478 novi $\to$ 1000 events (40 virt + 960 novi)}}} \\
\multicolumn{11}{@{}l@{}}{\makebox[0pt][l]{\footnotesize {\em mode F, Cs: Born / Real 281 / 253 batches, mean batch 288.6 / 1033.3 (mode C: each Real batch is split over 3 to 9 libraries)}}} \\
\multicolumn{11}{@{}l@{}}{\makebox[0pt][l]{\footnotesize {\em MINT2 jobs: 152 channels run (those with at least one event) out of 448}}} \\
\hline
mode & \tmtcell{F0} & \tmtcell{F1} & \tmtcell{F1u} & \tmtcell{F2} & \tmtcell{F} & \tmtcell{C} & \tmtcell{C} & \tmtcell{C} & \tmtcell{C} & \tmtcell{C} \\
cudacpp\_backend & \tmtcell{-} & \tmtcell{-} & \tmtcell{-} & \tmtcell{-} & \tmtcell{-} & \tmtcell{none} & \tmtcell{sse4} & \tmtcell{avx2} & \tmtcell{512y} & \tmtcell{512z} \\
cudacpp\_vecsize & \tmtcell{-} & \tmtcell{-} & \tmtcell{-} & \tmtcell{-} & \tmtcell{-} & \tmtcell{1} & \tmtcell{2} & \tmtcell{4} & \tmtcell{4} & \tmtcell{8} \\
\hline
WALL TOTAL & \tmtcell{733.8} & \tmtcell{731.6} & \tmtcell{945.7} & \tmtcell{736.8} & \tmtcell{903.0} & \tmtcell{742.3} & \tmtcell{561.8} & \tmtcell{428.8} & \tmtcell{418.8} & \tmtcell{381.4} \\
FORTRAN TOTAL & \tmtcell{726.8} & \tmtcell{726.8} & \tmtcell{940.7} & \tmtcell{733.1} & \tmtcell{899.2} & \tmtcell{738.7} & \tmtcell{558.1} & \tmtcell{425.1} & \tmtcell{415.2} & \tmtcell{377.7} \\
- Born & \tmtcell{50.3} & \tmtcell{53.4} & \tmtcell{190.7} & \tmtcell{53.5} & \tmtcell{77.7} & \tmtcell{56.3} & \tmtcell{44.8} & \tmtcell{17.6} & \tmtcell{16.4} & \tmtcell{10.1} \\
- Real & \tmtcell{327.8} & \tmtcell{332.2} & \tmtcell{332.4} & \tmtcell{333.1} & \tmtcell{458.9} & \tmtcell{358.2} & \tmtcell{189.5} & \tmtcell{83.7} & \tmtcell{75.1} & \tmtcell{44.3} \\
- Virtual & \tmtcell{292.2} & \tmtcell{282.5} & \tmtcell{282.7} & \tmtcell{283.2} & \tmtcell{282.8} & \tmtcell{284.7} & \tmtcell{284.3} & \tmtcell{284.4} & \tmtcell{284.2} & \tmtcell{283.9} \\
- PS\_Generation & \tmtcell{1.1} & \tmtcell{1.1} & \tmtcell{1.1} & \tmtcell{1.1} & \tmtcell{2.2} & \tmtcell{2.1} & \tmtcell{2.1} & \tmtcell{2.1} & \tmtcell{2.1} & \tmtcell{2.1} \\
- CT\_and\_MCsub & \tmtcell{19.6} & \tmtcell{21.4} & \tmtcell{97.7} & \tmtcell{28.0} & \tmtcell{39.4} & \tmtcell{1.3} & \tmtcell{1.3} & \tmtcell{1.3} & \tmtcell{1.3} & \tmtcell{1.3} \\
- Nbody\_prefactor & \tmtcell{0.3} & \tmtcell{0.3} & \tmtcell{0.3} & \tmtcell{0.1} & \tmtcell{2.2} & \tmtcell{0.1} & \tmtcell{0.1} & \tmtcell{0.1} & \tmtcell{0.1} & \tmtcell{0.1} \\
- Bridge\_crt\_del & \tmtcell{0.0} & \tmtcell{0.0} & \tmtcell{0.0} & \tmtcell{0.0} & \tmtcell{0.0} & \tmtcell{0.4} & \tmtcell{0.4} & \tmtcell{0.4} & \tmtcell{0.4} & \tmtcell{0.4} \\
- Misc & \tmtcell{35.4} & \tmtcell{35.8} & \tmtcell{35.9} & \tmtcell{34.2} & \tmtcell{36.0} & \tmtcell{35.5} & \tmtcell{35.5} & \tmtcell{35.5} & \tmtcell{35.5} & \tmtcell{35.5} \\
Job start/exit & \tmtcell{1.8} & \tmtcell{1.9} & \tmtcell{1.9} & \tmtcell{0.7} & \tmtcell{0.7} & \tmtcell{0.7} & \tmtcell{0.7} & \tmtcell{0.7} & \tmtcell{0.6} & \tmtcell{0.6} \\
Python outside jobs & \tmtcell{5.2} & \tmtcell{3.0} & \tmtcell{3.0} & \tmtcell{3.0} & \tmtcell{3.0} & \tmtcell{3.0} & \tmtcell{3.0} & \tmtcell{3.0} & \tmtcell{3.0} & \tmtcell{3.0} \\
\hline
\end{tabular}

\caption{Timing results (\vfoldscg)
for generating 1000 
\ppttjj\ events
with \ooo\ folding
using step17g in MINT2,
on the ``GOLD'' node.}
\label{tab:111GoldPPTTJJ}
}\end{center}
\end{table}

\begin{table}[p]
\begin{center}{
\hspace*{-2.3cm}
\small
\setlength\tabcolsep{5pt} 
\providecommand{\tmtcell}[1]{\makebox[\tmtwidth][r]{#1}}
\providecommand{\tmtcellw}[1]{\makebox[\dimexpr\tmtwidth+0.5em\relax][r]{#1}}
\begin{tabular}{lrrrrrrrrrrrrrrr}
\multicolumn{16}{@{}l@{}}{\makebox[0pt][l]{{\bf pp\_ttxjj $|$ gridpack $|$ NEVENTS=1000 $|$ folding=2,2,1 $|$ kbatch=128}}} \\
\multicolumn{16}{@{}l@{}}{\makebox[0pt][l]{\footnotesize {\em sborn 1: SBORN served from cudacpp, in mode C only}}} \\
\multicolumn{16}{@{}l@{}}{\makebox[0pt][l]{\footnotesize {\em gscale 1: Borns at the real-emission coupling $g'$ obtained by rescaling those at $g$, in mode C only}}} \\
\multicolumn{16}{@{}l@{}}{\makebox[0pt][l]{\footnotesize {\em timers: getrusage in F0, F1, F1u (a syscall per timer call), clock\_gettime in all other modes}}} \\
\multicolumn{16}{@{}l@{}}{\makebox[0pt][l]{\footnotesize {\em mode F0: 68736 kpoints = 3959 virt + 64777 novi $\to$ 1000 events (53 virt + 947 novi)}}} \\
\multicolumn{16}{@{}l@{}}{\makebox[0pt][l]{\footnotesize {\em mode F, F1, F2, Cs: 71783 kpoints = 3717 virt + 68066 novi $\to$ 1000 events (53 virt + 947 novi)}}} \\
\multicolumn{16}{@{}l@{}}{\makebox[0pt][l]{\footnotesize {\em mode F, Cs: Born / Real 672 / 619 batches, mean batch 282.8 / 1009.4 (mode C: each Real batch is split over 3 to 9 libraries)}}} \\
\multicolumn{16}{@{}l@{}}{\makebox[0pt][l]{\footnotesize {\em MINT2 jobs: 154 channels run (those with at least one event) out of 448}}} \\
\hline
mode & \tmtcellw{F0} & \tmtcellw{F1} & \tmtcellw{F1u} & \tmtcellw{F2} & \tmtcellw{F} & \tmtcellw{C} & \tmtcellw{C} & \tmtcell{C} & \tmtcell{C} & \tmtcell{C} & \tmtcell{C} & \tmtcell{C} & \tmtcell{C} & \tmtcell{C} & \tmtcell{C} \\
cudacpp\_backend & \tmtcellw{-} & \tmtcellw{-} & \tmtcellw{-} & \tmtcellw{-} & \tmtcellw{-} & \tmtcellw{none} & \tmtcellw{sse4} & \tmtcell{avx2} & \tmtcell{cuda} & \tmtcell{cuda} & \tmtcell{cuda} & \tmtcell{cuda} & \tmtcell{cuda} & \tmtcell{cuda} & \tmtcell{cuda} \\
cudacpp\_vecsize & \tmtcellw{-} & \tmtcellw{-} & \tmtcellw{-} & \tmtcellw{-} & \tmtcellw{-} & \tmtcellw{1} & \tmtcellw{2} & \tmtcell{4} & \tmtcell{32} & \tmtcell{64} & \tmtcell{128} & \tmtcell{256} & \tmtcell{512} & \tmtcell{1024} & \tmtcell{2048} \\
\hline
WALL TOTAL & \tmtcellw{2362.0} & \tmtcellw{2320.6} & \tmtcellw{3375.0} & \tmtcellw{2279.9} & \tmtcellw{2493.1} & \tmtcellw{1969.3} & \tmtcellw{1340.1} & \tmtcell{913.9} & \tmtcell{868.3} & \tmtcell{837.8} & \tmtcell{822.1} & \tmtcell{822.2} & \tmtcell{824.2} & \tmtcell{848.7} & \tmtcell{899.3} \\
FORTRAN TOTAL & \tmtcellw{2345.9} & \tmtcellw{2302.9} & \tmtcellw{3356.7} & \tmtcellw{2274.7} & \tmtcellw{2487.9} & \tmtcellw{1964.0} & \tmtcellw{1334.5} & \tmtcell{908.8} & \tmtcell{814.2} & \tmtcell{783.7} & \tmtcell{767.8} & \tmtcell{767.6} & \tmtcell{769.2} & \tmtcell{793.1} & \tmtcell{842.3} \\
- Born & \tmtcellw{145.5} & \tmtcellw{165.0} & \tmtcellw{877.4} & \tmtcellw{125.7} & \tmtcellw{144.6} & \tmtcellw{64.8} & \tmtcellw{55.5} & \tmtcell{28.6} & \tmtcell{61.9} & \tmtcell{45.5} & \tmtcell{33.1} & \tmtcell{30.4} & \tmtcell{27.1} & \tmtcell{29.9} & \tmtcell{33.9} \\
- Real & \tmtcellw{1360.7} & \tmtcellw{1464.2} & \tmtcellw{1459.9} & \tmtcellw{1470.5} & \tmtcellw{1603.3} & \tmtcellw{1318.8} & \tmtcellw{698.0} & \tmtcell{300.1} & \tmtcell{98.4} & \tmtcell{84.3} & \tmtcell{79.9} & \tmtcell{82.3} & \tmtcell{85.7} & \tmtcell{108.1} & \tmtcell{153.6} \\
- Virtual & \tmtcellw{671.2} & \tmtcellw{491.4} & \tmtcellw{492.4} & \tmtcellw{495.6} & \tmtcellw{494.6} & \tmtcellw{496.0} & \tmtcellw{496.4} & \tmtcell{495.5} & \tmtcell{497.7} & \tmtcell{497.7} & \tmtcell{498.4} & \tmtcell{498.4} & \tmtcell{499.6} & \tmtcell{498.2} & \tmtcell{498.3} \\
- PS\_Generation & \tmtcellw{5.1} & \tmtcellw{5.4} & \tmtcellw{5.4} & \tmtcellw{5.1} & \tmtcellw{8.4} & \tmtcellw{8.0} & \tmtcellw{8.0} & \tmtcell{8.0} & \tmtcell{8.0} & \tmtcell{8.0} & \tmtcell{8.0} & \tmtcell{8.0} & \tmtcell{8.0} & \tmtcell{8.0} & \tmtcell{8.0} \\
- CT\_and\_MCsub & \tmtcellw{89.3} & \tmtcellw{101.0} & \tmtcellw{445.5} & \tmtcellw{111.4} & \tmtcellw{156.0} & \tmtcellw{6.2} & \tmtcellw{6.2} & \tmtcell{6.2} & \tmtcell{6.1} & \tmtcell{6.1} & \tmtcell{6.1} & \tmtcell{6.1} & \tmtcell{6.1} & \tmtcell{6.1} & \tmtcell{6.1} \\
- Nbody\_prefactor & \tmtcellw{1.3} & \tmtcellw{1.3} & \tmtcellw{1.4} & \tmtcellw{0.4} & \tmtcellw{10.1} & \tmtcellw{0.3} & \tmtcellw{0.3} & \tmtcell{0.3} & \tmtcell{0.3} & \tmtcell{0.3} & \tmtcell{0.3} & \tmtcell{0.3} & \tmtcell{0.3} & \tmtcell{0.3} & \tmtcell{0.3} \\
- Bridge\_crt\_del & \tmtcellw{0.0} & \tmtcellw{0.0} & \tmtcellw{0.0} & \tmtcellw{0.0} & \tmtcellw{0.0} & \tmtcellw{0.7} & \tmtcellw{0.7} & \tmtcell{0.7} & \tmtcell{72.3} & \tmtcell{72.4} & \tmtcell{72.4} & \tmtcell{72.6} & \tmtcell{72.5} & \tmtcell{72.5} & \tmtcell{72.5} \\
- Misc & \tmtcellw{72.7} & \tmtcellw{74.5} & \tmtcellw{74.7} & \tmtcellw{66.0} & \tmtcellw{71.0} & \tmtcellw{69.3} & \tmtcellw{69.4} & \tmtcell{69.3} & \tmtcell{69.6} & \tmtcell{69.6} & \tmtcell{69.6} & \tmtcell{69.7} & \tmtcell{69.8} & \tmtcell{69.9} & \tmtcell{69.7} \\
Job start/exit & \tmtcellw{11.9} & \tmtcellw{13.5} & \tmtcellw{14.1} & \tmtcellw{1.0} & \tmtcellw{1.0} & \tmtcellw{1.0} & \tmtcellw{1.4} & \tmtcell{1.0} & \tmtcell{49.9} & \tmtcell{49.8} & \tmtcell{50.0} & \tmtcell{50.3} & \tmtcell{50.7} & \tmtcell{51.4} & \tmtcell{52.7} \\
Python outside jobs & \tmtcellw{4.2} & \tmtcellw{4.2} & \tmtcellw{4.2} & \tmtcellw{4.2} & \tmtcellw{4.2} & \tmtcellw{4.3} & \tmtcellw{4.2} & \tmtcell{4.2} & \tmtcell{4.2} & \tmtcell{4.2} & \tmtcell{4.2} & \tmtcell{4.3} & \tmtcell{4.3} & \tmtcell{4.3} & \tmtcell{4.3} \\
\hline
\end{tabular}

\caption{Timing results (\vfoldscg)
for generating 1000 
\ppttjj\ events
with \tto\ folding
using step17g in MINT2,
on the ``CUDA'' node.}
\label{tab:221CudaPPTTJJ}
\vspace*{1.2cm}
\hspace*{-0.8cm}
\small
\setlength\tabcolsep{5pt} 
\providecommand{\tmtcell}[1]{\makebox[\tmtwidth][r]{#1}}
\providecommand{\tmtcellw}[1]{\makebox[\dimexpr\tmtwidth+0.5em\relax][r]{#1}}
\begin{tabular}{lrrrrrrrrrr}
\multicolumn{11}{@{}l@{}}{\makebox[0pt][l]{{\bf pp\_ttxjj $|$ gridpack $|$ NEVENTS=1000 $|$ folding=2,2,1 $|$ kbatch=128}}} \\
\multicolumn{11}{@{}l@{}}{\makebox[0pt][l]{\footnotesize {\em sborn 1: SBORN served from cudacpp, in mode C only}}} \\
\multicolumn{11}{@{}l@{}}{\makebox[0pt][l]{\footnotesize {\em gscale 1: Borns at the real-emission coupling $g'$ obtained by rescaling those at $g$, in mode C only}}} \\
\multicolumn{11}{@{}l@{}}{\makebox[0pt][l]{\footnotesize {\em timers: getrusage in F0, F1, F1u (a syscall per timer call), clock\_gettime in all other modes}}} \\
\multicolumn{11}{@{}l@{}}{\makebox[0pt][l]{\footnotesize {\em mode F0: 68736 kpoints = 3959 virt + 64777 novi $\to$ 1000 events (53 virt + 947 novi)}}} \\
\multicolumn{11}{@{}l@{}}{\makebox[0pt][l]{\footnotesize {\em mode F, F1, F2, Cs: 71783 kpoints = 3717 virt + 68066 novi $\to$ 1000 events (53 virt + 947 novi)}}} \\
\multicolumn{11}{@{}l@{}}{\makebox[0pt][l]{\footnotesize {\em mode F, Cs: Born / Real 672 / 619 batches, mean batch 282.8 / 1009.4  (mode C: each Real batch is split over 3 to 9 libraries)}}} \\
\multicolumn{11}{@{}l@{}}{\makebox[0pt][l]{\footnotesize {\em MINT2 jobs: 154 channels run (those with at least one event) out of 448}}} \\
\hline
mode & \tmtcellw{F0} & \tmtcellw{F1} & \tmtcellw{F1u} & \tmtcellw{F2} & \tmtcellw{F} & \tmtcellw{C} & \tmtcell{C} & \tmtcell{C} & \tmtcell{C} & \tmtcell{C} \\
cudacpp\_backend & \tmtcellw{-} & \tmtcellw{-} & \tmtcellw{-} & \tmtcellw{-} & \tmtcellw{-} & \tmtcellw{none} & \tmtcell{sse4} & \tmtcell{avx2} & \tmtcell{512y} & \tmtcell{512z} \\
cudacpp\_vecsize & \tmtcellw{-} & \tmtcellw{-} & \tmtcellw{-} & \tmtcellw{-} & \tmtcellw{-} & \tmtcellw{1} & \tmtcell{2} & \tmtcell{4} & \tmtcell{4} & \tmtcell{8} \\
\hline
WALL TOTAL & \tmtcellw{1766.5} & \tmtcellw{1761.8} & \tmtcellw{2577.6} & \tmtcellw{1736.2} & \tmtcellw{1900.1} & \tmtcellw{1432.8} & \tmtcell{960.2} & \tmtcell{641.9} & \tmtcell{616.2} & \tmtcell{521.7} \\
FORTRAN TOTAL & \tmtcellw{1760.0} & \tmtcellw{1755.3} & \tmtcellw{2570.3} & \tmtcellw{1732.4} & \tmtcellw{1896.3} & \tmtcellw{1429.0} & \tmtcell{956.4} & \tmtcell{638.2} & \tmtcell{612.5} & \tmtcell{517.9} \\
- Born & \tmtcellw{107.9} & \tmtcellw{124.3} & \tmtcellw{672.9} & \tmtcellw{96.8} & \tmtcellw{107.6} & \tmtcellw{46.4} & \tmtcell{43.1} & \tmtcell{21.7} & \tmtcell{20.4} & \tmtcell{14.0} \\
- Real & \tmtcellw{1091.5} & \tmtcellw{1176.9} & \tmtcellw{1177.1} & \tmtcellw{1179.7} & \tmtcellw{1287.2} & \tmtcellw{999.1} & \tmtcell{530.2} & \tmtcell{233.4} & \tmtcell{209.0} & \tmtcell{121.3} \\
- Virtual & \tmtcellw{441.8} & \tmtcellw{324.5} & \tmtcellw{324.8} & \tmtcellw{325.3} & \tmtcellw{325.5} & \tmtcellw{326.2} & \tmtcell{326.0} & \tmtcell{325.8} & \tmtcell{325.8} & \tmtcell{325.4} \\
- PS\_Generation & \tmtcellw{3.3} & \tmtcellw{3.4} & \tmtcellw{3.5} & \tmtcellw{3.4} & \tmtcellw{5.9} & \tmtcellw{5.7} & \tmtcell{5.8} & \tmtcell{5.7} & \tmtcell{5.8} & \tmtcell{5.7} \\
- CT\_and\_MCsub & \tmtcellw{65.2} & \tmtcellw{74.7} & \tmtcellw{340.5} & \tmtcellw{82.8} & \tmtcellw{115.5} & \tmtcellw{4.2} & \tmtcell{4.2} & \tmtcell{4.2} & \tmtcell{4.2} & \tmtcell{4.2} \\
- Nbody\_prefactor & \tmtcellw{1.0} & \tmtcellw{1.1} & \tmtcellw{1.1} & \tmtcellw{0.2} & \tmtcellw{6.9} & \tmtcellw{0.2} & \tmtcell{0.2} & \tmtcell{0.2} & \tmtcell{0.2} & \tmtcell{0.2} \\
- Bridge\_crt\_del & \tmtcellw{0.0} & \tmtcellw{0.0} & \tmtcellw{0.0} & \tmtcellw{0.0} & \tmtcellw{0.0} & \tmtcellw{0.4} & \tmtcell{0.4} & \tmtcell{0.4} & \tmtcell{0.4} & \tmtcell{0.4} \\
- Misc & \tmtcellw{49.3} & \tmtcellw{50.5} & \tmtcellw{50.4} & \tmtcellw{44.2} & \tmtcellw{47.7} & \tmtcellw{46.7} & \tmtcell{46.7} & \tmtcell{46.7} & \tmtcell{46.8} & \tmtcell{46.7} \\
Job start/exit & \tmtcellw{3.4} & \tmtcellw{3.4} & \tmtcellw{4.2} & \tmtcellw{0.7} & \tmtcellw{0.7} & \tmtcellw{0.7} & \tmtcell{0.7} & \tmtcell{0.7} & \tmtcell{0.7} & \tmtcell{0.6} \\
Python outside jobs & \tmtcellw{3.1} & \tmtcellw{3.1} & \tmtcellw{3.1} & \tmtcellw{3.1} & \tmtcellw{3.1} & \tmtcellw{3.1} & \tmtcell{3.1} & \tmtcell{3.1} & \tmtcell{3.1} & \tmtcell{3.1} \\
\hline
\end{tabular}

\caption{Timing results (\vfoldscg)
for generating 1000 
\ppttjj\ events
with \tto\ folding
using step17g in MINT2,
on the ``GOLD'' node.}
\label{tab:221GoldPPTTJJ}
}\end{center}
\end{table}

\begin{table}[p]
\begin{center}{
\hspace*{-2.5cm}
\small
\setlength\tabcolsep{5pt} 
\providecommand{\tmtcell}[1]{\makebox[\tmtwidth][r]{#1}}
\providecommand{\tmtcellw}[1]{\makebox[\dimexpr\tmtwidth+0.5em\relax][r]{#1}}
\providecommand{\tmtcellww}[1]{\makebox[\dimexpr\tmtwidth+1.0em\relax][r]{#1}}
\begin{tabular}{lrrrrrrrrrrrrr}
\multicolumn{14}{@{}l@{}}{\makebox[0pt][l]{{\bf pp\_ttxjj $|$ gridpack $|$ NEVENTS=1000 $|$ folding=8,4,1 $|$ kbatch=64}}} \\
\multicolumn{14}{@{}l@{}}{\makebox[0pt][l]{\footnotesize {\em sborn 1: SBORN served from cudacpp, in mode C only}}} \\
\multicolumn{14}{@{}l@{}}{\makebox[0pt][l]{\footnotesize {\em gscale 1: Borns at the real-emission coupling $g'$ obtained by rescaling those at $g$, in mode C only}}} \\
\multicolumn{14}{@{}l@{}}{\makebox[0pt][l]{\footnotesize {\em timers: getrusage in F0, F1, F1u (a syscall per timer call), clock\_gettime in all other modes}}} \\
\multicolumn{14}{@{}l@{}}{\makebox[0pt][l]{\footnotesize {\em mode F0: 75613 kpoints = 2608 virt + 73005 novi $\to$ 1000 events (53 virt + 947 novi)}}} \\
\multicolumn{14}{@{}l@{}}{\makebox[0pt][l]{\footnotesize {\em mode F, F1, F2, Cs: 74780 kpoints = 3276 virt + 71504 novi $\to$ 1000 events (50 virt + 950 novi)}}} \\
\multicolumn{14}{@{}l@{}}{\makebox[0pt][l]{\footnotesize {\em mode F, Cs: Born / Real 1272 / 1199 batches, mean batch 1143.6 / 4118.4 (mode C: each Real batch is split over 3 to 9 libraries)}}} \\
\multicolumn{14}{@{}l@{}}{\makebox[0pt][l]{\footnotesize {\em MINT2 jobs: 154 channels run (those with at least one event) out of 448}}} \\
\hline
mode & \tmtcellww{F0} & \tmtcellww{F1} & \tmtcellww{F1u} & \tmtcellww{F2} & \tmtcellww{F} & \tmtcellww{C} & \tmtcellw{C} & \tmtcellw{C} & \tmtcellw{C} & \tmtcellw{C} & \tmtcellw{C} & \tmtcellw{C} & \tmtcellw{C} \\
cudacpp\_backend & \tmtcellww{-} & \tmtcellww{-} & \tmtcellww{-} & \tmtcellww{-} & \tmtcellww{-} & \tmtcellww{none} & \tmtcellw{sse4} & \tmtcellw{avx2} & \tmtcellw{cuda} & \tmtcellw{cuda} & \tmtcellw{cuda} & \tmtcellw{cuda} & \tmtcellw{cuda} \\
cudacpp\_vecsize & \tmtcellww{-} & \tmtcellww{-} & \tmtcellww{-} & \tmtcellww{-} & \tmtcellww{-} & \tmtcellww{1} & \tmtcellw{2} & \tmtcellw{4} & \tmtcellw{64} & \tmtcellw{128} & \tmtcellw{256} & \tmtcellw{512} & \tmtcellw{1024} \\
\hline
WALL TOTAL & \tmtcellww{15387.2} & \tmtcellww{14873.6} & \tmtcellww{23771.3} & \tmtcellww{14358.0} & \tmtcellww{15419.2} & \tmtcellww{11265.0} & \tmtcellw{6448.7} & \tmtcellw{3402.2} & \tmtcellw{1579.0} & \tmtcellw{1503.8} & \tmtcellw{1468.8} & \tmtcellw{1445.2} & \tmtcellw{1462.8} \\
FORTRAN TOTAL & \tmtcellww{15319.0} & \tmtcellww{14805.0} & \tmtcellww{23680.3} & \tmtcellww{14352.5} & \tmtcellww{15413.6} & \tmtcellww{11259.5} & \tmtcellw{6442.9} & \tmtcellw{3396.8} & \tmtcellw{1524.6} & \tmtcellw{1449.0} & \tmtcellw{1413.6} & \tmtcellw{1390.0} & \tmtcellw{1405.9} \\
- Born & \tmtcellww{1084.1} & \tmtcellww{1107.2} & \tmtcellww{7169.1} & \tmtcellww{684.2} & \tmtcellww{753.0} & \tmtcellww{78.6} & \tmtcellw{72.8} & \tmtcellw{51.9} & \tmtcellw{183.7} & \tmtcellw{142.1} & \tmtcellw{114.6} & \tmtcellw{85.8} & \tmtcellw{74.4} \\
- Real & \tmtcellww{12466.0} & \tmtcellww{11831.0} & \tmtcellww{11792.0} & \tmtcellww{11859.1} & \tmtcellww{12369.1} & \tmtcellww{10160.5} & \tmtcellw{5347.1} & \tmtcellw{2323.3} & \tmtcellw{277.4} & \tmtcellw{242.0} & \tmtcellw{234.9} & \tmtcellw{240.7} & \tmtcellw{265.1} \\
- Virtual & \tmtcellww{451.2} & \tmtcellww{547.8} & \tmtcellww{547.3} & \tmtcellww{548.9} & \tmtcellww{549.2} & \tmtcellww{550.2} & \tmtcellw{549.8} & \tmtcellw{549.8} & \tmtcellw{553.1} & \tmtcellw{553.5} & \tmtcellw{553.6} & \tmtcellw{553.3} & \tmtcellw{553.7} \\
- PS\_Generation & \tmtcellww{44.0} & \tmtcellww{43.0} & \tmtcellww{44.0} & \tmtcellww{40.1} & \tmtcellww{63.0} & \tmtcellww{61.0} & \tmtcellw{60.9} & \tmtcellw{60.8} & \tmtcellw{60.8} & \tmtcellw{60.9} & \tmtcellw{60.8} & \tmtcellw{60.7} & \tmtcellw{61.0} \\
- CT\_and\_MCsub & \tmtcellww{962.1} & \tmtcellww{968.4} & \tmtcellww{3813.6} & \tmtcellww{1009.3} & \tmtcellww{1381.3} & \tmtcellww{177.5} & \tmtcellw{180.7} & \tmtcellw{179.6} & \tmtcellw{145.7} & \tmtcellw{146.6} & \tmtcellw{146.2} & \tmtcellw{145.4} & \tmtcellw{146.9} \\
- Nbody\_prefactor & \tmtcellww{12.0} & \tmtcellww{11.8} & \tmtcellww{12.2} & \tmtcellww{2.9} & \tmtcellww{60.1} & \tmtcellww{2.1} & \tmtcellw{2.0} & \tmtcellw{2.0} & \tmtcellw{2.1} & \tmtcellw{2.0} & \tmtcellw{2.1} & \tmtcellw{2.0} & \tmtcellw{2.1} \\
- Bridge\_crt\_del & \tmtcellww{0.0} & \tmtcellww{0.0} & \tmtcellww{0.0} & \tmtcellww{0.0} & \tmtcellww{0.0} & \tmtcellww{0.7} & \tmtcellw{0.7} & \tmtcellw{0.7} & \tmtcellw{72.5} & \tmtcellw{72.5} & \tmtcellw{72.5} & \tmtcellw{72.9} & \tmtcellw{72.7} \\
- Misc & \tmtcellww{299.7} & \tmtcellww{295.9} & \tmtcellww{302.0} & \tmtcellww{207.9} & \tmtcellww{237.9} & \tmtcellww{229.0} & \tmtcellw{228.9} & \tmtcellw{228.7} & \tmtcellw{229.4} & \tmtcellw{229.3} & \tmtcellw{229.0} & \tmtcellw{229.1} & \tmtcellw{230.0} \\
Job start/exit & \tmtcellww{63.6} & \tmtcellww{64.2} & \tmtcellww{86.7} & \tmtcellww{1.1} & \tmtcellww{1.2} & \tmtcellww{1.1} & \tmtcellw{1.1} & \tmtcellw{1.0} & \tmtcellw{50.1} & \tmtcellw{50.5} & \tmtcellw{50.9} & \tmtcellw{51.0} & \tmtcellw{52.6} \\
Python outside jobs & \tmtcellww{4.5} & \tmtcellww{4.4} & \tmtcellww{4.4} & \tmtcellww{4.3} & \tmtcellww{4.4} & \tmtcellww{4.4} & \tmtcellw{4.7} & \tmtcellw{4.4} & \tmtcellw{4.3} & \tmtcellw{4.3} & \tmtcellw{4.3} & \tmtcellw{4.3} & \tmtcellw{4.3} \\
\hline
\end{tabular}

\caption{Timing results (\vfoldscg)
for generating 1000 
\ppttjj\ events
with \efo\ folding
using step17g in MINT2,
on the ``CUDA'' node.
The C/cuda columns
with \vecsize\ 32 and 2048
have been omitted
for lack of space.}
\label{tab:841CudaPPTTJJ}
\vspace*{1.2cm}
\hspace*{-0.8cm}
\small
\setlength\tabcolsep{5pt} 
\providecommand{\tmtcell}[1]{\makebox[\tmtwidth][r]{#1}}
\providecommand{\tmtcellw}[1]{\makebox[\dimexpr\tmtwidth+0.5em\relax][r]{#1}}
\providecommand{\tmtcellww}[1]{\makebox[\dimexpr\tmtwidth+1.0em\relax][r]{#1}}
\begin{tabular}{lrrrrrrrrrr}
\multicolumn{11}{@{}l@{}}{\makebox[0pt][l]{{\bf pp\_ttxjj $|$ gridpack $|$ NEVENTS=1000 $|$ folding=8,4,1 $|$ kbatch=64}}} \\
\multicolumn{11}{@{}l@{}}{\makebox[0pt][l]{\footnotesize {\em sborn 1: SBORN served from cudacpp, in mode C only}}} \\
\multicolumn{11}{@{}l@{}}{\makebox[0pt][l]{\footnotesize {\em gscale 1: Borns at the real-emission coupling $g'$ obtained by rescaling those at $g$, in mode C only}}} \\
\multicolumn{11}{@{}l@{}}{\makebox[0pt][l]{\footnotesize {\em timers: getrusage in F0, F1, F1u (a syscall per timer call), clock\_gettime in all other modes}}} \\
\multicolumn{11}{@{}l@{}}{\makebox[0pt][l]{\footnotesize {\em mode F0: 75613 kpoints = 2608 virt + 73005 novi $\to$ 1000 events (53 virt + 947 novi)}}} \\
\multicolumn{11}{@{}l@{}}{\makebox[0pt][l]{\footnotesize {\em mode F, F1, F2, Cs: 74780 kpoints = 3276 virt + 71504 novi $\to$ 1000 events (50 virt + 950 novi)}}} \\
\multicolumn{11}{@{}l@{}}{\makebox[0pt][l]{\footnotesize {\em mode F, Cs: Born / Real 1272 / 1199 batches, mean batch 1143.6 / 4118.4 (mode C: each Real batch is split over 3 to 9 libraries)}}} \\
\multicolumn{11}{@{}l@{}}{\makebox[0pt][l]{\footnotesize {\em MINT2 jobs: 154 channels run (those with at least one event) out of 448}}} \\
\hline
mode & \tmtcellww{F0} & \tmtcellww{F1} & \tmtcellww{F1u} & \tmtcellww{F2} & \tmtcellww{F} & \tmtcellw{C} & \tmtcellw{C} & \tmtcellw{C} & \tmtcellw{C} & \tmtcellw{C} \\
cudacpp\_backend & \tmtcellww{-} & \tmtcellww{-} & \tmtcellww{-} & \tmtcellww{-} & \tmtcellww{-} & \tmtcellw{none} & \tmtcellw{sse4} & \tmtcellw{avx2} & \tmtcellw{512y} & \tmtcellw{512z} \\
cudacpp\_vecsize & \tmtcellww{-} & \tmtcellww{-} & \tmtcellww{-} & \tmtcellww{-} & \tmtcellww{-} & \tmtcellw{1} & \tmtcellw{2} & \tmtcellw{4} & \tmtcellw{4} & \tmtcellw{8} \\
\hline
WALL TOTAL & \tmtcellww{12067.3} & \tmtcellww{11649.2} & \tmtcellww{18506.7} & \tmtcellww{11344.1} & \tmtcellww{12130.6} & \tmtcellw{8455.4} & \tmtcellw{4848.2} & \tmtcellw{2546.2} & \tmtcellw{2357.3} & \tmtcellw{1677.8} \\
FORTRAN TOTAL & \tmtcellww{12044.9} & \tmtcellww{11627.3} & \tmtcellww{18478.4} & \tmtcellww{11339.7} & \tmtcellww{12126.4} & \tmtcellw{8451.3} & \tmtcellw{4844.2} & \tmtcellw{2542.3} & \tmtcellw{2353.4} & \tmtcellw{1673.9} \\
- Born & \tmtcellww{796.3} & \tmtcellww{826.5} & \tmtcellww{5486.2} & \tmtcellww{548.9} & \tmtcellww{552.5} & \tmtcellw{56.9} & \tmtcellw{56.0} & \tmtcellw{39.0} & \tmtcellw{37.2} & \tmtcellw{37.6} \\
- Real & \tmtcellww{10002.5} & \tmtcellww{9484.5} & \tmtcellww{9484.9} & \tmtcellww{9504.6} & \tmtcellww{9926.5} & \tmtcellw{7690.9} & \tmtcellw{4083.5} & \tmtcellw{1799.4} & \tmtcellw{1613.4} & \tmtcellw{934.1} \\
- Virtual & \tmtcellww{296.6} & \tmtcellww{360.3} & \tmtcellww{361.4} & \tmtcellww{361.4} & \tmtcellww{361.2} & \tmtcellw{363.5} & \tmtcellw{362.6} & \tmtcellw{362.4} & \tmtcellw{362.0} & \tmtcellw{362.4} \\
- PS\_Generation & \tmtcellww{28.2} & \tmtcellww{27.4} & \tmtcellww{27.8} & \tmtcellww{27.2} & \tmtcellww{45.1} & \tmtcellw{44.7} & \tmtcellw{44.8} & \tmtcellw{44.7} & \tmtcellw{44.7} & \tmtcellw{44.7} \\
- CT\_and\_MCsub & \tmtcellww{707.6} & \tmtcellww{717.7} & \tmtcellww{2907.3} & \tmtcellww{757.2} & \tmtcellww{1036.2} & \tmtcellw{133.3} & \tmtcellw{135.4} & \tmtcellw{134.8} & \tmtcellw{134.3} & \tmtcellw{133.0} \\
- Nbody\_prefactor & \tmtcellww{9.2} & \tmtcellww{9.0} & \tmtcellww{9.3} & \tmtcellww{2.0} & \tmtcellww{41.6} & \tmtcellw{1.3} & \tmtcellw{1.3} & \tmtcellw{1.3} & \tmtcellw{1.3} & \tmtcellw{1.3} \\
- Bridge\_crt\_del & \tmtcellww{0.0} & \tmtcellww{0.0} & \tmtcellww{0.0} & \tmtcellww{0.0} & \tmtcellww{0.0} & \tmtcellw{0.4} & \tmtcellw{0.4} & \tmtcellw{0.4} & \tmtcellw{0.4} & \tmtcellw{0.4} \\
- Misc & \tmtcellww{204.6} & \tmtcellww{201.8} & \tmtcellww{201.5} & \tmtcellww{138.4} & \tmtcellww{163.4} & \tmtcellw{160.3} & \tmtcellw{160.2} & \tmtcellw{160.2} & \tmtcellw{160.1} & \tmtcellw{160.4} \\
Job start/exit & \tmtcellww{19.2} & \tmtcellww{18.7} & \tmtcellww{25.0} & \tmtcellww{0.8} & \tmtcellww{0.9} & \tmtcellw{0.8} & \tmtcellw{0.8} & \tmtcellw{0.8} & \tmtcellw{0.8} & \tmtcellw{0.8} \\
Python outside jobs & \tmtcellww{3.2} & \tmtcellww{3.2} & \tmtcellww{3.3} & \tmtcellww{3.5} & \tmtcellww{3.3} & \tmtcellw{3.3} & \tmtcellw{3.2} & \tmtcellw{3.1} & \tmtcellw{3.1} & \tmtcellw{3.1} \\
\hline
\end{tabular}

\caption{Timing results (\vfoldscg)
for generating 1000 
\ppttjj\ events
with \efo\ folding
using step17g in MINT2,
on the ``GOLD'' node.}
\label{tab:841GoldPPTTJJ}
}\end{center}
\end{table}

\begin{table}[t]
\hspace*{-1.5cm}
\centering
\begin{tabular}{|c|cccc|cc|cc|}
\hline
 & & & & & \multicolumn{2}{c|}{``CUDA'' node} & \multicolumn{2}{c|}{``GOLD'' node} \\
folding
 & $I$ [pb] & $J$ [pb] & $F$ [\%] & $C(F)$
 & F1 & C/cuda & F1 & C/512z \\
\hline
\ooo & $693.6 \pm 13.0$ & $4015 \pm 13$  & $41.4 \pm 0.2$ & $33.5 \pm 1.3$ &  1008.9s &  694.2s &   731.6s &  381.4s \\
\tto & $685.2 \pm 9.9$  & $3319 \pm 10$  & $39.7 \pm 0.1$ & $23.5 \pm 0.7$ &  2320.6s &  822.1s &  1761.8s &  521.7s \\
\efo & $684.9 \pm 8.7$  & $3123 \pm 8$   & $39.0 \pm 0.1$ & $20.8 \pm 0.5$ & 14873.6s & 1445.2s & 11649.2s & 1677.8s \\
\hline
\end{tabular}
\caption{Summary for \ppttjj\ with 
three different folding configurations.
Cross section $I$,
absolute cross section $J$,
negative weight fraction $\fneg=1/2(1-I/J)$
and relative cost $\cost(\fneg)=(J/I)^2$
are those of the MINT1 step 
of the gridpack creation
(a single run for each folding).
Times are the total wall-clock times
for the MINT2 generation 
of 1000 unweighted events from the gridpack
in mode F1 and in the fastest C/cuda 
configuration on the ``CUDA'' node,
and in modes F1 and C/512z 
on the ``GOLD'' node,
from Tables~\ref{tab:111CudaPPTTJJ} 
and~\ref{tab:111GoldPPTTJJ} (\ooo\ folding),
\ref{tab:221CudaPPTTJJ} 
and~\ref{tab:221GoldPPTTJJ} (\tto\ folding),
\ref{tab:841CudaPPTTJJ} 
and~\ref{tab:841GoldPPTTJJ} (\efo\ folding).}
\label{tab:ppttjjsummary}
\end{table}

\begin{itemize}

\item 
For this complex workflow
involving 150 Fortran 
applications, the one-off
overheads of GPU usage
of each application
sum up to very large times.
In all three foldings,
approximately 70s are spent
in GPU hardware initialization,
while another 50s are spent
on job start/exit
(most likely, in loading
or unloading CUDA libraries,
although this has not been
investigated further).
This is a very considerable
fraction of total time,
e.g. approximately 120s 
out of 700s for \ooo\ folding.
To reduce this, 
one could think of using
a single GPU-server application,
with a single GPU initialization
and a single load/unload overhead,
communicating with its 
channel-specific clients
via network packets;
however, this is 
almost certainly unnecessary,
if one takes into account
that real production jobs
would probably generate
much more than 1000 events
and run for much longer
than a few hours.
In the following comparison 
of SIMD and GPU results,
in any case,
I will not subtract
these one-off GPU overheads.

\item 
Through CUDACPP,
\mvf\ provides significant
speedups of Born and Real,
both using GPUs and SIMD.
Thanks to the various
optimizations targeting Born
other than vectorization,
however, the Born speedups
achieved through SIMD
(avx2 on the CUDA node,
512z on the GOLD node)
are comparable to
or better than
those achieved on the GPU.
In the absence of folding,
it turns out that 
event generation
on the CUDA node
is faster using AVX2
rather than using the GPU
(even if, again,
one should remember
that the GPU times
include a large 
one-off overhead,
and that no attempt
was made to have work 
done simultaneously
on the CPU and on the GPU).
The interest of the GPU,
however, appears very clearly
for \efo\ folding:
in this case, in fact,
the GPU yields a much better
speedup of Reals,
and as a consequence
of the whole workflow.
On the CUDA node,
for instance,
when moving from F1
to the best GPU
or SIMD result,
the total time
decreases from 1009s 
to 627s(1.6x) with C/avx2
and 694s(1.5x) with C/cuda
for \ooo\ folding,
and from 14874s
to 3402s(4.4x) with C/avx2
and 1445s(10.3x) with C/cuda
for \efo\ folding.
For comparison,
on the GOLD node,
the total time
decreases from 732s 
to 381s(1.9x) with C/512z
for \ooo\ folding,
and from 11649s
to 1678s(6.9x) with C/512z
for \efo\ folding.

\item
After accelerating
Born and Real in \mvf,
the Virtual contribution
becomes the bottleneck
of the full workflow
with \ooo\ (431.9s
out of 694.2s
with C/cuda/512 
on the CUDA node,
283.9s out
of 381.4s
with C/512z on 
the GOLD node)
and \tto\ folding
(498.4s
out of 822.1s
with C/cuda/128 
on the CUDA node,
325.4s out
of 521.7s
with C/512z on 
the GOLD node),
and to some extent
also with \efo\ folding
on the GPU
(553.3s
out of 1445.2s
with C/cuda/512 
on the CUDA node).
Using SIMD 
(C/512z on 
the GOLD node),
conversely,
Reals are still 
the bottleneck
(934.1s out of 1677.8s)
and a much larger
contribution than
Virtuals (362.4s).
\end{itemize}

The total times spent 
in the F1 mode and 
in the fastest 
GPU (for the CUDA node) and 
SIMD (for the GOLD node) modes
from these six tables
are also summarized
in Table~\ref{tab:ppttjjsummary},
which also gives
the cross section,
absolute cross section,
negative weight fraction $\fneg$
and corresponding cost $\cost(\fneg)$ 
for these six configurations.
The table shows that folding
clearly helps in reducing 
negative weights,
but has a diminishing
marginal utility:
\begin{itemize}
\item 
Moving from \ooo\ to 
\tto\ folding
reduces negative weight 
costs by a factor 1.4
from 33.5 to 23.5.
Using F1 mode,
which has the same
optimizations as stock MadGraph7, 
it only costs
a factor 2.3 (2321s/1009s)
or 2.4 (1762s/732s)
in event generation time
on the CUDA and GOLD nodes.
Using \mvf,
the folding cost 
in event generation time
is reduced to a 
factor 1.2 (822s/694s) with C/cuda
or 1.4 (522s/381s) with C/512z:
in this respect,
both of these 
\mvf\ configurations
already 
achieve a break-even point
even without considering
detector simulation
and other downstream costs.
\item 
Costs are much higher
in stock MadGraph7
for \efo\ folding,
which (with respect to \ooo)
reduces negative weight costs 
by a factor 1.6
from 33.5 to 20.8.
Using stock MadGraph7 
(F1 mode), this configuration,
which only brings a 
moderate marginal improvement
with respect to \tto,
is however much more expensive,
as it costs
a factor 14.7 (14874s/1009s)
or 15.9 (11649s/732s)
in event generation time
on the CUDA and GOLD nodes.
Using \mvf, however,
the folding cost 
in event generation time
is significantly reduced to a 
factor 2.1 (1445s/694s) with C/cuda
or 4.4 (1678s/381s) with C/512z.
Unlike the situation 
in \tto\ folding,
which seemed to provide
a sweet spot with 
a good compromise
between benefits
and generation costs,
these numbers show
that vectorized folding
does not reach 
a break-even point
on event generation costs alone
with \efo\ folding:
nevertheless,
I would still imagine
that this configuration
would reduce overall costs
if detector simulation
and other downstream costs
were included.

\item
Finally,
I rephrase
some of the observations
in the last bullet:
\efo\ folding in \ppttjj,
which gives a 1.6x reduction
in the cost of negative weights,
costs as much as 14.7x
in generation time 
using stock MadGraph7,
but only 2.1x using
\mvf\ on a GPU:
this proves that 
vectorized folding
and its implementation 
in \mvf\ are tools
that can be very useful
in addressing the
issue of NLO negative weights
in the LHC experiments.

\end{itemize}

\section{Conclusions}
\label{sec:conc}
In this paper,
I have proposed “vectorized folding” 
as a new idea to reduce negative weights 
in NLO simulations, 
by speeding up folding 
via the data-parallel
calculation of matrix elements
for different folds.
I have presented 
a first implementation
of NLO fold vectorization 
for \mgamc\ and MadGraph7 in 
a new package \mvf,
which offloads these calculations
to the CUDACPP software, 
previously developed for LO simulations.
This work initially targeted
the vectorization of folding,
but was later extended
to the vectorization
of NLO event generation 
in the absence of folding.
\mvf\ supports 
automatic code generation
in MadGraph7
and gridpack generation
for a large family
of NLO physics processes,
including complex
processes like \ppttjj\ that
are directly relevant 
to the LHC physics experiments.
In this specific case,
preliminary tests
show overall NLO speedups
over MadGraph7 of
6.9x and 10.3x for AVX512 and GPUs
with \efo\ folding
and of 1.9x and 1.5x 
without it.
To my knowledge,
this is the first demonstration 
of vectorized folding, 
and of an end-to-end speedup 
of the full MC@NLO 
event generation workflow 
with folding, 
on SIMD CPUs and GPUs.
The integration and synergies
with other ongoing efforts
on NLO vectorization 
in the MadGraph7 project 
are currently
being discussed.
This has already resulted
in the inclusion into MadGraph7
of some optimizations
partially derived from
those developed in \mvf.
Pull requests have 
also been created
to add \mvf\ support
in CUDACPP~\cite{bib:pr1097}
and MadGraph7~\cite{bib:pr451}.
This work is based on a test-centric, 
LLM-assisted software development process,
which has been described in detail;
a more thorough review 
of the LLM-developed code
would in any case be appropriate
before any production use
of this software.

\section*{Acknowledgements}
I thank the full \mgamc\ team
for their support and valuable input
on many aspects of this work.
In particular, I am indebted
to Stefano Frixione 
for his enlightening clarifications
about the different mechanisms
leading to negative weights
in NLO simulations:
this project could not have 
started without our 
fruitful and pleasant
discussions on these issues
over the last three years.
I warmly thank Marco Zaro
for his help and support
in understanding the internals
of the MINT-based Fortran 
codebase of \mgamc, 
which is where most of the
developments in this work
have concentrated;
I also thank him 
for the MINT1 data dumps,
which were one of his earliest
additions to the FVZ codebase
and which I later hijacked
for my tests in this project.
More generally, I thank
Stefano, Marco
as well as Veronica Fossa
and Luca Beccatini
for our pleasant collaboration
on investigating new
Machine Learning based
techniques for reducing 
negative weights:
the work I present in this
paper represents a spin-off
of our joint research
on those topics.
I thank Olivier Mattelaer
for useful discussions
on the possible integration
of \mvf\ in MadGraph7.
I 
thank Daniele Massaro
and Rikkert Frederix
for useful discussions
about various aspects
of the \mgamc\ and
MadGraph7 software.
As discussed extensively
in this paper,
this research has significantly
benefitted from the use
of Large Language Models
(initially Sonnet5
and later Opus5.5)
through the Claude Code
user interface,
for refactoring Fortran code
and for developing
CUDA/C++ optimizations.
These LLMs
have also been used 
to write the user guide 
in the appendix
and to provide feedback 
on my own text
in the main body
of this paper
(from abstract to
references included).
Finally, I thank
my colleagues and management
in the EP-LCB group
for providing a friendly
and relaxed atmosphere
to work on these topics,
alongside my other commitments
on LHCb simulation software activities.

\newcommand{\arxiv}[1]{\url{https://arxiv.org/abs/#1}}

\appendix
\providecommand{\vfenv}[1]{\texttt{ENV\_\allowbreak{}VFOLD\_\allowbreak{}#1}}
\providecommand{\vfvar}[1]{\texttt{#1}}
\providecommand{\vftest}[1]{\texttt{ENV\_\allowbreak{}VFOLDTEST\_\allowbreak{}#1}}
\providecommand{\vfenvnb}[1]{\mbox{\texttt{ENV\_VFOLD\_#1}}}
\providecommand{\vftestnb}[1]{\mbox{\texttt{ENV\_VFOLDTEST\_#1}}}
\providecommand{\vfdesc}[2]{\multicolumn{#1}{@{\hspace{5mm}}p{\dimexpr\textwidth-5mm\relax}@{}}{#2}}

\section{A brief user guide}
\label{app:guide}
\setlength{\emergencystretch}{2em}

This appendix describes
how to generate and run
a process directory with \mvf,
which environment variables
control it at runtime,
how to reproduce
the timing tables of this paper
with the \mvf\ test infrastructure,
and how to create and run
the NLO gridpacks
used for $pp\to t\bar{t}jj$.
It refers to the current versions of \mvf,
based on \mgamc\ v3.8.1
(development branch 
\texttt{vfold5381cg} of \mgamc)
and on MadGraph7
(development branch 
\texttt{vfold7cg} ---
now replaced by
\texttt{vfold7dev}).
The instructions are 
essentially the same for both,
but are 
given only for MadGraph7
in the following,
except when an important
difference must be noted.
Specifically, the following tags 
and branches have been defined 
for the convenience of users
in my private forks
of the 
madgraph4gpu 
and MadGraph7
code repositories:
\begin{itemize}

\item CUDACPP tag
\texttt{valassi\_cudacpp\_for3.8.1\_v1.02.01\_vfold7on20261127}~\cite{bib:cudacpptag}
represents the 
public release of
the CUDACPP plugin for \mvf.
As mentioned below and in the 
relevant
\texttt{README\_\allowbreak{}CUDACPP.md}
file in \mvf,
this is the tarball that 
should be installed
in the \texttt{PLUGIN}
directory of MadGraph7.

\item MadGraph7 tag
\texttt{valassi\_vfold7on20261127}~\cite{bib:vfold7tag}
represents the 
public release of
the \mvf\ code generation
infrastructure for MadGraph7.
This is what interested 
users should download.

\item MadGraph7 branch
\texttt{vfold7main}~\cite{bib:vfold7main}
is the user-facing 
branch which will include any 
future changes to
the \mvf\ code generation
infrastructure for MadGraph7.
It coincides 
with tag 
\texttt{valassi\_vfold7on20261127}
at the time of writing.

\item MadGraph7 tag
\texttt{valassi\_vfold7on20261127dev}~\cite{bib:vfold7devtag}
represents a frozen version
of the development branch
\texttt{vfold7dev}
based on the above release
of the code generation infrastructure.
This includes the 
automatically generated
physics process code 
and the results of the tests
described in this paper.

\item MadGraph7 branch
\texttt{vfold7dev}~\cite{bib:vfold7dev}
is the development 
branch
(replacing \texttt{vfold7cg})
which may include any 
future changes to
the \mvf\ code generation
infrastructure,
automatically generated
physics process code 
and test results for MadGraph7.
It coincides 
with tag 
\texttt{valassi\_vfold7on20261127dev}
at the time of writing.

\end{itemize}

\subsection{Installation and code generation}
\label{app:install}

\mvf\ is kept in a parallel structure
inside the MadGraph7 
repository,
which is designed not 
to interfere with the stock code.
All of its code is kept in its own
\texttt{madvfold} directory,
as two overlays
of the files of the 
standard MadGraph7 
templates and Python modules
(modified copies of a few 
stock Fortran and Python files,
and new files of its own),
which are installed in a generated process directory
only if this is requested.
Outside the 
\texttt{madvfold} directory,
\mvf\ only changes a few files 
of the stock code generator:
\texttt{madgraph/\allowbreak{}interface/\allowbreak{}amcatnlo\_\allowbreak{}interface.py}
(the new options of the \texttt{output} command),
\texttt{madgraph/\allowbreak{}iolibs/\allowbreak{}export\_\allowbreak{}fks.py}
(the installation of the overlays),
and, in \mgamc\ only,
\texttt{madgraph/\allowbreak{}iolibs/\allowbreak{}file\_\allowbreak{}writers.py}
(a one-line fix of a regular expression,
also proposed upstream).
Without the new options,
\texttt{output} generates
exactly the same process directory
as the stock MadGraph7
and the stock templates are never modified.
The other side of this design is that
the modified copies of the stock files in the overlays
follow the stock versions they were taken from:
\mvf\ currently supports only the specific versions
of \mgamc\ (v3.8.1) and MadGraph7
(the commit used as the base 
of the MadGraph7 tags above)
on which it was developed,
and it would need to be ported,
merging the changes of the stock files into their \mvf\ copies,
to follow any later changes of the stock code.
Mode C also needs the \mvf\ version
of the CUDACPP plugin (version 1.02.01),
whose tarball \texttt{cudacpp.tar.gz}
is available from the release
mentioned above~\cite{bib:cudacpptag}
(the same tarball is used for \mgamc\ v3.8.1 and for MadGraph7).
This must be installed in the
\texttt{PLUGIN/\allowbreak{}CUDACPP\_\allowbreak{}OUTPUT} directory
(as described in \texttt{madvfold/\allowbreak{}README\_\allowbreak{}CUDACPP.md};
in MadGraph7 this must be 
done using the script
\texttt{madvfold/\allowbreak{}cudacpp\_\allowbreak{}mg7/\allowbreak{}installPlugin.sh}).
MadGraph7 needs Python 3.12 or higher.
The external libraries used for the virtual corrections
(Ninja, COLLIER) are installed by default
in the \texttt{HEPTools} directory
the first time that a process with virtual corrections is generated,
as for any NLO installation 
of MadGraph7. 
Pythia8, which is needed only 
for \amcDELTA,
is not installed automatically:
it must be installed before 
the first run with \amcDELTA,
with the command
\texttt{install pythia8}
at the MadGraph7 prompt.

A process directory for \mvf\ is generated
by adding one of two options to the usual \texttt{output} command:
\begin{verbatim}
  generate e+ e- > b b~ b b~ [QCD]
  output eebbbb --vfold             # for users
  output eebbbb --vfold-with-tests  # for developers
\end{verbatim}
With either option,
the Fortran code of the process directory
includes the vectorized folding of \mvf,
whose behaviour is chosen at runtime
by the environment variables described below,
and the CUDACPP libraries for mode C
are generated and built
for all CPU backends
(and for the GPU, if CUDA is installed)
in the \texttt{CUDACPP} subdirectory
of the process directory
(this may take several minutes),
and then copied to its \texttt{lib/\allowbreak{}cudacpp} subdirectory.
The \texttt{--vfold-with-tests} option
also installs the test infrastructure
described in Sec.~\ref{app:tests},
and enables the test hooks of the Fortran code
(which are compiled out with \texttt{--vfold}).
The script \texttt{madvfold/\allowbreak{}DEV/\allowbreak{}genProc.sh}
generates a process directory with \texttt{--vfold-with-tests}
for a given \texttt{generate} command or
for one of a few predefined aliases
(e.g. \texttt{genProc.sh eebbbb} creates
\texttt{madvfold/\allowbreak{}DEV/\allowbreak{}eebbbb\_\allowbreak{}vfgen}).
The \texttt{<proc>\_\allowbreak{}vfgen} directories
are the reference process directories
committed in the repository:
users should give their own process directories
another name as second argument,
e.g. \texttt{genProc.sh eebbbb madvfold/\allowbreak{}DEV/\allowbreak{}eebbbb\_\allowbreak{}user}
(at the same depth \texttt{madvfold/\allowbreak{}DEV/\allowbreak{}<name>},
where the test scripts expect them).

\subsection{Running a production}
\label{app:run}

A process directory generated with \texttt{--vfold}
is run in the same way as a standard one,
e.g. with \texttt{./\allowbreak{}bin/\allowbreak{}generate\_\allowbreak{}events},
after setting the \texttt{ENV\_\allowbreak{}VFOLD\_\allowbreak{}*}
environment variables listed in Table~\ref{tab:envrun}.
These are the only switches added by \mvf\ for users.
The folding parameters are not among them:
they are taken as usual from the
\texttt{folding} entry of the run card
(e.g. \texttt{8, 4, 1}).
The variables can be set to example values with
\begin{verbatim}
  source vfold/setupVfoldDefaults.sh
  ./bin/generate_events
\end{verbatim}
where \texttt{setupVfoldDefaults.sh}
keeps the variables already set, and the example values
(mode C on CPU with the AVX2 backend, kpoint batching,
SBORN and its rescaled values served from CUDACPP)
should be tuned to the user's hardware and process.
Conversely,
\texttt{source vfold/\allowbreak{}unsetVfold.sh}
unsets all the \mvf\ environment variables
(including those of the tests, Table~\ref{tab:envtest}),
to go back to the defaults of the code.
In practice,
a user should at least:
\begin{itemize}
\item set \vfenv{ME\_\allowbreak{}MODE}, which has no default
(mode C for production, F0 to reproduce standard \mgamc);
\item in mode C, choose the backend
with \vfenv{CUDACPP\_\allowbreak{}BACKEND}
(e.g. \texttt{avx2} on CPUs with AVX2,
\texttt{cuda} on NVidia GPUs)
and, on GPUs, \vfenv{VECSIZE\_\allowbreak{}USED};
\item choose \vfenv{KPOINT\_\allowbreak{}BATCH},
which in this paper was 64 for folding \efo,
128 for folding $2\times2\times1$
and 512 for no folding (\ooo);
\item 
use \mvf\ only for processes 
whose Born has a single 
coupling combination: 
processes with several 
(e.g. mixed QCD and 
electroweak contributions) 
are not supported, 
and mode C stops the run 
with an error 
(or gives wrong results); 
these processes should be 
generated with stock MadGraph instead.
\end{itemize}
Modes F0, F1, F2 and F
are Fortran-only modes, which are mainly useful for testing:
F0 is the standard \mgamc\ code,
F1 adds the slot-based random numbers that make
all other modes give the same results
(and a diagram-level test of the good helicities of the Born,
see \vfenv{BORN\_\allowbreak{}DIAGGOODHEL} in Table~\ref{tab:envrun}),
F2 adds the optimizations of the scalar Fortran code
(e.g. the Born deduplication),
and F evaluates the matrix elements in batches
like mode C, but in Fortran.
With \mgamc\ v3.8.1,
modes F1, F2 and F give bit-by-bit identical results,
which differ from those of mode C
only by rounding.
With MadGraph7,
whose standard Fortran code reuses
the Born amplitudes computed at a different strong coupling
(by rescaling them),
the results of the three Fortran modes
differ by rounding
(they are bit-by-bit identical with
\vfenv{MG7BORNNOOPT}=1,
a developer switch which disables this reuse).

\begin{table}[tbh]
\begin{center}{\small\sloppy\setlength{\tabcolsep}{6pt}
\begin{tabular}{@{}ll@{}}
\hline
Variable & Values (default) \\
\vfdesc{2}{Description} \\
\hline
\vfenvnb{ME\_MODE} & F0, F1, F2, F, C (no default: required) \\
\vfdesc{2}{Mode of the matrix element calculation
  (C: CUDACPP matrix elements).} \\
\hline
\vfenvnb{CUDACPP\_BASE} & path (no default: required in mode C) \\
\vfdesc{2}{Directory of the CUDACPP libraries
  (\texttt{<procdir>/lib/cudacpp}).} \\
\hline
\vfenvnb{CUDACPP\_BACKEND} & none, sse4, avx2, 512y, 512z, cuda (default: 512y) \\
\vfdesc{2}{CUDACPP backend in mode C (SIMD on CPU, or GPU).} \\
\hline
\vfenvnb{VECSIZE\_USED} & integer (default: 128) \\
\vfdesc{2}{Events per CUDACPP call: the SIMD width on CPU
  (1, 2, 4, 4, 8 for none to 512z),
  a multiple of 32 on GPU.} \\
\hline
\vfenvnb{KPOINT\_BATCH} & integer (default: 1) \\
\vfdesc{2}{Kpoints (with all their folds) evaluated in one batch
  (capped so that kpoints times folds fit the buffers).} \\
\hline
\vfenvnb{BATCH\_PERVN} & 0, 1 (default: 0) \\
\vfdesc{2}{With kpoint batches, keep one batch per MINT2 queue
  of candidates (1) rather than regather it (0).} \\
\hline
\vfenvnb{SBORN\_BRIDGE} & 0, 1, 2 (default: 0) \\
\vfdesc{2}{Mode C: serve the Born (and the colour-linked Borns) from CUDACPP
  (2: also compare to the Fortran Born, for debugging).} \\
\hline
\vfenvnb{SBORN\_GSCALE} & 0, 1 (default: 0) \\
\vfdesc{2}{With \vfenvnb{SBORN\_BRIDGE}: serve the Born
  at another strong coupling by rescaling.} \\
\hline
\vfenvnb{CPUTIME} & $-2$ to 2 (default: 0) \\
\vfdesc{2}{Implementation of the \mgamc\ timers
  (0: \texttt{getrusage}; 1, 2: faster wall-clock timers;
  $-1$, $-2$: 0 in F0, F1 and 1, 2 otherwise).} \\
\hline
\vfenvnb{RNG\_SHAREDFKS} & 0, 1 (default: 0) \\
\vfdesc{2}{Modes F1 to C: the random numbers drawn once per FKS configuration
  within a fold (e.g. the colour partners of the MC counterterms)
  are independent for each FKS configuration
  (0, from a counter-based hash of the fold slot)
  or shared by all of them (1, as in earlier versions).} \\
\hline
\vfenvnb{BORN\_DIAGGOODHEL} & 0, 1 (default: 0 in F0, 1 otherwise) \\
\vfdesc{2}{Good-helicity test of the Born: a helicity is good if its Born
  is not zero (0, as in stock MadGraph7), or if its Born
  or its sum over diagrams of $|A|^2$ is not zero (1),
  so that the multichannel weights do not depend on rounding
  when the diagrams cancel (e.g. in $gg\to t\bar{t}$).} \\
\hline
\vfenvnb{QUIETWARN} & 0, 1 (default: 0) \\
\vfdesc{2}{1: print only the occurrences 1--9, 10, 20, \ldots, 100, 200, \ldots\
  of the most frequent stock warnings
  (their totals are printed at the end of each job).} \\
\hline
\texttt{CUDACPP\_RUNTIME\_BLASCOLORSUM} & unset, 1 (default: unset) \\
\vfdesc{2}{On GPU, colour sums with cuBLAS
  (\texttt{CUDACPP\_RUNTIME\_BLASCOLORSUM\_VFOLD<lib>}
  overrides it for one library, e.g. \texttt{BORN}, \texttt{REAL1}).} \\
\hline
\texttt{CUDACPP\_RUNTIME\_HELPAIR} & unset, 0, 1 (default: unset) \\
\vfdesc{2}{1: on CPU, in the real-emission libraries,
  compute only one of two good helicities related by flipping all helicities
  if their $|M|^2$ are equal, with weight 2
  (as the \texttt{T\_IDENT} logic of the stock Fortran).} \\
\hline
\end{tabular}}
\end{center}
\vspace*{-3mm}
\caption{Environment variables read at runtime
by a process directory generated with \texttt{--vfold}
(or \texttt{--vfold-with-tests}).
The defaults are those of the code when a variable is not set:
\texttt{vfold/\allowbreak{}setupVfoldDefaults.sh} sets
mode C,
the AVX2 backend and its SIMD width,
\vfenv{KPOINT\_\allowbreak{}BATCH}=64,
1 for
\vfenv{BATCH\_\allowbreak{}PERVN},
\vfenv{SBORN\_\allowbreak{}BRIDGE},
\vfenv{SBORN\_\allowbreak{}GSCALE},
\vfenv{QUIETWARN} and
\texttt{CUDACPP\_\allowbreak{}RUNTIME\_\allowbreak{}HELPAIR},
and (in MadGraph7)
\vfenv{CPUTIME}=$-1$, as in the tests;
\texttt{vfold/\allowbreak{}unsetVfold.sh} unsets them all.}
\label{tab:envrun}
\end{table}

\subsection{Test infrastructure}
\label{app:tests}

A process directory generated with \texttt{--vfold-with-tests}
also includes the test scripts used in this work,
in its \texttt{vfold} subdirectory,
and the MINT0/MINT1 grids for the MINT2-only tests,
if they exist for that process
(in \texttt{madvfold/\allowbreak{}test/\allowbreak{}grids/\allowbreak{}<proc>/\allowbreak{}folding\_\allowbreak{}<f1>\_\allowbreak{}<f2>\_\allowbreak{}<f3>}
of the repository, e.g. \texttt{<proc>} is \texttt{epem\_\allowbreak{}bbbxbx} for \eebbbb;
they can be created with \texttt{madvfold/\allowbreak{}test/\allowbreak{}makeGrids.sh}).
The main scripts are:
\begin{itemize}
\item \texttt{vfold/\allowbreak{}myLaunch.sh [-o] [-x]}:
one run in the mode chosen by \vfenv{ME\_\allowbreak{}MODE},
with the process-specific test settings
(a reduced number of MINT0/MINT1 iterations and points,
a single integration channel)
unless these are overridden by the variables
of Table~\ref{tab:envtest};
with \texttt{-o}, only MINT2 is executed,
from the MINT0/MINT1 grids of the test;
with \texttt{-x}, the existing binaries are used.
The run is kept in \texttt{Events/\allowbreak{}run\_\allowbreak{}01}.
\item \texttt{vfold/\allowbreak{}myBackends.sh}:
the runs of all modes and backends
(F0, F1, F2, F, then C with all CPU backends,
and cuda if a GPU is found, with a scan of
\vfenv{VECSIZE\_\allowbreak{}USED}),
and the comparison of their results.
\item \texttt{vfold/\allowbreak{}myStepTest.sh --ref REF<n>/\allowbreak{}<tag> <step>}:
the runs of \texttt{myBackends.sh}
for a predefined set of tests (\texttt{--tsts}),
such as \texttt{n200f841k64g1} for
200 events with folding \efo, \vfenv{KPOINT\_\allowbreak{}BATCH}=64
and \vfenv{SBORN\_\allowbreak{}GSCALE}=1,
kept in \texttt{Events/\allowbreak{}REF<n>/\allowbreak{}<tag>/\allowbreak{}<test>/\allowbreak{}<step>*};
then the comparison of their results
and their timing table.
\item \texttt{vfold/\allowbreak{}myTimingTable.py <prefix>}:
the timing table of the runs of one test,
as text or LaTeX.
\end{itemize}
The environment variables specific to the tests
are listed in Table~\ref{tab:envtest}.
They are read either by the test scripts,
or by the test hooks that \texttt{--vfold-with-tests}
adds to the Python and Fortran code:
they have no effect in a process directory
generated with \texttt{--vfold}.
The test scripts also set the variables
of Table~\ref{tab:envrun}
(e.g. \vfenv{CUDACPP\_\allowbreak{}BACKEND}=avx2
and \vfenv{CPUTIME}=$-1$),
unless they are set in the environment:
\texttt{source vfold/\allowbreak{}unsetVfold.sh}
clears the variables left over from previous tests.

\begin{table}[tbh]
\begin{center}{\small\sloppy\setlength{\tabcolsep}{6pt}
\begin{tabular}{@{}ll@{}}
\hline
Variable & Read by \\
\vfdesc{2}{Description} \\
\hline
\vftestnb{NEVENTS} & \texttt{myLaunch.sh} \\
\vfdesc{2}{Number of events (\texttt{nevents} in the run card).} \\
\hline
\vftestnb{FOLDING} & \texttt{myLaunch.sh} \\
\vfdesc{2}{Folding (\texttt{folding} in the run card, default 8,4,1).} \\
\hline
\vftestnb{MCATNLODELTA} & \texttt{myLaunch.sh} \\
\vfdesc{2}{1: MC@NLO-$\Delta$ matching.} \\
\hline
\vftestnb{PHYSICS} & \texttt{myLaunch.sh} \\
\vfdesc{2}{1: physics run, with the standard MINT0/MINT1 parameters
  and all channels.} \\
\hline
\vftestnb{REQACC} & \texttt{myLaunch.sh} \\
\vfdesc{2}{Required accuracy (\texttt{req\_acc} in the run card).} \\
\hline
\vftestnb{ISEED} & \texttt{myLaunch.sh} \\
\vfdesc{2}{Random seed (default 84100;
  \texttt{myStepTest.sh} uses 84118 for $gg\to t\bar{t}g$,
  so that its short tests include one event
  with the virtual corrections).} \\
\hline
\vftestnb{MLREDUCTION} & \texttt{myLaunch.sh} \\
\vfdesc{2}{MadLoop reduction library: ninja (default) or collier.} \\
\hline
\vftestnb{POLYCACHE} & \texttt{myLaunch.sh} \\
\vfdesc{2}{0: no cache of the compilation of the MadLoop
  \texttt{polynomial.f}.} \\
\hline
\vftestnb{CHANNELS} & Python \\
\vfdesc{2}{Integration channels (e.g. \texttt{10.0}, or \texttt{all});
  each process has a default for the tests.} \\
\hline
\vftestnb{MINT0\_NITERS} & \multirow{3}{*}{Python} \\
\vftestnb{MINT0\_NPOINTS} & \\
\vftestnb{MINT0\_ACC} & \\
\vfdesc{2}{MINT0 iterations, points per iteration and target accuracy.} \\
\hline
\vftestnb{MINT1\_NITERS} & \multirow{2}{*}{Python} \\
\vftestnb{MINT1\_NPOINTS} & \\
\vfdesc{2}{MINT1 iterations and points per iteration.} \\
\hline
\vftestnb{NOSHUFFLE} & Python \\
\vfdesc{2}{Keep the order of the MINT2 events in \texttt{events.lhe}
  (set by \texttt{myLaunch.sh -o} and \texttt{testGridpack.sh},
  for event-by-event comparisons).} \\
\hline
\vftestnb{MINT0\_SAVE} & \multirow{2}{*}{Python} \\
\vftestnb{MINT0\_FROM} & \\
\vfdesc{2}{Directory where the MINT0 results of all channels are saved
  after MINT0, or from which they are copied instead of running MINT0
  (used by \texttt{genGridpack.sh -mint0save} and \texttt{-mint0from}).} \\
\hline
\texttt{ENV\_FVZ\_NDUMP} & \multirow{2}{*}{Fortran} \\
\texttt{ENV\_FVZ\_ITMAX} & \\
\vfdesc{2}{Dumps of the MINT1 integrand for a few points
  (\texttt{myLaunch.sh} enables them in the tests, not in physics runs).} \\
\hline
\end{tabular}}
\end{center}
\vspace*{-3mm}
\caption{Environment variables of the test infrastructure,
which only exist in process directories generated with
\texttt{--vfold-with-tests}:
all have the prefix \texttt{ENV\_\allowbreak{}VFOLDTEST\_}
or \texttt{ENV\_\allowbreak{}FVZ\_},
which distinguishes them from the variables of Table~\ref{tab:envrun}.
The Python ones are read by the modified \mgamc\ Python modules
which run the integration jobs,
the Fortran ones by the test hooks of the Fortran code.}
\label{tab:envtest}
\end{table}

\subsection{Reproducing the timing tables}
\label{app:ref7}

The following instructions
create a new set of timing tables,
called REF7 below,
like those in the 
results sections of this paper,
from process directories generated with MadGraph7,
on a machine called \texttt{GOLD} below.
\begin{enumerate}
\item
Clone the MadGraph7 repository
(tag \texttt{valassi\_\allowbreak{}vfold7on20261127}),
install the CUDACPP plugin
and the external libraries
as described in Sec.~\ref{app:install},
and make sure that \texttt{python3}
is Python 3.12 or higher.
\item
Generate the process directory
with the test infrastructure:
\begin{verbatim}
  ./madvfold/DEV/genProc.sh eebbbb madvfold/DEV/eebbbb_user
  cd madvfold/DEV/eebbbb_user
\end{verbatim}
\item
Run the MINT2-only tests
for folding \efo\ and for no folding
(on a 4-core CPU node with AVX2,
the MINT2-only runs of all modes take
about 20 minutes for the first test
and 10 minutes for the second, after the build;
on a GPU node, the cuda runs with a scan
of \vfenv{VECSIZE\_\allowbreak{}USED} are added automatically,
and on CPUs with AVX512 the 512y and 512z backends):
\begin{verbatim}
  ./vfold/myStepTest.sh --ref REF7/GOLD \
    --tsts 'n200f841k64g1 n2000f111k512g1' step15b
\end{verbatim}
Each test runs all modes,
compares the events and the results
of F1, F2, F and C
(see Sec.~\ref{app:run} for what is expected),
and writes the text timing table, e.g. in
\texttt{Events/\allowbreak{}REF7/\allowbreak{}GOLD/\allowbreak{}n200f841k64g1/\allowbreak{}timingTable\_\allowbreak{}step15b.txt}.
The logs of the runs and of the comparisons are
\texttt{log\_\allowbreak{}step15b\_\allowbreak{}launch.txt}
and \texttt{log\_\allowbreak{}step15b\_\allowbreak{}diff.txt} in the same directory.
\item
Create the LaTeX tables
(with one decimal digit, as in this paper;
\texttt{tex2} gives two):
\begin{verbatim}
  mkdir -p Events/TEX7
  for t in n200f841k64g1 n2000f111k512g1; do
    ./vfold/myTimingTable.py -f tex1 \
      Events/REF7/GOLD/$t/step15b
  done
\end{verbatim}
This writes
\texttt{Events/\allowbreak{}TEX7/\allowbreak{}GOLD\_\allowbreak{}n200f841k64g1\_\allowbreak{}step15b.tex}
and
\texttt{Events/\allowbreak{}TEX7/\allowbreak{}GOLD\_\allowbreak{}n2000f111k512g1\_\allowbreak{}step15b.tex},
which can be included in a paper
with \verb|\input|
(they need the length \verb|\tmtwidth|
to be defined in the main file).
\item
Optionally,
check the physics results
with a full run of all integration channels
in mode C, e.g. without folding:
\begin{verbatim}
  ENV_VFOLDTEST_PHYSICS=1 ENV_VFOLDTEST_REQACC=0.003 \
    ENV_VFOLDTEST_FOLDING=1,1,1 \
    ENV_VFOLD_KPOINT_BATCH=512 ENV_VFOLD_SBORN_GSCALE=1 \
    ENV_VFOLD_ME_MODE=C ./vfold/myLaunch.sh
\end{verbatim}
and compare the cross section in
\texttt{Events/\allowbreak{}run\_\allowbreak{}01/\allowbreak{}LOG.txt}
to that of mode F1
(or to that of standard \mgamc, i.e. mode F0).
\end{enumerate}

\subsection{Gridpacks}
\label{app:gridpacks}

The timing tables of $pp\to t\bar{t}jj$
in this paper
were not produced with the tests
of Sec.~\ref{app:ref7},
but with gridpacks.
Unlike LO, \mgamc\ and MadGraph7
have no gridpack mode at NLO:
a \mvf\ gridpack is a tarball
of a process directory
where MINT0 and MINT1 have been run once
(with \texttt{nevents=0} and an explicit
\texttt{req\_\allowbreak{}acc} in the run card),
from which events are generated
by running only MINT2,
as many times as needed,
for any number of events and any seed.
A gridpack is created with
\begin{verbatim}
  ./madvfold/DEV/genGridpack.sh [-mode <mode>] \
    [-backend <backend>] [-reqacc <acc>] \
    [-folding <f1,f2,f3>] [-j <ncores>] \
    [-savegrids|-usegrids] [-mint0save|-mint0from] \
    <procdir>|<alias> [<gridpack.tar.gz>]
\end{verbatim}
from a process directory generated with
\texttt{--vfold} or \texttt{--vfold-with-tests},
or from one of the aliases of \texttt{genProc.sh}
(e.g. \texttt{pptt}, \texttt{ppttj} or \texttt{ppttjj}
for $pp\to t\bar{t}$, $pp\to t\bar{t}j$ or $pp\to t\bar{t}jj$,
in which case the process directory
\texttt{madvfold/\allowbreak{}DEV/\allowbreak{}<alias>\_\allowbreak{}vfgen}
is generated if it does not exist).
The defaults are mode F0, a required accuracy of 0.01,
folding \efo\ and all the cores of the machine.
The folding is fixed in the gridpack,
while the mode (and the backend in mode C)
are only those used for MINT0 and MINT1:
the gridpack contains the binaries,
the CUDACPP libraries of all the backends that were built
and the external libraries (Ninja, COLLIER),
and it can be run in any mode and backend.
It uses the built-in \texttt{nn23nlo} PDF,
so that it does not depend on LHAPDF.
MINT0 is always run without folding:
with \texttt{-mint0save},
its results are saved
in \texttt{madvfold/\allowbreak{}test/\allowbreak{}grids/\allowbreak{}<proc>},
and with \texttt{-mint0from}
they are reused for another folding,
instead of running MINT0 again
(this needs \texttt{--vfold-with-tests}).
With \texttt{-savegrids},
the MINT0/MINT1 grids are also saved in
\texttt{madvfold/\allowbreak{}test/\allowbreak{}grids/\allowbreak{}<proc>/\allowbreak{}folding\_\allowbreak{}<f1>\_\allowbreak{}<f2>\_\allowbreak{}<f3>};
with \texttt{-usegrids},
MINT0 and MINT1 are not run,
the binaries are only compiled
and the grids are copied from there
(the mode, required accuracy and seed are then those of the grids).

In the unpacked gridpack
(which can be moved anywhere),
events are generated with
\begin{verbatim}
  ./vfold/runGridpack.sh [-mode <mode>] [-backend <backend>] \
    [-vecsize <n>] [-kbatch <n>] [-seed <iseed>] [-j <ncores>] \
    <nevents>
\end{verbatim}
where the defaults are mode C and the AVX2 backend,
and all the other variables of Table~\ref{tab:envrun}
take the values of \texttt{vfold/\allowbreak{}setupVfoldDefaults.sh},
unless they are already set.
The events and the logs are written in
\texttt{Events/\allowbreak{}run\_\allowbreak{}<nn>}.
To compare several modes,
\begin{verbatim}
  ./madvfold/DEV/testGridpack.sh [-j <ncores>] [-kbatch <n>] \
    [-seed <iseed>] [-channels "<P>/GF<c> ..."] \
    <gridpack.tar.gz> <outdir> <nevents> <mode1> [<mode2> ...]
\end{verbatim}
unpacks the gridpack in a new directory for each mode
(F0, F1, F2, F, or C followed by the backend,
e.g. \texttt{Cavx2}, \texttt{C512z} or \texttt{Ccuda512}
for the cuda backend with \vfenv{VECSIZE\_\allowbreak{}USED}=512),
runs it with \texttt{runGridpack.sh}
with the same number of events and seed (default 1001),
keeps the output in \texttt{<outdir>/\allowbreak{}<mode>},
and compares the timings and the events of all modes
with \texttt{myGridpackTable.py},
which can also produce the LaTeX tables of this paper.
For instance,
the tables for $pp\to t\bar{t}jj$ with folding \efo\
can be reproduced
from the MINT0/MINT1 grids committed in the repository
(which took about 21 hours of MINT0 on 4 cores
and 4 hours of MINT1 on 32 cores to compute)
with
\begin{verbatim}
  ./madvfold/DEV/genGridpack.sh -folding 8,4,1 -usegrids ppttjj
  ./madvfold/DEV/testGridpack.sh -j 1 -kbatch 64 \
    madvfold/DEV/ppttjj_gridpack_Cavx2_f841.tar.gz ppttjj_841 \
    1000 F0 F1 F2 F Cavx2 C512z Ccuda512
  ./madvfold/overlay_test/Template/NLO/vfold/myGridpackTable.py \
    -f tex1 ppttjj_841/{F0,F1,F2,F,Cavx2,C512z,Ccuda512}
\end{verbatim}
on a single core (\texttt{-j 1}),
as in this paper,
using \texttt{-kbatch 128} and \texttt{-kbatch 512}
for the foldings $2\times2\times1$ and \ooo.

\end{document}